\documentclass[iop]{emulateapj}
\pdfoutput=1
\usepackage{graphicx}
\usepackage{pdflscape}
\usepackage{subfigure}

\makeatletter

\newcommand{\Rmnum}[1]{\expandafter\@slowromancap\romannumeral #1@}
\makeatother

\def\Lsun{\hbox{\it L$_\odot$}}

\def\Msun{\hbox{\it M$_\odot$}}

\newcommand{\etal}{\mbox{et al.}}

\def\ca{\citeauthor}
\def\cy{\citeyear}

\begin{document}
\title{Red Eyes on Wolf-Rayet Stars: 60 New Discoveries via Infrared Color Selection}

\shorttitle{}

\author{Jon C. Mauerhan\altaffilmark{1}, Schuyler D. Van Dyk\altaffilmark{1}, Patrick. W. Morris\altaffilmark{2}}

\altaffiltext{1}{Infrared Processing and Analysis Center, California Institute of Technology, Pasadena, CA 91125; mauerhan@ipac.caltech.edu}
\altaffiltext{1}{NASA Herschel Science Center, California Institute of Technology, Pasadena, CA 91125}

\begin{abstract}
We have spectroscopically identified 60 Galactic Wolf-Rayet stars (WRs), including 38 nitrogen types (WN) and 22 carbon types (WC).
Using photometry from the \textit{Spitzer}/GLIMPSE and Two-Micron All-Sky Survey (2MASS) databases, the new WRs were selected via a method we have established that exploits their unique infrared colors, which is mainly the result of excess  radiation generated by free-free scattering within their dense ionized winds. The selection criterion has been refined since the last report, resulting in a WR detection rate 
of $\approx$20\% in spectroscopic follow-up of candidates that comprise a broad color space defined by the color distribution of all known WRs having $B>14$ mag. However, there are smaller regions within 
this color space which yield WRs at a rate of $>$50\% in spectroscopic follow-up. Candidates which are not WRs are mainly Be stars, which is possibly attributable to the physical similarities between the free-free emission parameters of Be disks and WR winds. As an additional selection experiment, the list of WR candidates was cross-correlated with archival X-ray point-source catalogs, 
which increases the WR detection rate of the broad color space to $\approx$40\%; ten new WR X-ray sources have been found, in addition to a previously unrecognized X-ray counterpart to a known WR. 
The extinction values, distances, and galactocentric radii of all new WRs are calculated using the method of spectroscopic parallax.  Although the majority of the new WRs have no obvious association with stellar clusters, 
two WC8 stars reside in a previously unknown massive-star cluster, in which five OB supergiants were also identified. The new system lies at an estimated distance of $\approx6.1$ kpc, near the intersection of the Scutum-Centaurus Arm 
with the Galaxy's bar. In addition, two WC and four WN stars, all but one of which are X-ray sources, were identified in association with the stellar clusters Danks 1 and 2.  A WN9 star has also been  
associated with the cluster [DBS2003] 179.  This work brings the total number of known Galactic WRs to 476, or $\approx$7\%--8\% of the total empirically estimated population. An examination 
of their Galactic distribution reveals an approximate tracing of spiral arms and an enhanced WR surface density toward several massive-star formation sites. 

 \end{abstract}
\section{Introduction}
Wolf-Rayet (WR) stars are the evolved descendants of the most massive stars ($M>25$ \Msun; see Crowther 2007 and references therein). They are highly luminous ($L\sim10^5$--$10^6 \Lsun$) and drive ferocious winds, which, when integrated over the entire WR phase, can deposit an amount of mechanical energy comparable to that released by a supernova blast.  As the strong winds of WRs eject hydrogen and helium from their atmospheres, deeper and hotter layers containing the byproducts of core nuclear burning are exposed. The high temperatures and fast winds result in the development of strong, velocity-broadened emission lines of helium, carbon, nitrogen, and oxygen.  Based on the observed relative strengths of these lines, WRs are classified into spectral subtype categories of nitrogen type (WN), carbon type (WC), and oxygen type (WO). These subtypes are believed to follow the general evolutionary sequence: WN$\rightarrow$WC$\rightarrow$WO (e.g., \citealt{langer94}; \citealt{nor98}). As WRs are hydrogen deficient upon death, they are suspected to be the progenitors of Type Ib and Ic core-collapse supernovae and long-duration gamma-ray bursts. With ages of $<$10 Myr, WRs are excellent markers of recent massive star formation. Their population demographics provide insight on the formation mode and distribution of the most massive stars in the Galaxy, while the observed frequencies of WR subtypes as a function of environmental metallicity provide a useful testing ground for evolutionary models (Meynet \& Maeder 2005).

Prior to this work, there were upward of 400 cataloged WRs. Those that lie relatively close to the Sun have bright optical counterparts \citep{vdh01} and this local population is estimated to be complete to $B<14$ mag (Shara et al. 1999). Based on the local WR space density, a simple extrapolation implies that $\approx$6500 WRs populate the Galaxy (\citealt{vdh01}; \citealt{shar99,shar09}). Since the lifetime of WR precursors is $<$10 Myr, they are unable to drift far from their birthplaces in molecular clouds  before the end of their lives. Hence, from the vantage point of Earth, the vast majority of Galactic WRs lie in regions of  high optical obscuration. Fortunately, infrared observations have the ability to unveil WRs in such environments. Large populations of obscured WRs have been revealed in regions of the Galaxy for which astronomical interest is relatively biased, such as the Galactic center  (\citealt{fig99,fig02}; \citealt{mau07,mau09, mau10b}) and extraordinary starburst clusters such as Westerlund 1 \citep{crow06}.  More recently, infrared-based searches for WRs have been extended to the greater Galactic plane. \citet{shar09} have identified 41 obscured WRs via infrared narrow-band imaging and follow-up spectroscopy. Meanwhile, complementary efforts by \citet{had07} and \citet{mau09, mau10a} have utilized the 2MASS and \textit{Spitzer}/GLIMPSE point-source catalogs to develop a very efficient color-based criterion for selecting WR candidates from the field.  This work is a continuation of that effort. We report the detection of 60 new WRs, which brings the current total number of WRs selected by our method to $\approx$90. The color-based criterion has been refined since our last major discovery report in Mauerhan et al. (2009), leading to an enhanced detection rate for WRs and a better suppression of  non-WR ``contaminating" sources. 

\section{WR Selection Criterion}
\subsection{Infrared Color and Brightness} 
WRs are significant sources of infrared excess. Figure 1 demonstrates that they are clearly separable from the vast majority of field stars in $(H-K_s)$ versus $(J-H)$, $(K_s-[8.0])$ versus $(J-K_s)$, and $([3.6]-[8.0])$ versus $([3.6]-[4.5])$ color-color diagrams.  The color excesses are mainly the result of free-free emission from \textit{e}$^-$ scattering off of helium ions in the outer regions of their dense winds (Wright \& Barlow 1975; Cohen et al. 1975).  Broad emission lines and thermal emission from hot dust for a subset of WC stars can provide additional color excess (e.g., see Williams et al. 1987).  To illustrate the extent by which line emission contributes to the infrared excess of WRs, we removed the emission lines from the flux-calibrated spectra of selected objects in our sample and used the STSDAS SYNPHOT package in IRAF\footnote{IRAF is distributed by the National Optical Astronomy Observatories, which are operated by the Association of Universities for Research in Astronomy, Inc., under cooperative agreement with the National Science Foundation.} to generate synthetic $JHK_s$ photometry from the pure-continuum photometry. Figure \ref{fig:cc_lines} illustrates the resulting shift in $(H-K_s)$ and $(J-H)$ color that several WRs experience after their line emission was removed. Selected WN stars from our sample do not undergo a significant color shift as a result of emission lines, which probably explains why WN stars form a relatively tight trend in the $(H-K_s)$ versus $(J-H)$ color space of Figure 1. However, WC stars undergo a significant shift in $H-K_s$ color if their emission lines are removed. The effect is mainly driven by very strong emission from C {\sc iv} and C {\sc iii} in the $K$-band. Relative to the continuum, WR line strengths decrease with increasing wavelength, so, line emission has a much smaller effect on broad-band colors of WRs in the mid-infrared. 

The infrared excesses of WRs shown forms the basis of a selection criterion within the area of sky covered by the 2MASS and \textit{Spitzer}/GLIMPSE surveys of the inner Galaxy ($l=295^{\circ}$--$0^{\circ}$, $0^{\circ}$--$65^{\circ}$; $b=\pm1^{\circ}$; Benjamin et al. 2003). The color space used in this work has been expanded to encompass more distant and highly obscured WR candidates that lie along the reddening vector. Furthermore, we determined that the inclusion of color-magnitude restrictions provides an additional ``sieve" for separating WRs from non-WR ``contaminants".  The color spaces bounded by solid lines in Figure 1 were defined by the color distributions of all known WRs having $B>14$ mag (the Galactic WR sample is already complete to $B<14$ mag; Shara et al. 1999)  We will hereafter refer to the spaces defined by the black lines in Figure 1, collectively, as the \textit{broad color space}. Known WRs of all subtypes occupy this space, including dusty WC stars. There are $\approx$6600 unidentified WR candidates in the space having $K_s<13.1$ mag; these have become targets for spectroscopic characterization. Focusing on the brightest sources with $K\le11.5$ mag, we have experienced a $\approx$20\% detection rate for WRs in the broad color space. In an attempt to increase the WR detection rate, we experimented with selecting candidates from smaller color subspaces in the $(K_s-[8.0])$ versus $(J-K_s)$, and $([3.6]-[8.0])$ versus $([3.6]-[4.5])$ spaces where non-WR ``contaminants" are relatively sparse; we will refer to these color subspaces, collectively, as the ``\textit{WR sweet spot}". As expected, the WR detection rate increased significantly to upward of $\approx$50\%, and the rate of WC type detection also increased notably. 

It is not entirely clear what infrared spectral and extinction properties drive WR colors into the sweet spot or what subtype selection biases result from selecting candidates from this color region. A clue might lie in the fact that the detection rate for WC stars in the sweet spot increased relative to the detection rate using the broad color criterion. The strong line emission from C {\sc iii} and C {\sc iv} in the $K_s$-band could explain the sweet spot in the ($K_s-$[8.0]) versus ($J-K_s$) color space (moving strong-lined WC stars blueward in $K_s$-[8.0] axis and redward in $J-K_s$ axis), but emission-line effects are probably less significant for the $[3.6]-[8.0]$ versus $[3.6]-[4.5]$, because the line strengths of WRs relative to the continuua decreases significantly in the mid-infrared. It is also plausible that the WR sweet spot is produced, in part, by the relative strength of free-free emission for particular subtypes, which could be the cause of the enhanced detection rate of WC types in this subspace.  Whatever the case, it should be noted that representatives from most WN and WC subtypes are found within the WR sweet spot, which suggests that there is no exclusive subtype bias in this particular color space with respect to the broader color space. However, the color distribution of dusty WC stars in Figure 1 suggests that there will be some bias against these particular types in the WR sweet spot. 

Additional work is needed to establish an improved understanding of the WR phenomenology in the infrared and the selection biases and limitations of our survey.  The main goal of this work was to maximize the detection rate for Galactic WRs, in order to best exploit the limited observing time made available for our program. We have, thus far, given precedence to the brightest WRs candidates in our sample ($K\le11.5$ mag), in order to expedite the discovery of new WRs. As a result, we have limited the Galactic volume probed by our survey. For instance, by focusing the spectroscopic sample to objects having $K_s\le11.5$ mag, we are mainly sensitive to WRs lying at distances of $\lesssim$8 kpc, assuming an average $A_{K_s}=2$ and an average absolute magnitude of $M_{K_s}=-5$ for WRs (see Crowther et al. 2006). According to \citet{shar09}, $\approx$2500 Galactic WRs ($\approx$40\%) are expected to have $K>13$ mag. Most of these fainter WRs, which we have not yet sampled in our survey, will lie on the far half of the Galaxy. Thus our survey is strongly biased towards WR candidates on the near side of the Galaxy. 

Most of the non-WR stars selected in our survey exhibit the spectral characteristic of Be stars. These objects exhibit H {\sc i} Paschen and Brackett series emission in their spectra, shown in Figure \ref{fig:be}, which is consistent with the $H$ and $K$ spectral atlases of Be III--V stars from Clark \& Steele (2000) and Steele \& Clark (2001). We suspect the reason that WRs and Be stars have such similar infrared colors is a result of the similarities between the sources of their infrared excesses, which, in the case of Be stars, is free-free emission generated by their ionized circumstellar disks. For both WR winds and Be disks, the free-free flux is a function of electron density ($N_e$), electron temperature ($T_e$), the physical size of the emitting source ($R_{\textrm{\tiny{ff}}}$), and the wavelength at which the emission becomes optically thick ($\lambda_t$). All of these parameters are practically identical for WRs (see Cohen et al. 1975; Mauerhan et al. 2010b) and Be stars (see Gehrz et al. 1974), having typical values on the order of $N_e\sim10^{11}$ cm$^{-3}$, $T_e\sim10^4$ K, $R_{\textrm{\tiny{ff}}}=10^{12}$--10$^{13}$ cm, and $\lambda_t\approx5$ {\micron}. These physical similarities provide a simple and natural explanation for the infrared color spaces that WRs and Be stars share. In addition to Be stars, our method has also selected several candidate evolved B[e] and O supergiants, and we suspect that these objects were also selected because of similarities in the free-free parameters. The discovery of these stars will be presented in a forthcoming paper.

The refinement of the selection method in our most recent survey resulted in a significant improvement over our previous WR return rate of $<$10\% reported in \citet{mau09}. It appears that by including the recent color-magnitude restriction, the frequency of cool, late-type ``contaminants"  has dropped significantly (to $<5$\%) since the surveys of Hadfield et al. (2007) and Mauerhan et al. (2009). The broad color space, thus, returns early-type emission-line stars almost entirely, at a rate of $\approx$95\%.  

\subsection{X-ray Emission: An Additional Selection Tool} 
WRs are commonly X-ray sources (see Skinner et al. 2010 and references therein). The high-energy emission is generated via shocks that form from line-driven instabilities within their winds (Lucy \& White 1980), or from the collision between opposing winds in a massive binary (Luo et al. 1990). 
Thus, as a complementary WR selection technique, we cross-correlated our list of infrared color-selected WR candidates from the broad color space of Figure 1 with archival X-ray 
catalogs that were compiled from multiple Galactic plane observations made with the \textit{Chandra X-ray Observatory} (Evans et al. 2010) and \textit{XMM-Newton} (Hands et al. 2004). We recovered
43 WR-candidate/X-ray matches, including several known WRs that were not previously recognized as X-ray sources.  We obtained spectra for 23 of the candidates and recovered 10 WRs, for a  return rate of $\approx$40\%,  a factor-of-two increase in the return rate using the broad infrared color space alone. The non-WRs in this particular sample are, still, mainly Be stars, but we also identified several candidate YSOs, and late-type stars with infrared excess; these objects will be the subject of a forthcoming paper. The basic X-ray properties of these new WRs are presented in Section 6.

Although X-ray emission proves to be an effective tool for selecting WRs from the list of infrared-color candidates and improving the WR detection rate, there are certainly biases towards certain WR subtypes. For instance, although X-ray emission appears to be ubiquitous property of WN stars (Skinner et al. 2010), most WC stars are undetectable as X-ray sources (Oskinova et al. 2003), unless they are members of colliding-wind binary systems. Furthermore, the X-ray emission from single WRs is typically dominated by a soft X-ray photons having energies below 1 keV, which are heavily absorbed by the interstellar medium. Thus, X-ray selection will be biased toward WRs in regions of lower extinction, or those that emit harder X-rays (e.g., colliding-wind binaries) which are less attenuated by dust and gas along the intervening ISM.  

\section{Infrared Spectroscopy}
Near-infrared spectra of WR candidates was obtained with the Palomar Hale 5 m telescope and the TripleSpec (TSPEC) near-infrared spectrograph during 2009 July 13--15 and 2010 June 21 (UT), the NASA Infrared Telescope Facility (IRTF) and SpeX instrument during 2009 August 4--6 (UT), and the Southern Observatory for Astrophysical Research (SOAR) and OSIRIS Spectrograph during 2010 May 23--27 (UT).  TSPEC provides simultaneous coverage of the $J$, $H$, and $K$ bands ($\lambda=1.0$--2.4 {\micron}) and produces a moderate-resolution spectrum ($R\approx2500$--2700) through a $1\arcsec \times 30\arcsec$ slit. SpeX and OSIRIS provide wavelength coverage similar to TSPEC, although at respective resolutions of $R\approx1200$ through a 0\farcs5 slit and $R\approx800$ through a 1{\arcsec} slit. All spectra were obtained utilizing an ABBA 
telescope nodding sequence for the subtraction of sky emission and to suppress bad pixels. For the TSPEC data, wavelength calibration was performed using the OH emission lines detected in the background sky spectra. For the SpeX and OSIRIS data, spectra from arc lamps internal to the instruments were used for wavelength calibration. TSPEC and OSIRIS images were reduced using standard IRAF routines, and the spectra were extracted using the IRAF package APALL. The SpeX data were reduced using the IDL-based package Spextool \citep{cush04}. For all observations A0V standard stars were observed to obtain a telluric absorption spectrum at an airmass difference of $\lesssim0.1$ relative to the science target.
 The science spectra were corrected for telluric absorption using the IDL program {\it xtellcor} (Vacca et al. 2003), which removes model H {\sc i} absorption lines from the A0{\sc V} spectra before application to the science data. All spectra were flux normalized after modeling the continuum by the fitting of a low-order polynomial.  

Table 1 lists the 2MASS near-infrared positions and photometry of the new WR stars, and the mid-infrared photometry from the \textit{Spitzer}/GLIMPSE database. Table 2 lists the instruments used to obtain the spectra and the dates of observation. 

\section{Analysis and WR Classification}
Figures \ref{fig:wnl_med}--\ref{fig:wc_low} show the NIR spectra of the new WRs; Figures \ref{fig:wnl_med}--\ref{fig:wc_med} present the medium-resolution TSPEC and SpeX spectra, and Figures \ref{fig:wnl_low}--\ref{fig:wc_low} present the lower-resolution OSIRIS spectra. Owing to the relatively low signal-to-noise ratio ($S/N$) of the $J$-band OSIRIS data, we present only $H$ and $K$ spectra for the stars observed with this instrument. WR subtypes were assigned following the criteria of \citet{fig97} and \citet{crow06}, who demonstrated that WN subtypes can be classified quantitatively via the equivalent width (EW) ratios of the following helium emission lines: He {\sc ii} ($\lambda$1.012 {\micron})/He {\sc i} ($\lambda$ 1.083 {\micron}) in the $Y$ band, and He {\sc ii} ($\lambda$2.189 {\micron})/He {\sc ii}$+$Br$\gamma$ ($\lambda$ 2.165 {\micron}) and He {\sc ii} ($\lambda$2.189 {\micron})/He {\sc i} ($\lambda$ 2.112 {\micron}) in the $K$-band. For WN subtypes of WN6 or earlier, the emission of high-ionization N {\sc v}, relative to He {\sc i}/C {\sc iii}/N {\sc ii},  can also be used as a subtype diagnostic \citep{crow06}. WC stars and their subtypes are classified in a manner similar to that of the WN stars via the EW ratios of the following carbon emission lines: C {\sc iii} ($\lambda$0.971 {\micron})/C {\sc ii} ($\lambda$ 0.990 {\micron}) in the $Y$ band, and  C {\sc iv} ($\lambda$2.076 {\micron})/C {\sc iii} ($\lambda$ 2.110 {\micron}) in the $K$ band. For WN stars, decreasing subtype is associated with an increase in the strength of He {\sc ii} emission relative to He {\sc i}; while for WC stars, decreasing subtype is associated with an increase in the strength of C {\sc iv} emission relative to C {\sc iii}, and C {\sc iii} relative to C {\sc ii}. We measured EW values using the SPLOT routine in IRAF, whereby the individual flux contributions of blended emission lines may be recovered if the central wavelengths of the emission are known. However, emission-line ``deblending" was only necessary for the extraction of carbon line strengths in the $K$-band spectra of the WC stars  that exhibit very strong emission lines of C {\sc iv} and C {\sc iii}. Tables 3 and 4 list the EW values for prominent emission lines in the spectra of WN and WC stars, respectively, and the resulting subtype classifications.  

The new sample contains a total 38 WN stars and 22 WC stars. The majority of WN stars have subtypes in the WN6--7 range, exhibiting He {\sc ii} ($\lambda$2.189 {\micron})/He {\sc ii}$+$Br$\gamma$ ($\lambda$2.165 {\micron}) ratios in the range of $\approx$0.4--1.2. Earlier WN stars, such as MDM23 and MDM52, exhibit slightly higher ratios of 1.6 and 2.5, respectively, consistent with earlier subtypes in the range of WN4--5. For the latest WN subtypes, He {\sc ii} is substantially weaker than for the earlier WN types, as indicated by a comparison of their relatively low He {\sc ii} ($\lambda$2.189 {\micron})/He {\sc ii}$+$Br$\gamma$ ($\lambda$2.165 {\micron}) ratios of $\lesssim$0.2. 

Stars MDM8 and MDM19 have very weak features of He {\sc ii} in their spectra. In the $K$-band, MDM8 appears to have this transition in absorption, which may be regarded as a feature more indicative of an O supergiant. However, in the $H$-band He {\sc ii} possibly appears as a weak P Cygni profile at $\lambda$1.692 {\micron}, which would warrant the WN classification. For star MDM19, He {\sc ii} at $\lambda$2.189 {\micron} appears to have a definite absorption component, while there might be some indication of an emission component having a P Cygni profile, although it is too near the noise level to be regarded as a secure detection. Thus, the spectral properties of MDM8 and MDM19 are mutually consistent with hot emission-line supergiants, possibly in transition to late WN stars. As an additional diagnostic, the strength of the Br$\gamma$ emission may help discriminate between O and WN types. Spectroscopic analysis of O and hydrogen-rich WNh and WNha stars in Bohannan \& Crowther (1999) suggest that O supergiants, such as HD 151804, have EW (Br$\gamma$) $\lesssim10$ {\AA}, whereas WN9ha and WN9h stars exhibit higher strengths of $\approx$20 {\AA} and $\approx$50 {\AA}, respectively. Following this rationale, we tentatively classify MDM8 as a WN9h star and MDM19 as a WN9ha star, albeit with caution. Higher $S/N$ data with increased spectral resolution is necessary to assign these stellar classifications with more confidence. 

Star MDM50 has a particularly unique spectrum, which is dominated by emission lines of helium that are relatively narrow, compared with the other WRs. He~{\sc i} at $\lambda$2.058 {\micron} is the strongest line, with FWHM$\approx$300 km s$^{-1}$, just above the 250 km s$^{-1}$ velocity resolution of the $R\approx1200$ spectrum. The lower FWHM values of the other lines in the spectrum are consistent with the velocity resolution limit of the data, which suggests that the lines could be nebular in origin.  In the $K$-band, two resolved lines are apparent near $\lambda$2.165 {\micron}; their centroids lie at $\lambda\lambda$2.1622, 2.1654 {\micron}, which suggests that they could both be He transitions (see \citealt{fig97}, their Table 2), specifically He~{\sc i} $7^{1}L$--$4^{1}D$ at $\lambda$2.1622 {\micron}, and He {\sc i}  $7^{1}D$--$4^{1}F$ at $\lambda$2.1653 {\micron}, although we identify these particular transitions with caution. There is also significant He {\sc ii} emission at $\lambda$2.189 {\micron} and $\lambda$1.012 {\micron}. The $H$-band spectrum of MDM50 appears relatively devoid of any hydrogen emission features, which might also be an indication that the $K$-band emission feature at $\lambda$2.1654 {\micron} is a transition of He and not Br$\gamma$. There is, however, a potential emission line of H {\sc i} ($n$=10-4) at $\lambda$1.736 {\micron}. There are also weak, but significant, features of low-ionization emission lines of Na {\sc i} at $\lambda$2.206 {\micron} and Mg {\sc ii} at $\lambda$2.138 and $\lambda$2.144 {\micron}, implying that the emission at  $\lambda$1.736 {\micron} could also have contribution from Fe {\sc ii} at the same wavelength (e.g., see Morris et al. 1996).  The presence of low-ionization features, accompanied by high-ionization He {\sc ii}, is somewhat puzzling. The diagnostic line ratios we have used for the other WRs in our sample would imply a WN8 spectral type for MDM50. However, the overall character of its spectrum and the narrowness of its spectral lines are not characteristic of any known WN8 star. The near-infrared spectrum is more similar to WR122 (aka NaSt1), classified as Ofpe/WN9 star by Figer et al. (1997), but re-classified as a hotter WR cloaked in an ionized nebula by Crowther \& Smith (1999). The spectrum of MDM50 is inconsistent with either of these phenomena. Ofpe/WN9 stars typically exhibit much stronger emission from H {\sc i}, particularly for the Paschen series in the $H$ band \citep{pmor96}. No H {\sc i} emission has been detected from MDM50. Moreover, the cloaked WR NaSt1 exhibits a very strong thermal excess at near- and mid-infrared wavelengths; MDM50 does not. Finally, as we will demonstrate in the next section, the photometry for this source is also inconsistent with the supergiant luminosity class. 

\subsection{WC Candidates for Thermal Dust Emission}
Late-type WC stars are commonly associated with thermal emission from hot circumstellar dust ($T\sim500$--$1000$ K), which supplements the free-free emission produced by the stellar wind (Cohen et al. 1975; Williams et al. 1987). The significant near-infrared continuum excess can reduce the strengths of emission lines, particularly in the $K$ band. Stars having C {\sc iii} (0.97 {\micron})/ C {\sc iii} (2.11 {\micron}) ratios that are relatively high ($\approx$8--10) may be regarded as candidates for hot dust emission (WCd stars), since there will be more dilution in the $K$ band than in the $Y$ band. If $Y$ band data are not available, stars that have generally weak lines in the $K$ band, as quantified by the sum of He {\sc i} (2.06 \micron) C {\sc iv}$(2.08 \micron)+$C {\sc iii}$ (2.11 \micron)+$He {\sc i} (2.06 {\micron}), can also be regarded as candidates for emission-line dilution by dust. However, stellar continuum emission from an unresolved OB companion can also dilute emission lines of the WR. Thus, the sum of $K$-band line strengths is not a definitive indicator of dust dilution; the C {\sc iii} (0.97 {\micron})/ C {\sc iii} (2.11 {\micron}) ratio is a much more reliable diagnostic. Seven of the WC8--9 stars in our sample have the $Y$-band coverage necessary for the calculation of the C {\sc iii} (0.97 {\micron})/ C {\sc iii} (2.11 {\micron}) ratio. MDM40, MDM49, and MDM60 have relatively high values of 9.8, 8.9, and 8.7. Of those, MDM40 and MDM60 also appear to be outliers having additional excess in the ($K_s-[8.0]$) versus ($J-K_s$) color space shown in Figure 1. There is also marginal evidence of this additional excess in the ($H-K_s$) versus ($J-H$) color space. MDM40 and MDM60 exhibit relatively weak emission lines in the $K$ band, shown in Figure 6, potentially the result of dilution from the thermal dust excess. MDM49 exhibits very strong lines in the $K$ band, however. MDM18, MDM24, MDM26, and MDM58 also have relatively weak lines, yet, examining their colors with reference to Figure 1, these stars do not exhibit excesses with strength comparable to confirmed WCd stars; MDM40 and MDM60 do, however. In conclusion, MDM40 and MDM60 are the only newly identified WC stars to exhibit strong evidence for thermal dust emission, so, we classify them as WC8--9d stars.  For HDM18, HDM24, HDM26, HDM49, and HDM58, the evidence for hot-dust emission is less convincing, so, we classify them as WCd? stars. 

\section{Extinction and Distance}
Using the photometric data from Table 1 and the spectral type classifications in Tables 3 and 4, we computed the interstellar reddening of each source by adopting intrinsic $(J-K_s)_0$, $(H-K_s)_0$, and  $M_{K}$ values for WR subtypes at known distances from Crowther et al. (2006). We calculated the average values for $(J-K_s)_0$, $(H-K_s)_0$, and  $M_{K}$ for four WN7 stars that reside within the cluster Westerlund 1: Stars B, D, G, and P, obtaining $(J-K_s)_0=0.37$, $(H-K_s)_0=0.09$, and  $M_{K}=-5.18$ mag.  The same calculation was performed for the WN6 stars Q and U in that cluster, yielding $(J-K_s)_0=0.37$, $(H-K_s)_0=0.11$, and  $M_{K}=-4.65$ mag. We applied these values to calculate the extinction and distance values of the WN6 and WN7 stars in our sample. For all other WR subtypes we adopted the calibration photometry values from Crowther et al. (2006; their Table A1). Table \ref{tab:phot2} includes the adopted color and $M_{K_s}$ values, which were applied to the observed photometry to calculate the color excesses $E_{J-K_s}$ and $H_{H-K_s}$ caused by interstellar extinction. Employing the extinction relation of Indebetouw et al. (2005), two values of $A_{K_s}$ were calculated, given by \[  A_{K_{s}} = 1.82^{+0.30}_{-0.23} E_{ H-K_{s}} \] and \[  A_{K_{s}} = 0.67_{-0.06}^{+0.07} E_{J-K_{s}}, \] where $E(J-K_{s})=(J-K_{s})_{\textrm{\scriptsize{obs}}}-(J-K_{s})_{0}$, and similarly for the $H-K_s$ colors. The two results were averaged to obtain a final $\overline{A_{K_s}}$ estimate. Assuming values of $M_{K}$, based upon WR subtype, we calculated the distance modulus, DM, and the physical distance, $D$, to each WR star by using the following expressions \[\textrm{DM}=K_s-M_{K_s}-A_{K_s} \] and \[ D~(\rm{pc})=10^{(DM+5)/5}. \] \noindent{Assuming} a distance to the Galactic center of $R_0=8$ kpc (Reid 1993), we also calculated the galactocentric distance of each star, given by 

 \[  {R_G}^2=R_0^2+D^2~cos^2(b)-2R_0~D~cos(l)~cos(b), \] 
 
\noindent {where} $l$ and $b$ are the Galactic longitude and latitude, respectively. The results of these calculations are listed in Table \ref{tab:phot2}. The uncertainties are significant. Following a rationale similar to that made by Shara et al. (2009), we attribute the bulk of the distance errors to the adopted average values of $M_{K_s}$ from Crowther et al. (2006), which, although assigned to given subtype groups, exhibit significant scatter within those groups ($\approx$0.5--1 mag, depending on subtype). There is additional uncertainty as a result of scatter in the adopted intrinsic colors (which is related to the intrinsic scatter in absolute magnitude for each subtype bin, and is a well known problem for hot stars with strong winds) and in the adopted extinction law. It can be argued that the extinction properties along any highly reddened individual sightline are not well represented by a mean extinction curve; however, the differences in the redundant calculations of $A_{K_s}$,  derived separately for the $J-K_s$ and $H-K_s$ colors, can be used as an indicator of the accuracy of our results. The average difference between these two values of $A_{K_s}$ is $\approx$0.14 mag for WN stars and $\approx$0.23 mag for WC stars, which is significantly less than the potential scatter in the adopted values of $M_{K_s}$. Based on these considerations, the distance errors are, in general, estimated to be $\approx$25\%, although they might be as high as $\approx$40\% in some cases. The WC9 stars, in particular, may have the largest spread in absolute photometry if any of these stars are accompanied by a hot-dust excess. However, most of the WC8--9 stars that we have found lie in the WR sweet spot. Dusty WC stars should exhibit an additional color shift in these diagrams. Therefore, it is reasonably safe to assume that many of the WC8--9 stars we have found can be regarded as not significantly dusty. We adopt $M_{K_s}=-6.3$ mag for non-dusty WC9 stars, which is consistent with the measured values for Star E in the cluster Westerlund 1 (Crowther et al. 2006b) and qF 76 in the Quintuplet cluster (Figer et al. 1999). Still, we present the distances for all WC9 stars with caution. 

The derived extinction and distance for MDM50 is inconsistent with its preliminary classification as a WR. Although the spectral morphology and line ratios suggest a late WN spectral type, the large distance of $\approx$12 kpc that results from assigning the associated luminosity is very suspect, as it is grossly inconsistent with the presence of the bright optical counterpart ($B=14.2$ mag) and the low value of extinction  ($A_{K_s}=0.37$ mag) derived for this source. Instead, the relatively blue color suggests that this object lies much closer to Earth and, as such, is intrinsically much less luminous than a late-type WN star. We will resume discussion of this object in Section 7.2.  

\section{X-ray Detections}
Nine of the WRs presented in this work are associated with X-ray point sources found in the \textit{Chandra} archive (Evans et al. 2010). Table \ref{tab:x} lists the X-ray sources and their broadband fluxes. Four sources lie within the field containing the young clusters Danks 1 and 2 (Danks et al. 1984; Georgelin et al. 1988); this field also contains the known colliding-wind binary WR48a (Zhekov et al. 2011). The WN stars HDM9 from Hadfield et al. (2007) and WR19a (van der Hucht et al. 2001) were also identified in the \textit{Chandra} archive, and their basic X-ray data are presented in Table 6 as well.  

WN-type stars typically exhibit an admixture of emission components from cool ($kT<1$ keV) and hot ($kT>2$ keV) thermal plasmas (Skinner et al. 2010). The cool plasma is predicted by wind shock models (Lucy \& White 1980), while the hotter component requires wind collision in a binary (Luo et al. 1990) or some unidentified mechanism. WC stars in binaries are typically X-ray sources, while in solitude they are almost never detected, probably as a result of high opacity of WC winds to softer X-rays (Oskinova et al. 2003; Skinner et al. 2006, 2009). With the exception of the brightest colliding-wind binaries, the derived X-ray luminosities for WN stars approximately scale with the stellar bolometric luminosity ($L_{\textrm{\tiny{X}}}\sim10^{-7}L_{\textrm{\tiny{bol}}}$) and, perhaps more importantly, with wind momentum (Skinner et al. 2010); the establishment of this relationship suggests that intrinsic X-ray generation is a ubiquitous feature of WN stars. Bona fide colliding-wind binaries, however, commonly exhibit X-ray luminosities above the trend, sometimes by $\approx$1--2 orders of magnitude, and typically exhibit a harder X-ray spectral component (e.g., see Zhekov et al. 2011, Anderson et al. 2011).

To estimate the X-ray luminosities for the WR X-ray sources of our sample, we first converted the derived value of $A_{K_s}$ from Table \ref{tab:phot2} to an equivalent hydrogen absorption column, using the relation from Cardelli, Clayton, \& Mathis (1989; $A_V=8.9A_K$), and the relation from Predehl \& Schmitt (1995; $N_{\textrm{\tiny{H}}}/A_V=1.79\times10^{21}$ cm$^{-2}$ mag$^{-1}$). We then assumed that the X-ray emission from these WRs is thermal, having an average temperature somewhere between 0.6 and 2 keV, consistent with range exhibited by WRs (e.g., see Skinner et al. 2010 and references therein).
Using the PIMMS\footnote{http://heasarc.gsfc.nasa.gov/Tools/w3pimms.html} software package, we then calculated the unabsorbed 0.5--7 keV X-ray flux for the WRs and derived the resulting luminosity by applying our derived physical distances from Table \ref{tab:phot2}. The results are presented in Table \ref{tab:x}.  Most of the WRs have X-ray luminosities in the range of $L_{\textrm{\tiny{X}}}\sim10^{31}$--$10^{33}$ erg s$^{-1}$, which is typical (Mauerhan et al. 2010b, Skinner et al. 2010). 

We also derived the X-ray luminosity for WR48a, obtaining $L_{\textrm{\tiny{X}}}=(3$--$19)\times10^{34}$ erg s$^{-1}$. This is comparable to the value independently obtained by Zhekov et al. (2011), who modeled the X-ray spectrum of this object and derived $L_{\textrm{\tiny{X}}}=(8$--$34)\times10^{34}$ erg s$^{-1}$ from a fit to a 2-component thermal plasma having $kT_{\textrm{\tiny{cool}}}=1$ keV ($\approx$60\% of the unabsorbed flux) and $kT_{\textrm{\tiny{hot}}}=3$ keV ($\approx$40\% of the unabsorbed flux). Their model result implies that the average plasma temperature is indeed between 1--2 keV, as we have assumed. The mutual consistency implies that our simple method of approximating X-ray luminosities is sound.

An X-ray counterpart to the WN7h star WR19a (van der Hucht 2001) was also identified in our cross-correlation of WR candidates with the \textit{Chandra} archive. Using the 2MASS photometry ($J=9.08$, $H=8.13$, $K_s=7.50$ mag) and intrinsic photometry from Crowther et al. (2006; $[J-K_s]_0=0.13$, $[H-K_s]_0=0.11$, $M_{K_s}=-5.92$ mag), we calculated an extinction of $A_{K_s}=0.96$ mag ($A_V=8.6$ mag) and a spectrophotometric distance of 3.1 kpc, which implies $L_{\textrm{\tiny{X}}}=(2$--$10)\times10^{32}$ erg s$^{-1}$. Adopting a bolometric correction of BC$_{K_s}=-4.4$ mag, which is the implied average value for two  WN7--8h stars in Martins et al. (2008), we obtained $L_{\textrm{\tiny{bol}}}=10^{6.0}$ {\Lsun} and Log($L_{\textrm{\tiny{X}}}/L_{\textrm{\tiny{bol}}}$) between $-6.6$ and $-7.3$.  Shara et al. (1991) measured a visual extinction of $A_V=8.1$ mag for WR19a, similar to our value, but calculated much larger possible distances of 11.2 and 15.2 kpc, with their uncertainty being mainly the result of the $M_V$ and $B-V$ colors that they adopted. Applying their distances does not have a significant effect on the resulting $L_{\textrm{\tiny{X}}}/L_{\textrm{\tiny{bol}}}$ ratio, but implies an X-ray luminosity that is $\approx$16--25 times higher than our value in Table \ref{tab:x}, which is still within the range of X-ray luminosities exhibited by WRs. Thus, the X-ray properties do not effectively constrain the distance to WR19a. Still, the discrepancy between our distance to WR19a and that of Shara et al. (1991) is somewhat puzzling.
To derive our distance value, we adopted $M_{K_s}=-5.92$ mag from Crowther et al. (2006), which is the average value of four WN7--9h stars that exhibit individual $M_{K_s}$ values between $-4.89$ and $-6.83$ mag. If either of the distances derived by Shara et al. (1991) are correct, it would imply $M_{K_s}=-8.71$ or $-9.38$ mag; both values are anomalously luminous for a WR in the infrared. So, either WR19a is unusually luminous for a WN7h star, or the distances in Shara et al. (1991) are significantly overestimated.

\section{Discussion}
\subsection{WR Distribution and Association}
Figure \ref{fig:wrdist} illustrates the galactocentric distribution of all known WRs with respect to the Galactic model of \ca{church09} (\cy{church09}); Figure \ref{fig:lplot} illustrates the longitudinal distribution of these WRs; and, Figure \ref{fig:rg} illustrates their galactocentric radial distribution. Table \ref{tab:wr_census} lists the current census of the Galactic WRs included in these figures. The Galactic distribution in Figure \ref{fig:wrdist} is likely to be affected by significant scatter, which is certainly a partial consequence of the distance uncertainties. Nonetheless, an overall concentration toward major spiral arms can be seen, while several regions of enhanced WR density also appear to punctuate the longitudinal distribution. However, we stress caution while interpreting Figures \ref{fig:wrdist}--\ref{fig:rg}, since the coverage of WR surveys is not evenly distributed across the Galactic plane. Still, it is noteworthy that, although we (i.e., Hadfield et al. 2007, Mauerhan et al. 2009, and this work) have sampled more of our WR candidates from the side of the Galactic plane that is north of the Galactic center ($l=10^{\circ}$--$65^{\circ}$), we have experienced a slightly, but significantly, higher WR return rate for the southern side of the Galactic center. We attribute this to the Galactic spiral arm distribution with respect to our line of sight through the Galactic plane. For example, the Scutum-Centaurus Arm, one of the two main Galactic arms illustrated in Figure \ref{fig:wrdist}, and the Sagittarius Arm, both run approximately parallel to our line of sight between $l\approx300^{\circ}$--$330^{\circ}$ and $l\approx280^{\circ}$--$295^{\circ}$, respectively.  As a result of this, there is more spiral arm coverage per unit longitude on the southern side of the Galactic center than on the northern side, at least when considering the near-half of the Galaxy, to which our survey has mainly been limited. Since massive stars are presumably concentrated in spiral arms, we should expect a higher observed surface density of WRs on the southern side of the Galaxy than the northern side. Thus, although we approach the result with caution, owing to the non-uniformity and incompleteness of WR surveys across the Galactic plane, we are confident that the peaks of enhanced WR surface density in the longitudinal distribution of Figure \ref{fig:lplot} are manifestations of real physical structure in the Galaxy. 

The largest WR  concentration is apparently toward the direction of the starburst cluster Westerlund 1, near $l\approx335^{\circ}$--$340^{\circ}$; this line of sight pierces the Scutum-Centaurus Arm perpendicularly, and beyond that, runs approximately parallel to the Norma Arm of the Galaxy, which contains the $\approx$24 WRs within Westerlund 1 and $\approx$10 more-distant, field WRs that lie $\approx$3--4 kpc from the Galactic center. We experienced our highest WR return rate in this region of longitude; during a particularly remarkable interval of observing time, while surveying WR candidates from the WR color ``sweet spot," near $l\approx340^{\circ}$, we confirmed 6 WRs consecutively, without interruption from non-WR ``contaminants." Based on this experience, we regard our remaining $\approx$10 unobserved WR candidates from the sweet spot that lie near this longitude as particular promising.

Although not selected from our color-based survey, there is a strong concentration of WRs toward the Galactic center. The Central Molecular Zone, which comprises the central half-kiloparsec of the Galaxy, contains 92 known WRs (see Mauerhan et al. 2010a and references therein), $\approx$70\% of which are contained within the three known starburst clusters in the region: the Arches, Quintuplet, and Central clusters. The region as a whole contains $\approx$20\% of the known Galactic WR population. For this region of the Galaxy, \textit{Spitzer} and 2MASS observations suffer from extreme stellar confusion and very high extinction ($A_V>30$ mag), which often prevents the detection of a $J$-band counterpart in 2MASS. The resulting degradation of point-source photometry has limited our color-based selection to identify only one of the 92 WRs known in this region. Thus, our method of WR selection is currently not effective in the Galactic center or fields of comparable stellar confusion and extinction, such as, much of GLIMPSE-II.

For the Galactic plane on the northern side of the Galactic center there are also several noteworthy concentrations of WRs. There is a peak near $l\approx30^{\circ}$ in Figure \ref{fig:lplot}. This line of sight pierces the base of the Scutum-Centaurus Arm, where it intersects the Galaxy's Long Bar, and also contains the W43 starburst region. Vigorous star formation has occurred in this part of the Galaxy over the last $\approx$20 Myr, as evidenced by the presence of the WR-bearing cluster W43 (Blum et al 1999), four clusters of red supergiants between $l=24^\circ$--$29^\circ$ (\ca{fig06}~\cy{fig06}; \ca{dav07}~\cy{dav07}; \ca{al09}~\cy{al09}; \ca{cl09}~\cy{cl09}; \ca{neg10}~\cy{neg10}), and an additional WR-bearing cluster at $(l,  b)=(28\fdg46, 0\fdg32$; Mauerhan et al. 2010c). Thus, the relatively high concentration of WRs here is not surprising.

Figures \ref{fig:k1}--\ref{fig:k4} show the 2MASS $K_s$-band field images for new WRs that do not appear to be associated with a local ($\lesssim$2{\arcmin}) stellar density enhancement (exceptions will be discussed in Section 7). The relative ``isolation" of the new WRs motivates the question as to whether all WRs form exclusively in clusters and associations or form via an alternate mode of relatively isolated birth as well. It is possible that some of the WRs originated within clusters and associations, but were subsequently ejected dynamically, either by inter-cluster gravitational encounters or following the supernova explosion of an erstwhile companion (Moffat \& Isserstedt 1980; Moffat et al. 1998). Indeed, it has been noted that approximately half of all O stars exhibit kinematical evidence for runaway velocities (de Wit et al. 2004, 2005, and references therein; Dray et al. 2005), and the same might be true for some of the WRs presented here, since they are descended from O stars. 

We must also consider the possibility that stellar clusters of relatively low density and mass, when compared to massive, starburst systems, such as Westerlund 1, can persist undetected as a density enhancement because of background confusion. It is likely that many elusive clusters could populate the Milky Way, while being undetectable within existing near-infrared surveys, such as 2MASS. Such clusters have been referred to as ``false negative" clusters (Popescu \& Hanson 2009; Hanson et al. 2010b) In the following subsection, we present the discovery of WRs in one such false negative cluster that was not found by any previous cluster search algorithms. 

\subsubsection{A New Stellar Cluster at G37.51$-$0.46}
Figure \ref{fig:newcluster} shows the $K_s$-band field image of the WC8 stars MDM53 and MDM54, which are separated by $\approx$7{\arcsec}. Their respective distances of 6.2 and 6.0 kpc in Table 5 are consistent with equivalent distance, given the $\approx\pm$0.5 mag uncertainty in the adopted value of $M_{K_s}$ that we noted in Section 5, which results in a $\approx \pm$25\% uncertainty in distance. The average distance of $\approx$6.1 kpc, implies a physical separation of $\approx$0.2 pc. The stars lie within a slight stellar density enhancement at  $(l, b)=(37\fdg51, -0\fdg46)$  that we noticed in field acquisition images with TSPEC while observing at Palomar Observatory.  This possible cluster was not selected by any previous searches (i.e., Dutra \& Bica 2001; Mercer et al. 2005), and so the system qualifies as a false-negative cluster with respect to those searches. Motivated by the reasonable suspicion that these WC8 stars occupy a massive stellar cluster, we spontaneously acquired near-infrared spectra of the brightest neighboring field stars, which are also marked in Figure \ref{fig:newcluster}. Table \ref{tab:newclusterphot} lists the near-infrared photometry, estimated spectral types, extinction values, and distances for the surrounding stars. Figure \ref{fig:newclusterspec} presents their $K$-band spectra.  

Stars A, B, D, F, and G exhibit $K$-band spectral 
characteristics that are very similar to the B0I--B3I stars in cluster Sgr 1806$-$20 (Bibby et al. 2008), which have $(H-K_s)_0=-0.08$. Adopting this value for the stars in cluster  G37.51$-$0.46, we derived extinction values for stars A, B, D, F, and G, and  present them  in Table \ref{tab:newclusterphot}. The average of the spectroscopic parallactic distances of the neighboring WC8 stars MDM53 and MDM54 (6.1 kpc) implies $M_{K_s}$ values in the luminosity class of I--II for stars A, B, D, F, and G.  

Stars C, E, and H exhibit the CO bandhead characteristic of late-type M stars.  Their derived extinction values are significantly below those of the early-type stars, which implies that they lie in the foreground and are unrelated to the system. Thus, they are not likely to be cool supergiants within the cluster. Even if we were to assuming intrinsic colors of M0Ia stars, which are the bluest subtypes of the MI class, the extinction values still remain significantly lower than the values for the OB stars. Thus, the available data suggests that the late-type stars C, E, and H lie in the cluster foreground.

The distance of 6.1 kpc places the G37.51$-$0.46 cluster within the massive-star formation region near the intersection of the Scutum-Centaurus Arm and Galaxy's Long Bar (\ca{church09}~\cy{church09}), where we noted earlier the enhancement in the WR surface density and the presence of numerous massive stellar clusters. Additional 
spectroscopy of surrounding stars and photometric analysis of the field will be required to determine the full extent of the evolved massive-star population in this false-negative cluster.

\subsubsection{New WRs Near Danks 1 and 2 and Implications}
A relatively high concentration of WR candidates were found in the vicinity of the G305 H {\sc ii} region, which contains the known stellar clusters Danks 1 and 2 (Danks et al. 1984; Georgelin et al. 1988), shown in Figure \ref{fig:danks}. A spectacular three-lobed wind-blown bubble, known to surround the clusters, exhibits evidence for vigorous, triggered massive star formation in the form of  H$_2$O, OH, and methanol masers that are coincident with numerous, embedded, infrared sources (Clark \& Porter 2004). Several WRs are already known to exist within the G305 complex, including the colliding-wind binary WR48a (WC6+OB) and another WC8 star identified by Mauerhan et al. (2009). We obtained spectra of 7 additional candidates during May 2010, shown in Figure \ref{fig:danks}. Stars MDM4 and MDM6 have WC8 and WC7 spectral types, respectively, while stars MDM3, MDM5, MDM7, and MDM8 have spectral types in the WN8--9 range.  Only one of the candidates we observed in this region was revealed to be a  Be ``contaminant." Four additional candidates remain unobserved in the field covered by Figure \ref{fig:danks}, yet they are significantly fainter than the confirmed WRs in this region, which is somewhat inconsistent with them being WRs within the G305 complex. If they are members of this system, it is, perhaps, more likely that they are Be stars. They have the 2MASS designations J13120279$-$6243224 ($K_s=12.6$ mag), J13121640$-$6239543 ($K_s=12.4$ mag), J1321223498$-$621456 ($K_s=11.3$ mag), and J13125377$-$6238308 ($K_s=11.9$ mag). Of these unidentified WR candidates, we suspect that 2MASS J1321223498$-$621456  has the highest likelihood of being a WR, since it is the brightest of the remaining candidates and closest to the main cluster center in Figure \ref{fig:danks}.

The Danks 1 and 2 clusters lie at an estimated distance of $\approx$3.5 kpc, behind an extinction screen of $A_K\approx1$ mag (Bica et al. 2004). Our derived extinction and distances for the newly confirmed stars in Figure \ref{fig:danks} average out to $A_{K_s}=1.2\pm0.3$ mag and $D=3.6\pm1.1$ kpc, consistent with the previously published values for the cluster. However, several of the individual values differ significantly from the average. Our derived distance of 6.9 kpc for the WC7 star MDM6 is particularly deviant, having nearly twice the accepted value for the clusters; yet, it has an extinction value that is consistent with the value for the region. Thus, either MDM6 lies at a distance significantly beyond the clusters and is not physically associated with them, or the adopted value of $M_K=-4.59$ mag from Crowther et al. (2006) used for WC5--7 stars is unsuitably bright for this particular star. In fact, assuming that the infrared colors that we used are applicable, MDM6 is expected to have $M_K=-3.1$ mag if it lies at the accepted cluster distance of 3.5 kpc (Clark \& Porter 2004). The discrepancy cannot be the result of unaccounted-for dust emission, since the additional reddening and $K_s$-band flux that are associated with hot dust emission would lead to an underestimation of distance, not an overestimation. The same would be true if there were a bright, unresolved companion contributing to the flux. Thus, we are forced to consider the possibility that WC6 stars exhibit a broader range of $M_{K_s}$ values than we have anticipated. If we exclude MDM6 from the calculation of average distance to G305, then we obtain a value of 2.9$\pm$0.8 kpc ($\approx28$\%), which implies that our anticipated, average distance uncertainty of $\approx$30\% using the spectroscopic parallax method is a reasonable estimate.

Including the WC6$+$OB binary WR48a, the results of our survey bring the total number of known WRs associated with the G305 H {\sc ii} region to eight objects.  Clark \& Porter (2004) suggested that the Danks 1 and 2 clusters have succeeded in clearing away the majority of their natal material, implying that significant photon leakage is taking place in tandem with the ionization of the surrounding three-lobed bubble. This suggestion, coupled with the integrated radio flux emanating from the region, led those authors to the conclusion that a large population of massive stars having a total Lyman continuum (Lyc) photon output greater than the equivalent of 31 O7V stars is likely to populate this environment, supplying at least $3.2\times10^{50}$ Lyc s$^{-1}$. The new WRs presented here, in addition to massive stars presented in Clark et al. (2011), certainly represent a significant fraction of the population presumed to exist by Clark \& Porter (2004). In fact, by adopting $N_{\textrm{\tiny{Lyc}}}$ values of WR subtypes from Figer et al. (1999), we approximate that the eight WRs shown in Figure \ref{fig:danks} supply $3.3\times10^{50}$ Lyc photons s$^{-1}$ (averaging over subtypes from WRs in Figer et al. 1999, we find that log$~N_{\textrm{\tiny{Lyc}}}\approx49.8$, 49.7, and 49.5 for WC6--7, WC8, and WN8--9 stars, respectively). Therefore, the ionizing flux of G305 inferred by Clark \& Porter (2004) could now be fully accounted for. Of course, owing to the natural photon leakage out of the G305 complex and the obvious presence of candidate supergiant stars that remain unidentified within the clusters, there is undoubtedly more Lyc flux being generated in this region than we have currently estimated.

It is interesting to find that five of the WRs associated with G305 are relatively well separated from the centers of the Danks 1 and 2 clusters.  MDM3, 4, 5, 6, and WR48a, as a group, appear to be centered on Danks 1, and are separated from the cluster by $\approx$20{\arcsec}, implying a physical separation of $\approx$20 pc (assuming the cluster distance of 3.5 kpc). The distribution of these stars is rather puzzling. As the most highly evolved massive stars in the G305 system, these WRs presumably have the most massive progenitors, whose orbits should have relaxed near the cluster center as a result of dynamical friction. However, within the dense cluster centers there is a relatively high probability of gravitational encounters between other massive stars and binaries, which leads to an increased probability of orbital ``heating" or cluster ejection. This could explain the extended distribution of MDM3, MDM4, MDM5, MDM6, and WR48a.  We speculated earlier that such processes provide one potential explanation as to why so many of the new WRs presented in this work appear to be ``isolated" objects, possibly runaways from their birthplaces in clusters (Moffat \& Isserstedt 1980; Moffat et al. 1998). The WRs surrounding the Danks clusters are particularly interesting in this regard, as they could be objects in the early phase of running away from their birth sites. 
 
\subsubsection{A New WN9 Star Near Cluster [DBS2003] 179}
The WN8--9 star MDM32 is apparently associated with the stellar cluster [DBS2003] 179 (Dutra \& Bica 2003), which is separated from the star by $\approx$16{\arcsec}. The field shown in Figure \ref{fig:k3} includes stars MDM32 (near field center) and the cluster (lower-right of field center). Borissova et al. (2008) identified seven O stars in the cluster, having luminosity classes ranging from dwarfs to supergiants, in addition to an Ofpe/WN9 star (Object 4 from their paper). Those authors estimated a distance of 7.9 kpc and an average extinction value of $A_V=16.6$ mag, which corresponds to $A_K=1.9$ mag, assuming the Cardelli, Clayton, \& Mathis (1989) law. Although our extinction value of $A_K=2.0$ mag for MDM32 is consistent with the value estimated for the cluster by Borissova et al. (2008), our measured distance of $D=6.4$ kpc is significantly lower than theirs, which they measured by averaging spectroscopic parallax distances for the O supergiants in the cluster.   If their distance of $7.9\pm1.1$ kpc is the correct value for MDM32, then it should have $M_K\approx-7.3$ mag, which is notably higher than than the adopted value of $M_K=-6.87$ mag for the brightest late WN stars from Crowther et al. (2006). This discrepancy is not surprising, given the potential $\approx$0.5--1 mag uncertainty in $M_{K_s}$ that we discussed in Section 5.  The distances to the Ofpe/WN9 star Object 4 from Borissova et al. (2008) and MDM32 are likely to be similar; although saturated in $H$ and $K$, the $J$-band and mid-infrared photometry for Object 4 is very similar to MDM32, which is evidence favoring mutually close proximity to cluster [DBS20003] 179.  In conclusion, we physically associate MDM32 with this cluster, and our favor leans toward the distance of 6.4 kpc that we estimated for the system. However, more observations will be necessary to establish a more accurate distance.

\subsubsection{Identification of a WN6 Star in Cluster [MFD2008]}
The WN6 star MDM34 lies within the cluster [MFD2008], discovered by Messineo et al. (2008), who associated the system with the high-energy gamma-ray source HESS J1813$-$178. Those authors recently conducted a more thorough spectroscopic survey of candidate cluster members (Messineo et al. 2011), identifying MDM34 (their star \#7), in addition to almost two dozen other hot massive stars, and a red supergiant.  They classified MDM34 as a WN7 star, although the value EW (2.189 \micron)/ EW (2.166 \micron)=1.5 and the significant detection of N {\sc v} emission at $\lambda$2.10 {\micron} favor the earlier WN6 subtype that we have assigned.  Our measured extinction of $A_{K_s}$=0.88 mag and distance of 3.1$\pm$0.8 kpc to MDM34 is consistent with their extinction value and average spectro-photometric distance ($A_{K_s}=0.83$ mag; $3.6\pm0.7$). Messineo et al. (2011) also derive an X-ray luminosity of $1.3\times10^{32}$ erg s$^{-1}$ for MDM34, which is consistent with our estimated range of $(2$--$22)\times10^{31}$ erg s$^{-1}$. 

\subsection{The uncertain nature of MDM50}
The photometry of MDM50 implies that this star's light suffers relatively low extinction and, thus, lies at a relatively small distance, suggesting that MDM50 is significantly less luminous than a typical WR or evolved supergiant.  Furthermore, its narrow lines imply a relatively weak, low-velocity wind with $v\lesssim$250 km s$^{-1}$, set by the resolution limit of our spectrum. Yet, the significant detection of multiple transitions of He {\sc ii} presented in Figure 3 implies high excitation, and an effective temperature that would place MDM50 within the O spectral class. Yet, the detection of weak, low-ionization features implies a separate, cooler emission component, which could arise from circumstellar, nebular material. MDM50's location in the color-color and color-magnitude plots in Figure 1 suggest that, although this star is one of lowest extinction stars in our sample, its infrared excess
is consistent with the color distribution of the non-WRs (presumed Be stars) and WNs, although it lies on the blue edge of the broad color space. Thus, although we do not know the intrinsic color of MDM50, it is probably reasonable to presume that its intrinsic excess is similar to the bluest WRs in our sample, which have $J-{K_s}\approx0.2$ mag and $H-{K_s}\approx0.1$ mag. If so, then applying the method of extinction calculation from Section 6 yields ${A_{K_s}}=0.38$ mag, using both the $J-K_s$ and $H-K_s$ colors.  Assigning $M_K$ values ranging from O dwarfs ($\approx-4$ mag) to supergiants ($\approx-5.5$ mag) we obtain possible distances of 3.3 kpc and 6.5 kpc, respectively. Examining the distances of eight WRs in Table \ref{tab:phot2} that have $A_{K_s}\le1$ mag shows that they have an average distance of 2.7 kpc with a standard deviation of 0.8 kpc. Therefore, the photometry is inconsistent with the supergiant luminosity class, since an extinction of  ${A_{K_s}}=0.4$ mag would be anomalously low for a distance of 6.5 kpc.  The $M_K\approx-4$ mag is more appropriate, and is consistent with a mid-to-late OV star ($L\sim10^5$ \Lsun).

Moving forward with the O-star hypothesis, MDM50 is similar in $K$-band spectral morphology to the variable O6V:pe star HD 39680 (note that the ``pe" qualifier is He {\sc i} emission), which is dominated by He {\sc i} emission at $\lambda$2.058 {\micron}, while also exhibiting weak He {\sc ii} at $\lambda$2.189 {\micron} (Hanson et al. 1996). HD 39680 is a low-amplitude variable star, exhibiting small fluctuations of amplitude 0.13 mag (\ca{lef09} \cy{lef09}). MDM50 is also photometrically variable. Eight observations of calibration fields for the Sloan Digital Sky Survey spaced throughout 1 year revealed MDM50 to be a variable star exhibiting variations of $\approx$0.1--0.2 mag at $R$ band (Henden \& Stone 1998), similar to the variability amplitude of HD 39680.   It is, therefore, plausible that MDM50 is also an OV:pe star like HD 39680. If so, the reason for selection of this star by our color criteria is unclear. It may be the result of excess emission from an ionized, free-free emitting circumstellar disk similar to those surrounding the Be class (Negueruela et al. 2004). It is plausible that the low-ionization features of Na {\sc i} and Mg {\sc ii} that we detected in the MDM50's spectrum are generated in such a disk, especially since we do not expect these features to arise from the same emission component as the high-excitation He {\sc ii} lines. As there is currently no firm evidence for the existence of ionized disks around Oe stars (Vink et al. 2009), the confirmation of such a structure around MDM50 would be particularly interesting. 

We also noted in Section 4 the spectral similarity between MDM50 and NaSt1, the latter of which has been classified as WR cloaked in a massive $\eta$ Car-like nebula (Crowther \& Smith 1999). NaSt1 also exhibits relatively narrow, nebular emission lines, including He {\sc ii} emission. However, the mid-infrared SED of NaSt1 exhibits a very large thermal excess from dust, while MDM50 exhibits a far weaker excess component, which is inconsistent with the presence of a massive dusty envelope. 

The spectrum of MDM50 shares some similarities with the spectra of Luminous Blue Variable (LBV) stars in their hot state (e.g., see Clark et al. 2011 and references therein).  However, the absence of significant H {\sc i} emission features in our infrared spectra is somewhat inconsistent with a LBV classification. Still, given that MDM50 is a confirmed variable emission-line star, its connection to the LBV phenomenon is worthy of speculation, and we should not rule out this possibility. If it is a quiescent LBV, it would have to be one of the least luminous such stars known, having $L\sim10^{5}$ {\Lsun}. 

In conclusion, although we regard the Oe classification for MDM50 as most appropriate, given the available data,  the true nature of this star remains uncertain, and it deserves more detailed spectroscopic investigation and photometric monitoring. 

\section{Summary and Conclusions}
We have spectroscopically detected 60 new Galactic WRs using the method of infrared color selection, and also by cross-correlation with X-ray point source catalogs. Our WR detection rate of $\approx$20\%, using the broad-color zones defined in Figure 1, is a significant improvement over previous efforts, which had WR detection rates below 10\% (Hadfield et al. 2007; Mauerhan et al. 2009). The improvement is attributable to our refinement of the color selection criterion, including new color-magnitude restrictions, and the inclusion of X-ray emission, which is not a ubiquitous property of the Be ``contaminants" also selected by our color method. Including the Be stars, it is remarkable that the broad color space returns early-type emission-line stars almost entirely, at a rate of $\approx$95\%.  By selecting WR candidates from a smaller color space of the ``WR sweet spot," which contains significantly fewer Be stars, we experienced a detection rate of $>$50\%. For the X-ray emitting candidates, we detected 10 WR sources after observing 23 objects, implying a detection rate of $\approx$40\%. We presented their X-ray properties, which are consistent with known, solitary WRs, as well as  those in some colliding-wind binaries. 

Most of the new WRs appear ``isolated" in the field; some could be runaways, others might have formed in relative isolation, or could be members of ``false negative" clusters that are difficult to distinguish against the dense stellar background of the Galactic plane.  We have discovered at least one such cluster, containing two WC8 stars and several OB supergiants.  We also identified several WRs in the known stellar clusters Danks 1 and 2, [DBS2003] 179, and [MFD2008]. The results imply that our method of selecting candidate WRs from the field could provide the most effective means of detecting WR-bearing false-negative clusters in the Galaxy.

The latest results bring the total number of known Galactic WRs to 476, which is $\approx$7--8\% of the total expected population of the Galaxy, estimated to be as high as $\approx$6000-6500 objects (Shara et al. 2009). The current WN/WC ratio of the Galactic  sample is $\approx$1.5. Although WR subtype ratios are commonly referenced in tests of stellar evolutionary models (e.g., Meynet \& Maeder 2005), they can only have meaning for complete or subtype-unbiased samples of WRs.  There may be reason to suspect a subtype bias in the current sample. For example, even though the earliest WNE types (WN3--4) have dense winds (strong free-free excesses and broad He {\sc i} and He {\sc ii} lines), they are intrinsically the faintest of the WRs, by 1--2 mag (Crowther et al. 2006), we would expect surveys of the inner Galaxy to have more difficulty recovering these objects relative to, say, WN7--9 stars. Thus, the Galactic longitudinal and radial distributions of WRs as a function of subtype, shown in Figures 11b and 12, are simply the observed distributions, and should not be misconstrued or over-interpreted. On the other hand, the observed surface density of all known WRs with respect to Galactic longitude (see Figure 11a) does appear to reflect true physical structures of the Galaxy, including prominent massive-star formation sites, such as W43 and the Westerlund 1 region, and portions of the sky where major spiral arms run approximately parallel to our line of sight, e.g., the Scutum-Centaurus and Sagittarius arms in the range $l\approx280^{\circ}$--$315^{\circ}$. 

There are $\approx$4900 unobserved WR candidates remaining in our sample from the GLIMPSE-{\sc I} survey area ($l=295^{\circ}$--$350^{\circ}$, $10^{\circ}$--$65^{\circ}$; $|b|\le1^{\circ}$). If the overall broad-color detection rate of $\approx$20\% persists, than we expect that almost 1000 WRs could potentially be detected in future confirmation observations. However, the observations that we have  conducted thus far have given priority to bright sources  over fainter ones. $K_s$-band magnitudes are typically in the range of 8--11 for our sample, while average extinction values are $\approx$2 mag. Thus, assuming a generic absolute $K_s$ magnitude of $\approx-5$ for WRs implies that we are mainly sensitive to WRs lying at distances of $\lesssim$8 kpc from the Sun, on our side of the Galaxy. Expanding the survey volume to include WRs all the way to the Solar Circle on the far side of the Galaxy would require sampling candidates having $K_s\gtrsim13$ mag. However,  we presume that as these fainter candidates are targeted, the associated degradation in both their 2MASS and GLIMPSE photometric qualities and the increase in stellar confusion will result in a significantly decreased WR confirmation rate.  A decline in photometric quality is also expected for the inner GLIMPSE-{\sc II} survey area ($|l|\le10^{\circ}$ from the Galactic center), owing to the relatively high source confusion in the innermost regions of the Galaxy. There are $\approx$1700  candidates in this area, although the WR detection rate has not yet been determined here, since the earliest color-based surveys by Hadfield et al. (2007) failed to detect any WRs in the GLIMPSE-{\sc II} fields. Continued surveys in this region are thus necessary in order to estimate the number of WRs that are contained within the GLIMPSE-{\sc II} point-source catalog and the extent of the Galactic volume that our survey is sensitive to. 

The color-based search for Galactic WRs is currently the most efficient method for identifying these objects in the Galaxy. Although the observational prescription described in this work has resulted in the greatest WR confirmation rate produced thus far, the infrared phenomenology behind the 
color selection method deserves further study, in order to establish an improved understanding of the WR subtype selection biases inherent in this technique.

\acknowledgments
This publication makes use of data products from the Two-Micron All-Sky Survey (2MASS), which is a joint project of the University of Massachusetts and IPAC/Caltech, funded by NASA and the NSF. The research was based on observations made with the \textit{Spitzer Space Observatory}, the Hale 5 m Telescope at Palomar Observatory, Cerro Tololo International Observatory, and the NASA Infrared Telescope Facility (IRTF). We thank the referee for a very insightful report, which helped to improve the manuscript.

\newpage
\LongTables
\begin{landscape}
\setlength{\tabcolsep}{0.05in}
\renewcommand{\arraystretch}{1.05}
\begin{deluxetable}{lccrrrrrrrrccccc}
\tablecolumns{16}
\tablewidth{0pc}
\tabletypesize{\scriptsize}
\tablecaption{Infrared Photometry of New WRs}

\tablehead{
 \colhead{Star} & \colhead{Source} & \colhead{R.A.} & \colhead{Decl.} &\colhead{$l$} &\colhead{$b$} & \colhead{$J$} & \colhead{$H$} & \colhead{$K_s$} &  \colhead{[3.6]}  & \colhead{[4.5]}  & \colhead{[5.8]} &\colhead{[8.0]}  & \colhead{USNO opt.}  & \colhead{$B1$} & \colhead{$R1$} \\ [2pt]
 \colhead{MDM} & \colhead{(2MASS J)} & \multicolumn{2}{c}{(deg, J2000)} & \colhead{(deg)} & \colhead{(deg)} & \colhead{(mag)} & \colhead{(mag)} & \colhead{(mag)} & \colhead{(mag)} & \colhead{(mag)} & \colhead{(mag)} & \colhead{(mag)} & \colhead{(mag)} & \colhead{counterpart} & \colhead{mag}}
\startdata
  1& 10232880$-$5746294& 155.87000&$ -57.77481$& 284.22&$  -0.38$& 10.60&  9.20&  8.34&  7.53&  7.21&  7.00&  6.74     & \nodata & \nodata & \nodata \\
  2& 12151249$-$6246438& 183.80207&$ -62.77892$& 298.80&$  -0.20$& 13.13& 11.03&  9.84&  8.78&  8.37&  8.08&  7.78        & \nodata & \nodata & \nodata \\
  3& 13120905$-$6243267& 198.03780&$ -62.72412$& 305.30&    0.05  & 10.21&  8.57&  7.58&  6.83&  \nodata &  \nodata &  6.11&    \nodata       & \nodata & \nodata \\
  4& 13122130$-$6240125& 198.08882&$ -62.67021$& 305.33&$   0.10$& 10.75&  9.57&  8.77&  8.15&  7.91&  7.81&  7.71& 0273$-$0506079 & \nodata &   16.24 \\
  5& 13122546$-$6244417& 198.10621&$ -62.74498$& 305.33&$   0.03$&  9.81&  8.48&  7.65&  6.87&  6.50&  6.31&  5.96& 0272$-$0487929  & 20.13 &  15.26 \\
  6& 13122766$-$6244220& 198.11535&$ -62.73948$& 305.34&$   0.03$& 13.16& 11.82& 10.71& 10.18&  9.64&  9.48&  9.18& 0272$-$0487952 & \nodata & \nodata \\
  7& 13122850$-$6241509& 198.11882&$ -62.69752$& 305.34&$   0.08$&  9.81&  8.83&  8.15&  7.81&  7.49&  7.34&  7.12     & \nodata & \nodata & \nodata \\
  8& 13122855$-$6241438& 198.11906&$ -62.69554$& 305.34&$   0.08$&  8.26&  7.27&  6.61&  \nodata &  \nodata &  5.25&  4.94& 0273$-$0506154 &  15.62&   12.52 \\
  9& 13145704$-$6223533& 198.73773&$ -62.39817$& 305.65&$   0.35$& 13.82& 11.88& 10.67&  9.62&  9.04&  8.73&  8.49      & \nodata & \nodata & \nodata \\
 10& 13440695$-$6245021& 206.02895&$ -62.75055$& 308.92&$  -0.49$& 15.01& 12.77& 11.46& 10.38&  9.89&  9.55&  9.40        & \nodata & \nodata & \nodata \\
 11& 14011549$-$6238199& 210.31455&$ -62.63883$& 310.86&$  -0.85$& 14.87& 12.84& 11.56& 10.80& 10.28&  9.97&  9.79       & \nodata & \nodata & \nodata \\
 12& 14043667$-$6129165& 211.15279&$ -61.48787$& 311.55&$   0.15$& 15.58& 12.71& 11.00&  9.50&  8.89&  8.39&  8.12        & \nodata & \nodata & \nodata \\
 13& 14203074$-$6048221& 215.12811&$ -60.80609$& 313.58&$   0.21$& 13.62& 11.63& 10.38&  9.32&  8.78&  8.52&  8.21        & \nodata & \nodata & \nodata \\
 14& 15533185$-$5345443& 238.38267&$ -53.76225$& 327.88&$   0.05$& 15.16& 12.51& 10.97&  9.65&  9.00&  8.68&  8.37        & \nodata & \nodata & \nodata \\
 15& 15584971$-$5251324& 239.70707&$ -52.85898$& 329.07&$   0.23$& 15.94& 13.05& 11.36&  9.91&  9.30&  8.89&  8.58       & \nodata & \nodata & \nodata \\
 16& 15585493$-$5202455& 239.72882&$ -52.04592$& 329.60&$   0.84$& 15.64& 12.59& 10.91&  9.64&  9.07&  8.70&  8.43       & \nodata & \nodata & \nodata \\
 17& 16002326$-$5251423& 240.09689&$ -52.86171$& 329.24&$   0.08$& 15.56& 12.87& 11.31& 10.10&  9.45&  9.20&  8.85       & \nodata & \nodata & \nodata \\
 18& 16100625$-$5047585& 242.52600&$ -50.79953$& 331.74&$   0.61$& 15.62& 13.05& 11.53& 10.28&  9.71&  9.53&  9.26        & \nodata & \nodata & \nodata \\
 19& 16205143$-$5004033& 245.21424&$ -50.06755$& 333.48&$  -0.04$& 15.50& 12.28& 10.59&  9.26&  8.74&  8.35&  8.22        & \nodata & \nodata & \nodata \\
 20& 16242329$-$4921295& 246.09703&$ -49.35817$& 334.38&$   0.05$& 15.64& 12.90& 11.43& 10.28&  9.78&  9.50&  9.35      & \nodata & \nodata & \nodata \\
 21& 16285342$-$4833394& 247.22258&$ -48.56092$& 335.47&$   0.08$& 14.65& 12.80& 11.76& 10.84& 10.28&  9.96&  9.88        & \nodata & \nodata & \nodata \\
 22& 16322205$-$4747425& 248.09190&$ -47.79510$& 336.43&$   0.18$& 14.97& 12.96& 11.63& 10.50&  9.87&  9.60&  9.39 & 0422$-$0651194  & 20.05 &  18.95 \\
 23& 16331120$-$4819412& 248.29677&$ -48.32812$& 336.13&$  -0.28$& 15.00& 12.44& 10.90&  9.59&  9.00&  8.67&  8.28        & \nodata & \nodata & \nodata \\
 24& 16334543$-$4751290& 248.43937&$ -47.85793$& 336.54&$  -0.03$& 15.56& 12.98& 11.41&  9.86&  9.19&  8.84&  8.73       & \nodata & \nodata & \nodata \\ 
 25& 16345746$-$4704129& 248.73942&$ -47.07023$& 337.26&$   0.35$& 13.75& 11.70& 10.46&  9.41&  8.86&  8.57&  8.19        & \nodata & \nodata & \nodata \\
 26& 16350555$-$4717135& 248.77312&$ -47.28706$& 337.11&$   0.19$& 13.60& 11.62& 10.46&  9.30&  8.73&  8.56&  8.43      & \nodata & \nodata & \nodata \\
 27& 16353888$-$4709130& 248.91197&$ -47.15359$& 337.27&$   0.21$& 15.02& 12.85& 11.49& 10.33&  9.74&  9.56&  9.19       & \nodata & \nodata & \nodata \\
 28& 16355116$-$4719515& 248.96308&$ -47.33092$& 337.17&$   0.06$& 13.60& 11.67& 10.39&  9.27&  8.71&  8.53&  8.23       & \nodata & \nodata & \nodata \\
 29& 16405078$-$4551231& 250.21166&$ -45.85646$& 338.84&$   0.41$& 14.84& 12.00& 10.42&  9.09&  8.55&  8.25&  7.94        & \nodata & \nodata & \nodata \\
 30& 16474603$-$4559049& 251.94186&$ -45.98473$& 339.53&$  -0.58$& 11.12&  9.84&  9.10&  8.43&  8.18&  7.94&  7.71&    0440$-$0524226 & \nodata &  \nodata \\
 31& 16482758$-$4609271& 252.11498&$ -46.15760$& 339.47&$  -0.79$& 14.14& 11.85& 10.59&  9.48&  8.90&  8.65&  8.41        & \nodata & \nodata & \nodata \\
 32& 17113303$-$3910400& 257.88770&$ -39.17783 $& 347.58&$  0.18 $&12.37&  10.43&    9.19&   8.27&   7.63 & 7.47& 7.08     &  \nodata & \nodata & \nodata \\ 
 33& 18124110$-$1826304& 273.17134&$ -18.44182$&  12.18&$  -0.12$& 13.80& 11.97& 10.92& 10.09&  9.57&  9.38&  9.19       & \nodata & \nodata & \nodata \\
 34& 18132248$-$1753503& 273.34378&$ -17.89736$&  12.74&$  -0.01$& 10.31&  9.27&  8.66&  8.03&  7.70&  7.49&  7.30 & 0721$-$0782909  & 18.63 &  16.57 \\ 
 35& 18255361$-$1250031& 276.47347&$ -12.83424$&  18.62&$  -0.27$& 11.53& 10.13&  9.28&  8.46&  8.04&  7.86&  7.51 & 0771$-$0547089  & 20.65 &  17.43 \\ 
 36& 18260611$-$1304104& 276.52555&$ -13.06962$&  18.44&$  -0.43$&  9.59&  8.82&  8.30&  7.71&  7.39&  7.16&  6.85 & 0769$-$0553041  & 15.86 & \nodata \\
 37& 18305320$-$1019370& 277.72174&$ -10.32701$&  21.41&$  -0.19$& 10.68&  9.91&  9.43&  8.70&  8.53&  8.38&  8.16&    0796$-$0404402  & 17.24 &  14.41 \\
 38& 18311653$-$1009250& 277.81885&$ -10.15694$&  21.61&$  -0.20$&  9.09&  8.29&  7.63&  7.17&  6.83&  6.66&  6.32&    0798$-$0402571  & 16.79 & \nodata \\
 39& 18334763$-$0923077& 278.44855&$  -9.38552$&  22.58&$  -0.39$& 12.18& 10.52&  9.36&  8.53&  8.03&  7.78&  7.51&     \nodata & \nodata & \nodata \\
 40& 18375149$-$0608417& 279.46452&$  -6.14489$&  25.92&$   0.21$& 12.73& 10.74&  9.31&  7.76&  7.26&  6.92&  6.73        & \nodata & \nodata & \nodata \\
 41& 18382716$-$0710447& 279.61319&$  -7.17912$&  25.07&$  -0.40$& 13.64& 11.73& 10.59&  9.48&  9.04&  8.81&  8.48       & \nodata & \nodata & \nodata \\
 42& 18393458$-$0544230& 279.89407&$  -5.73971$&  26.47&$   0.01$& 11.83& 10.49&  9.56&  8.81&  8.31&  8.18&  7.90& 0842$-$0402447  & 18.36 &  16.96 \\
 43& 18413407$-$0504012& 280.39198&$  -5.06695$&  27.30&$  -0.12$& 10.22&  9.21&  8.51&  7.85&  7.43&  7.20&  6.95 & 0849$-$0402164  & 19.18 &  15.71 \\
 44& 18420255$-$0356263& 280.51059&$  -3.94062$&  28.35&$   0.29$& 12.86& 10.82&  9.66&  8.56&  8.06&  7.82&  7.49        & \nodata & \nodata & \nodata \\
 45& 18433257$-$0404190& 280.88570&$  -4.07191$&  28.41&$  -0.10$& 15.15& 12.55& 11.14& 10.02&  9.64&  9.41&  9.23       & \nodata & \nodata & \nodata \\
 46& 18435803$-$0245172& 280.99176&$  -2.75475$&  29.63&$   0.40$& 10.00&  9.16&  8.53&  7.84&  7.49&  7.32&  6.97& 0872$-$0545092  & 17.96 &  14.30 \\
 47& 18445157$-$0321496& 281.21485&$  -3.36375$&  29.19&$  -0.07$& 14.22& 12.23& 11.13& 10.10&  9.69&  9.33&  9.20        & \nodata & \nodata & \nodata \\
 48& 18445161$-$0327437& 281.21499&$  -3.46214$&  29.10&$  -0.12$& 13.81& 11.87& 10.73&  9.75&  9.24&  9.07&  8.77       & \nodata & \nodata & \nodata \\
 49& 18493230$-$0224270& 282.38454&$  -2.40750$&  30.57&$  -0.68$& 14.13& 12.13& 10.66&  9.72&  9.17&  8.86&  8.65        & \nodata & \nodata & \nodata \\
 50& 18521900$-$0011114& 283.07912&$  -0.18649$&  32.87&$  -0.28$&  9.75&  9.29&  8.98&  8.69&  8.47&  8.23&  7.93& 0898$-$0397750  & 14.21 &  12.32 \\
 51& 18524369$+$0008415& 283.18205&$   0.14495$&  33.21&$  -0.22$& 12.49& 10.68&  9.61&  8.63&  8.19&  7.99&  7.68        & \nodata & \nodata & \nodata \\
 52& 18540312$+$0124508& 283.51297&$   1.41415$&  34.49&$   0.06$& 12.81& 11.06&  9.99&  9.02&  8.62&  8.37&  8.09        & \nodata & \nodata & \nodata \\
 53& 19012661$+$0351553& 285.36087&$   3.86544$&  37.51&$  -0.46$& 14.70& 12.67& 11.25& 10.05&  9.51&  9.27&  9.06        & \nodata & \nodata & \nodata \\
 54& 19012711$+$0351542& 285.36295&$   3.86514$&  37.51&$  -0.46$& 14.25& 12.27& 10.99&  9.77&  9.34&  9.09&  8.88        & \nodata & \nodata & \nodata \\
 55& 19083809$+$0928210& 287.15875&$   9.47248$&  43.31&$   0.53$& 12.22& 10.81&  9.94&  9.04&  8.64&  8.45&  8.16      & \nodata & \nodata & \nodata \\
 56& 19161838$+$1246493& 289.07663&$  12.78035$&  47.12&$   0.39$& 12.95& 11.82& 10.77& 10.22&  9.72&  9.50&  9.24        & \nodata & \nodata & \nodata \\
 57& 19174121$+$1129190& 289.42182&$  11.48859$&  46.13&$  -0.51$& 15.30& 13.00& 11.53& 10.53&  9.91&  9.59&  9.33        & \nodata & \nodata & \nodata \\
 58& 19202932$+$1412061& 290.12221&$  14.20167$&  48.85&$   0.16$& 14.34& 11.98& 10.63&  9.46&  8.93&  8.65&  8.35        & \nodata & \nodata & \nodata \\
 59& 19225361$+$1408498& 290.72343&$  14.14711$&  49.07&$  -0.38$& 12.67& 10.91&  9.68&  8.82&  8.23&  8.04&  7.73       & \nodata & \nodata & \nodata \\
 60& 19225446$+$1411280& 290.72697&$  14.19109$&  49.11&$  -0.36$& 12.83& 11.02&  9.58&  8.05&  7.61&  7.20&  7.02       & \nodata & \nodata & \nodata \\
 61& 19300530$+$1746010& 292.52211&$  17.76694$&  53.08&$  -0.18$& 12.71& 11.07& 10.12&  9.14&  8.72&  8.45&  8.13        & \nodata & \nodata & \nodata 
\enddata
\tablecomments{Photometric data is from the 2MASS ($JHK_s$) and \textit{Spitzer}/GLIMPSE (3.6--8.0 {\micron}) catalogs. MDM38 was independently discovered by Motch et al. (2010). MDM34 was independently identified in Messineo et al. (2011).}
\label{tab:phot}
\end{deluxetable}
\clearpage
\end{landscape}
\newpage

\begin{deluxetable}{lcl}
\tablecolumns{4}
\tablewidth{0pc}
\tabletypesize {\scriptsize}
\tablecaption{Spectroscopic Observations of WRs}
\tablehead{
\colhead{Star} & \colhead{Observation Date} & \colhead{Telescope/Instrument}  \\
\colhead{MDM} & \colhead{(UT)} & \colhead{} 
}
\startdata
1  & 2010-05-23~ 23:09 & SOAR/OSIRIS \\
2  & 2010-05-24~ 00:02 & SOAR/OSIRIS \\
3  & 2010-05-27~ 00:16 & SOAR/OSIRIS \\
4  & 2010-05-24~ 22:38 & SOAR/OSIRIS \\
5  & 2010-05-24~ 23:28 & SOAR/OSIRIS \\
6  & 2010-05-24~ 23:39 & SOAR/OSIRIS \\
7  & 2010-05-25~ 00:00 & SOAR/OSIRIS \\
8  & 2010-05-25~ 00:13 & SOAR/OSIRIS \\
9  & 2010-05-23~ 00:27 & SOAR/OSIRIS \\
10 & 2010-05-24~ 01:51 & SOAR/OSIRIS \\
11 & 2010-05-23~ 01:03 & SOAR/OSIRIS \\
12 & 2010-05-23~ 03:44 & SOAR/OSIRIS \\
13 & 2010-05-25~ 01:04 & SOAR/OSIRIS \\
14 & 2010-05-25~ 01:52 & SOAR/OSIRIS \\
15 & 2010-05-26~ 03:32 & SOAR/OSIRIS \\
16 & 2010-05-23~ 06:06 & SOAR/OSIRIS \\
17 & 2010-05-26~ 03:14 & SOAR/OSIRIS \\
18 & 2010-05-26~ 03:52 & SOAR/OSIRIS \\
19 & 2010-05-26~ 04:10 & SOAR/OSIRIS \\
20 & 2010-05-26~ 04:28 & SOAR/OSIRIS \\
21 & 2010-05-27~ 04:38 & SOAR/OSIRIS \\
22 & 2010-05-27~ 05:59 & SOAR/OSIRIS \\
23 & 2010-05-27~ 05:24 & SOAR/OSIRIS \\
24 & 2010-05-27~ 05:40 & SOAR/OSIRIS \\
25 & 2010-05-26~ 09:05 & SOAR/OSIRIS \\
26 & 2010-05-26~ 09:18 & SOAR/OSIRIS \\
27 & 2010-05-27~ 07:38 & SOAR/OSIRIS \\
28 & 2010-05-27~ 08:15 & SOAR/OSIRIS \\
29 & 2010-05-27~ 09:01 & SOAR/OSIRIS \\
30 & 2010-05-26~ 05:11 & SOAR/OSIRIS \\
31 & 2010-05-27~ 09:24 & SOAR/OSIRIS \\
32 & 2010-05-25~ 06:21 & SOAR/OSIRIS \\
33 & 2009-08-06~ 07:42 & IRTF/SPEX \\
34 & 2009-08-05~ 06:15 & IRTF/SPEX \\
35 & 2009-07-13~ 08:42 & HALE/TSPEC \\
36 & 2009-07-13~ 08:25 & HALE/TSPEC \\
37 & 2009-07-14~ 06:03 & HALE/TSPEC \\
38 & 2009-07-14~ 05:13 & HALE/TSPEC \\
39 & 2009-07-15~ 04:52 & HALE/TSPEC \\
40 & 2009-07-15~ 09:38 & HALE/TSPEC \\
41 & 2009-08-06~ 07:10 & IRTF/SPEX \\
42 & 2009-07-15~ 05:14 & HALE/TSPEC \\
43 & 2009-07-14~ 06:50 & HALE/TSPEC \\
44 & 2009-07-15~ 09:16 & HALE/TSPEC \\
45 & 2009-08-04~ 05:41 & IRTF/SPEX \\
46 & 2009-07-14~ 08:52 & HALE/TSPEC \\
47 & 2009-08-06~ 06:12 & IRTF/SPEX \\
48 & 2009-08-06~ 05:40 & IRTF/SPEX \\
49 & 2009-08-06~ 06:53 & IRTF/SPEX \\
50 & 2009-08-05~ 09:22 & IRTF/SPEX \\
51 & 2009-07-15~ 10:40 & HALE/TSPEC \\
52 & 2009-07-15~ 07:10 & HALE/TSPEC \\
53 & 2010-06-21~ 10:23 & HALE/TSPEC \\
54 & 2010-06-21~ 10:42 & HALE/TSPEC \\
55 & 2009-07-15~ 06:32 & HALE/TSPEC \\
56 & 2009-07-15~ 06:51 & HALE/TSPEC \\
57 & 2010-06-21~ 04:50 & HALE/TSPEC \\
58 & 2010-06-21~ 04:37 & HALE/TSPEC \\
59 & 2009-08-06~ 09:30 & IRTF/SPEX \\
60 & 2009-07-15~ 11:20 & HALE/TSPEC \\
61 & 2009-07-14~ 10:50 & HALE/TSPEC 
\enddata
\tablecomments{Facilities have the following spectral resolutions ($\lambda/\delta\lambda$): SOAR/OSIRIS$\approx$800;  IRTF/SPEX$\approx$1200; HALE/TSPEC$\approx$2600.}
\label{tab:obs}
\end{deluxetable}

\newpage
\LongTables
\begin{landscape}
\begin{deluxetable}{lcccccccccccccllll}
\tablecolumns{18}
\setlength{\tabcolsep}{0.02in}
\tablewidth{0pc}
\tabletypesize{\tiny}
\tablecaption{Equivalent Widths of Prominent Emission Lines for WN Stars}
\tablehead{ \colhead{Star}  & \colhead{He {\sc ii}} & \colhead{He {\sc i}} & \colhead{He {\sc ii}$+$P$\gamma$} & \colhead{He {\sc ii}} & \colhead{He {\sc ii}$+$P$\beta$} & \colhead{He {\sc ii}} & \colhead{He {\sc ii}} & \colhead{He {\sc i}} &  \colhead{He {\sc i}} & \colhead{N {\sc v}} & \colhead{He {\sc i}/N{\sc iii}} & \colhead{He {\sc ii}$+$Br$\gamma$} & \colhead{He {\sc ii}} & \colhead{1.012/} & \colhead{2.189/} & \colhead{2.189/} & \colhead{Type}\\ [2pt]
\colhead{MDM}&\colhead{1.012} &\colhead{1.083} &\colhead{1.093} &\colhead{1.163} &\colhead{1.281} &\colhead{1.476} &\colhead{1.692} &\colhead{1.700} &\colhead{2.058} &\colhead{2.110} &\colhead{2.115} &\colhead{2.165} &\colhead{2.189} & \colhead{1.083} &\colhead{2.165} &\colhead{2.115} & \colhead{}}
\startdata
1&	\nodata &\nodata	&\nodata	&\nodata&30	&\nodata	&\nodata	&\nodata	&\nodata&\nodata&3&18	&5	&\nodata&0.28 \tiny{(WN8)} &1.67 \tiny{(WN7)} & WN7--8	\\
2&	\nodata &\nodata	&\nodata	&\nodata&42	&6	&3&	9	&2	&\nodata	&11	&37	&6	&\nodata	&0.16 \tiny{(WN8)} &0.55 \tiny{(WN8)} & WN8	\\
3& \nodata &\nodata	&\nodata	&\nodata&45	&5	&2&	9	&3	&\nodata	&20	&42	&4	&\nodata	&0.10 \tiny{(WN8--9)}	&0.20 \tiny{(WN8--9)} & WN8--9		\\
5& \nodata &\nodata	&\nodata	&\nodata&15	&5	&2	&17	&8	&\nodata	&22	&54	&4	&\nodata	&0.07 \tiny{(WN9)} & 0.18 \tiny{(WN8--9)} & WN9 \\		
7&	\nodata &\nodata	&\nodata	&\nodata&19	&2	&$<$1&	2&$<$1	&\nodata	&9&17	&$-$1&\nodata	&0.05 \tiny{(WN9)} &0.1 \tiny{(WN8--9)} & WN9 \\
8&	 \nodata &\nodata &\nodata&	\nodata &55& $<$1 &$<$1& 8	&6&\nodata&10& 62	& $<$1 &\nodata&$<0.02$\tiny{ (WN$\ge9$)} & $<0.1$ \tiny{(WN$\ge$9)} & WN9h\tablenotemark{a}	 \\
9&	\nodata	&\nodata&\nodata&	\nodata&	88	&11	&9	&31	&10	&\nodata	&25	&83	&11	&\nodata	&0.13 \tiny{(WN8)} & 0.44 \tiny{(WN8)} & WN8 \\
10&		\nodata	&\nodata&\nodata&	\nodata&	\nodata	&\nodata	&13	&40	&14	&\nodata	&37	&46	&26	&\nodata&	0.57 \tiny{(WN6--7)}&	0.71 \tiny{(WN7)}&	 WN7	\\
14  &	\nodata	&\nodata&\nodata&	\nodata&	91	&108	&87	&26	&29	&5	&54	&49	&106	&\nodata&	2.16 \tiny{(WN3--5b)} &	1.96 \tiny{(WN5--7)} &WN5b		\\
15  &	\nodata	&\nodata&\nodata&	\nodata&	\nodata	&238	&102	&blend	&20	&14	&57	&84	&93	&\nodata&	1.11 \tiny{(WN6--7b)} &	1.63 \tiny{(WN5--7)} & WN6b		\\
16&	\nodata	&\nodata&\nodata&	\nodata&	51	&31	&8	&44	&17	&\nodata&	48	&54	&29	&\nodata&	0.54 \tiny{(WN6--7)} &0.6	\tiny{(WN7--8)}	& WN7\\
19&\nodata&\nodata&\nodata&\nodata&\nodata&\nodata&	\nodata&	2	&1.2	&\nodata&	6&	26	&0.9&	\nodata&	0.03 \tiny{(WN9)} 	&0.15 \tiny{(WN8--9)} & WN9ha\tablenotemark{a}		\\
20&\nodata	&\nodata&\nodata&	\nodata&	\nodata&	27?	& \multicolumn{2}{c}{$\leftarrow$28 (blend)$\rightarrow$}	&4	&\nodata&	19	&32	&18	&\nodata&	0.56	\tiny{(WN6--7)} &0.95 \tiny{(WN7)} & WN7		\\
21 &\nodata	&\nodata&\nodata&	\nodata&	42&	51	&\multicolumn{2}{c}{$\leftarrow$30 (blend)$\rightarrow$}	&29	&20	&24	&37&	69	&\nodata&	1.86 \tiny{(WN5--7b)} &	6.27	\tiny{(WN4--6)} & WN5b\tablenotemark{b}	\\
23  &	\nodata&\nodata&	\nodata&	\nodata&	\nodata	&115& \multicolumn{2}{c}{$\leftarrow$86 (blend)$\rightarrow$}	& 24& 	16	&47	&26	&229&	\nodata	&8.81 \tiny{(WN4b)}	&4.87 \tiny{(WN4--6)} & WN4--5b\tablenotemark{b}		\\
25& \nodata&\nodata&\nodata&	\nodata&	83&	119	&\multicolumn{2}{c}{$\leftarrow$92 (blend)$\rightarrow$}&	17&	\multicolumn{2}{c}{$\leftarrow$55 (blend)$\rightarrow$}&	78&	113	&\nodata&1.44 \tiny{(WN5--7b)} &	2.05 \tiny{(WN5--6)} &WN5--6b\\
27 &	\nodata&\nodata&\nodata&	\nodata&	85&	66	&\multicolumn{2}{c}{$\leftarrow$88 (blend)$\rightarrow$}&	54&	6	&66	&75&	64	&\nodata&	0.85 \tiny{(WN6--7)} &	0.97 \tiny{(WN7)} &WN6\\
29&	\nodata&\nodata&\nodata&	\nodata&	44&	13	&\multicolumn{2}{c}{$\leftarrow$19 (blend)$\rightarrow$}&	10&	\nodata	&23	&35&	3	&\nodata&	0.08 \tiny{(WN9)}&	0.13 \tiny{(WN8--9)} &WN9\\
30&\nodata&\nodata&\nodata&	\nodata&	15&	24	&14	&7&	4&	1	&14	&17&	32	&\nodata&	1.88 \tiny{(WN6--7)} &	2.29 \tiny{(WN5--6)} &WN6\tablenotemark{b}  \\
31 &\nodata&\nodata&\nodata&	\nodata&	46&	80	&32	&23&	26&	5	&28	&41&	90	&\nodata&	2.20 \tiny{(WN5)} &	3.21 \tiny{(WN5--6)} &WN5\tablenotemark{b}  \\
32  & \nodata&\nodata&\nodata&\nodata & 81 & 47  & \multicolumn{2}{c}{$\leftarrow$40 (blend)$\rightarrow$} & 182 &\nodata & 18 &40&3 & \nodata & 0.08 \tiny{(WN9)} & 0.17 \tiny{(WN8--9)} &WN9 \\
33&	8	&544	&40&	15&	54&	14&	5&	61&	62	&\nodata&	58&	70	&18	&0.02 \tiny{(WN9)} &	0.25	\tiny{(WN8)} &0.31 \tiny{(WN8)} &WN8--9\\
34&	81	&145	&10&	61&	27&	29&	13&	21&	6	&2	&24	&29&	43	&0.56 \tiny{(WN6--7)	}&1.48 \tiny{(WN6)}	&1.79 \tiny{(WN6--7)} & WN6\\
35&	64	&516	&65&	40&	53&	17&	10&	52&	24	&\nodata&	59&	62	&27	&0.12 \tiny{(WN8)}	&0.44 \tiny{(WN8)}	&0.46 \tiny{(WN8)} & WN8 \\
36&	98	&375	&24&	48&	60&	34&	14&	39&	14	&\nodata&	47&	55	&41	&0.26 \tiny{(WN7)}	&0.75 \tiny{(WN6--7)}	&0.87 \tiny{(WN6--7)} & WN7 \\
37&	32	&33	&5&	21&	12&	11&	5&	8&	10	&\nodata&	6&	13	&15	&0.97 \tiny{(WN6)}	&1.15 \tiny{(WN6--7)}	&2.50 \tiny{(WN5--6)} & WN6\\ 
38&	58	&172	&19&	39&	46&	25&	\multicolumn{2}{c}{$\leftarrow$333 (blend)$\rightarrow$}&	19	&\nodata&	43&	47	&28	&0.34 \tiny{(WN7)}	&0.6 \tiny{(WN6--7)}	&0.65 \tiny{(WN7--8)} &WN7\\ 
41&	186	&517	&27&	111&	85&	44&	8&	49&	17	&\nodata&	67&	65	&46	&0.36 \tiny{(WN7)}	&0.71 \tiny{(WN6--7)	}&0.69 \tiny{(WN7)} & WN7\\
43&	93	&1026	&190&	63&	54&	40&	\multicolumn{2}{c}{$\leftarrow$46 (blend)$\rightarrow$}&	18	&\nodata&	61&	55	&44	&0.09 \tiny{(WN7--8)}	&0.80 \tiny{(WN6--7)}	&0.72 \tiny{(WN7)} & WN7\\ 
44&	11	&503	&22&	21&	68&	\nodata&	2&	32	&41	&\nodata&	40	&74	&11	&0.02 \tiny{(WN9)}	&0.15 \tiny{(WN8)}	&0.28 \tiny{(WN8--9)} &WN8--9  \\
45 & \nodata	&\nodata&	\nodata&	\nodata	&\nodata&	11	&4	&15	&5&	\nodata	&20	&22	&17	&\nodata	&0.77 \tiny{(WN6--7)} &	0.85 \tiny{(WN7--8)} & WN7\\  
46&	93	&494	&28&	57&	73&	40&	13&	57&	18	&\nodata&	57&	64	&39	&0.20 \tiny{(WN7--8)}	&0.61 \tiny{(WN6--7)} &	0.68 \tiny{(WN7--8)} & WN7 \\
47&150	&482	&6&	          102&	59&	32&	14&	43&	13	&\nodata&	46&	59	&39	&0.31\tiny{ (WN7)}	&0.66 \tiny{(WN6--7)} &	0.85	\tiny{(WN7--8)}	& WN7\\
48& 132	&570	&35&	60&	80&	29&	12&	47&	18	&\nodata&	42&	56	&32	&0.23 \tiny{(WN7)}	&0.57 \tiny{(WN6--7)}&	0.76	\tiny{(WN7--8)} & WN7	\\
50&		9	&29&	2&	4&	9&	3&	4&	15	&73	&\nodata&	20	&16	&4	&0.31 \tiny{(non-WR?)}&	0.25 \tiny{(non-WR?)}&	0.20 \tiny{(non-WR?)}	& non-WR?\tablenotemark{c}	\\
51&		12	&149	&22&	7&	49&	4&	3&	15&	7	&\nodata&	22&	55	&3	&0.08 \tiny{(WN8)}	&0.05 \tiny{(WN9)} &	0.14 \tiny{(WN8--9)}& WN8--9	\\	
52&		110	&118	&10&	86&	21&	35&	23&	10&	2	&6	&19&	21&	59	&0.93 \tiny{(WN6)}	&2.81 \tiny{(WN5)}	&3.11 \tiny{(WN5--6)}& WN5--6		\\
55&		37	&175	&21&	19&	49&	11&	6&	18&	7	&\nodata&	21&	53	&15	&0.20 \tiny{(WN7--8)}	&0.28 \tiny{(WN8)} &	0.71\tiny{ (WN7--8)} & WN8 \\		
61&	110	&345	&24&	82&	63&	39&	13&	32&	15	&\nodata&	42&	61	&53	&0.32 \tiny{(WN7)}	&0.87 \tiny{(WN7)}&	1.26	\tiny{(WN7)} & WN7	
\enddata
\tablecomments{EW values are in units of Angstroms. Uncertainties are $\approx$10\%. The last three columns list the EW ratios of diagnostic lines, with the implied subtype following in parentheses. The final subtype is given in the last column. The 1.012/1.083 and 2.189/2.165 ratio diagnostics are from Crowther et al. (2006, C06), and the 2.189/2.115 ratio diagnostic is from Figer et al. (1997, F97). Final classifications respect both criterion.  Stars are assigned the broad designation ("b") if the FWHM of He {\sc ii} at 2.189 \micron is above 130 {\AA}, following C06. }
\tablenotetext{a}{Spectra consistent with both O supergiants and hydrogen-rich WN stars (WNh or WNha) were further classified  by the overall strength of Br$\gamma$ emission at $\lambda2.165$ {\micron} (see text).}
\tablenotetext{b}{For these particular stars, the ratio of N {\sc v} (2.10 \micron)/He {\sc i}/N{\sc iii} (2.115 \micron) is also used as a diagnostic, following C06. MDM21, for example, has a possible subtype range of WN4--7, given the C06 and F97 criterion. However, the spectrum exhibits 2.100/2.115=0.8, which implies WN5, according to C06. So, WN5 was assigned.}
\tablenotetext{c}{The nature of MDM50 is uncertain; the subtype diagnostics implied by the line ratios are not considered applicable.}
\label{tab:wnspec}
\end{deluxetable}
\clearpage
\end{landscape}

\newpage
\LongTables
\begin{landscape}
\begin{deluxetable}{llcccccccccccclllll}
\tablecolumns{19}
\tablewidth{0pc}
\tabletypesize{\tiny}
\tablecaption{Equivalent Widths of Prominent Emission Lines for WC Stars}
\tablehead{ \colhead{Star}  & \colhead{C {\sc iii}} & \colhead{C {\sc ii}} & \colhead{He {\sc ii}} & \colhead{He {\sc i}} & \colhead{C {\sc iv}} & \colhead{C {\sc iii}} & \colhead{He {\sc i}/C {\sc ii}} & \colhead{C {\sc iv}} &   \colhead{He {\sc i}} & \colhead{C {\sc iv}} & \colhead{C {\sc iii}} & \colhead{He {\sc ii}/C {\sc iii}} & \colhead{He {\sc ii}} & \colhead{0.971/} & \colhead{2.076/} & \colhead{0.971/} & \colhead{$\Sigma$$K$} &\colhead{Type}\\ [2pt]
\colhead{MDM}  &\colhead{0.971} &\colhead{0.990} &\colhead{1.012} &\colhead{1.083} &\colhead{1.191} &\colhead{1.198} &\colhead{1.700} &\colhead{1.74} &\colhead{2.058} &\colhead{2.076} &\colhead{2.110} &\colhead{2.165} & \colhead{2.189} &\colhead{0.990} & \colhead{2.110} & \colhead{2.110} &\colhead{} &\colhead{}}
\startdata
4& \nodata	&\nodata&	\nodata&	\nodata&	\nodata&	\nodata& weak&	89&	\nodata&	340	&118&	10&	5	&\nodata&	2.9 \tiny{(WC8)} & \nodata & 458 & WC8 	\\
6&	\nodata&	\nodata&	\nodata&	\nodata&	\nodata&	\nodata&	40	&501&	\nodata&	1862 &	385	&\nodata&	80	&\nodata&	4.8  \tiny{(WC7)} & \nodata & 2247 & WC7 \\	
11&	\nodata&	\nodata&	\nodata&	\nodata&	\nodata&	\nodata&	8	&97	&\nodata	&486&	106	&20	&34	&\nodata	&4.6 \tiny{	(WC7)} & \nodata & 592 &WC7 \\
12	&	\nodata&	\nodata&	\nodata&	\nodata&	\nodata&	\nodata&	22	&108&	85&	256&	149&	18	&24	&\nodata	&1.7  \tiny{(WC8)}& \nodata & 490 &WC8 \\
13	&	\nodata&	\nodata&	\nodata&	\nodata&	\nodata&	\nodata&	56	&70	&109	&161	&172&	51	&51&	\nodata&	0.9  \tiny{(WC9)} & \nodata & 442 &WC9 \\	
17	&	\nodata&	\nodata&	\nodata&	\nodata&	\nodata&	\nodata&	87	&184&	\nodata&	556	&256&	114&	57&	\nodata	&2.2  \tiny{(WC8)}& \nodata  & 812 &WC8 \\	
18	&	\nodata&	\nodata&	\nodata&	\nodata&	\nodata&	\nodata&	64	&32	&54	&45&	80&	73	&19	&\nodata&	0.6  \tiny{(WC9)} & \nodata &	179 (d?) &WC9d? \\
22	&	\nodata&	\nodata&	\nodata&	\nodata&	\nodata&	\nodata&	77	&36	&362	&79	&135&	88&	21&	\nodata&	0.6  \tiny{(WC9)} & \nodata & 576 &WC9 \\	
24	&	\nodata&	\nodata&	\nodata&	\nodata&	\nodata&	\nodata&	25	&47	&25&	82	&84	&24	&28	&\nodata	&1.0  \tiny{(WC8--9)} & \nodata & 191	 (d?) &WC8--9d? \\
26    &	\nodata&	\nodata&	\nodata&	\nodata&	\nodata&	\nodata&	29	&11	&76&	31	&65	&34	&26	&\nodata	&0.5	 \tiny{(WC9)} & \nodata & 172 (d?) &WC9d? \\
28   &	\nodata&	\nodata&	\nodata&	\nodata&	\nodata&	\nodata&	82	&134&	\nodata	&352&	195&	68&	59&	\nodata	&1.8  \tiny{(WC8)} & \nodata & 486 &WC8	\\
39 &	714	&71	&80	&537	& \multicolumn{2}{c}{$\leftarrow$202 (blend)$\rightarrow$} &	84	&109	&\nodata	&303	&173	&67	&60	&10  \tiny{(WC8)} &	1.8  \tiny{(WC8)} & 4.1& 476 &WC8\\ 
40 &	616	&64	&26	&282	&30	&114	&25	&19	&68	&44	&63	&20	&19	&9.6  \tiny{(WC9)}	&0.7  \tiny{(WC9)} & 9.8 (d) & 175 (d?) &WC9d \\ 
42 &	743	&58	&93	&600	&262	&69	&100	&169	&\nodata	&433	&260	&65	&57	&12.8  \tiny{(WC8)}	&1.7  \tiny{(WC8)} & 2.9 &693 &WC8	\\ 
49 &	2460	&\nodata&	257	&315	& \multicolumn{2}{c}{$\leftarrow$390 (blend)$\rightarrow$}	&12	&290	&\nodata&	1073	&276	&95	&43&	\nodata&	3.9  \tiny{(WC8)} & 8.9 (d?)	& 1349 &WC8d?  \\ 
53 &	\nodata	&\nodata	&\nodata	&\nodata	&\nodata	&\nodata	&91	&212	&\nodata&	560&	227&	52&	57&	\nodata&	2.5  \tiny{(WC8)} & \nodata & 787 &WC8 \\
54 &	\nodata	&\nodata	&\nodata	&\nodata	&\nodata	&\nodata	&49	&116	&\nodata&	374&	152&	35&	42&	\nodata&	2.4  \tiny{(WC8)} & \nodata & 526 &WC8 \\
56 &	1001	&89	&138	&215	& \multicolumn{2}{c}{$\leftarrow$329 (blend)$\rightarrow$}	&63	&311	&\nodata	&1202	&422	&\nodata&	132&	\nodata&	2.8  \tiny{(WC8)} & 2.4  & 1624 &WC8 \\ 
57 &	\nodata	&\nodata	&\nodata&	\nodata	&132	&78	&65	&266	&\nodata&	1036	&226	&105&	83&	\nodata&	4.6  \tiny{(WC7)} & \nodata & 1367 &WC7 \\
58 &	\nodata	&\nodata	&\nodata&	358	&26	&25	&53	&38	&\nodata	&71	&69	&60	&29&	\nodata&	1.0  \tiny{(WC8--9)} & \nodata & 200 (d?) &WC8-9d? \\
59&	897	&26	&15	&550	&155	&16	&95	&203	&\nodata	&766	&159	&54	&81	&34.5  \tiny{(WC5--7)}	&4.8  \tiny{(WC7)} & 5.6& 925 &WC7 \\ 
60&		980	&49	&73	&324	&149	&12	&37	&108	&\nodata	&299	&113	&21	&22	&\nodata&	2.6  \tiny{(WC8)} & 8.7 (d?) & 412 (d?) &WC8d 
\enddata
\tablecomments{Equivalent widths are given in units of Angstroms. The uncertainties are $\approx$10\% of the given value. Columns 15 and 16 list the equivalent width ratios of selected diagnostic lines from Crowther et al. (2006), and the implied WC subtype follows in parentheses. The final assigned subtype is given in the last column. The ratio of C {\sc iii} lines given in column 17 is used as a diagnostic to select candidates for thermal dust emission, which can dilute line strengths in the $K$-band relative to the $Y$ band.  Column 18 lists $\Sigma$$K$, which is defined as the sum of the equivalent widths of He {\sc i} (2.06 \micron), C {\sc iii} (2.11 \micron), and C {\sc iv} (2.08 \micron) in the $K$ band. For WC8--9 stars, this value may be relatively low for thermal dust emitters, but this does not necessarily imply dust emission (see text).}
\label{tab:wcspec}
\end{deluxetable}
\clearpage
\end{landscape}

\begin{deluxetable}{llcccccccc}
\tablecolumns{10}
\tablewidth{0pc}
\tabletypesize{\scriptsize}
\tablecaption{Extinction and Absolute Photometry for New WRs}
\tablehead{ \colhead{Star} & \colhead{Type} & \colhead{${(J-K_s)}_0$} & \colhead{${(H-K_s)}_0$} & \colhead{$A^{J-K_{s}}_{K_{s}}$} & \colhead{$A^{H-K_{s}}_{K_{s}}$} & \colhead{$\overline{A_{K_{s}}}$} & \colhead{$M_{K_s}$} & \colhead{$R$} & \colhead{$R_G$}  \\ [2pt]
\colhead{MDM} &\colhead{} &\colhead{(mag)} &\colhead{(mag)}  &\colhead{(mag)}  &\colhead{(mag)}  &\colhead{(mag)}  &\colhead{(mag)} &\colhead{(kpc)} &\colhead{(kpc)} }
\startdata
 1&  WN7--8& 0.13& 0.11& 1.43& 1.36& 1.39&$ -5.92$&  3.7&  8.0 \\ 
 2&    WN8 & 0.13& 0.11& 2.12& 1.97& 2.04&$ -5.92$&  5.5&  7.2 \\ 
 3&    WN8--9 & 0.13& 0.11& 1.68& 1.60& 1.64&$ -5.92$&  2.4&  6.9 \\ 
 4&     WC8& 0.43& 0.38& 1.04& 0.76& 0.90&$ -4.65$&  3.2&  6.7 \\ 
 5&  WN9& 0.13& 0.11& 1.36& 1.31& 1.33&$ -5.92$&  2.8&  6.8 \\ 
 6&     WC7& 0.62& 0.58& 1.23& 0.96& 1.10&$ -4.59$&  6.9&  6.9 \\ 
 7&    WN9 & 0.13& 0.11& 1.03& 1.04& 1.04&$ -5.92$&  4.0&  6.6 \\ 
 8&  WN9h& 0.13& 0.11& 1.02& 0.99& 1.00&$ -5.92$&  2.0&  7.0 \\ 
 9&     WN8& 0.13& 0.11& 2.02& 2.00& 2.01&$ -5.92$&  8.3&  7.4 \\ 
10&     WN7& 0.37& 0.09& 2.13& 2.21& 2.17&$ -5.18$&  7.9&  6.8 \\ 
11&     WC7& 0.62& 0.58& 1.83 & 1.27 & 1.55 &$ -4.59$& 8.3 & 2.6   \\ 
12&     WC8& 0.13& 0.11& 2.98& 2.91& 2.95&$ -5.92$&  6.2&  6.1 \\ 
13&     WC9& 0.23& 0.26& 2.01& 1.80& 1.91&$ -6.30$&  9.0&  6.8 \\ 
14&    WN5b& 0.37& 0.27& 2.55& 2.31& 2.43&$ -4.77$&  4.6&  4.8 \\ 
15&    WN6b& 0.37& 0.27& 2.82& 2.58& 2.70&$ -4.77$&  4.9&  4.6 \\ 
16&     WN7& 0.37& 0.09& 2.92& 2.89& 2.90&$ -5.18$&  4.3&  4.8 \\ 
17&     WC8& 0.43& 0.38& 2.56& 2.15& 2.36&$ -4.65$&  5.3&  4.4 \\ 
18&     WC9d?& 0.23& 0.26& 2.59& 2.30& 2.45&$ -6.30$& 11.9\tablenotemark{b}&  6.2\tablenotemark{b} \\ 
19&  WN9ha& 0.13& 0.11& 3.21& 2.89& 3.05&$ -5.92$&  4.9&  4.2 \\ 
20&     WN7& 0.37& 0.09& 2.57& 2.52& 2.55&$ -5.18$&  6.5&  3.5 \\ 
21&    WN5b& 0.37& 0.27& 1.69& 1.39& 1.54&$ -4.77$&  10.0&  4.3 \\ 
22&     WC8--9d?& 0.23& 0.26& 2.08& 1.95& 2.01&$ -6.30$& 15.3&  8.5 \\ 
23&  WN4--5b& 0.37& 0.27& 2.50& 2.30& 2.40&$ -4.77$&  4.5&  4.3 \\ 
24&     WC8--9d?& 0.23& 0.26& 2.62& 2.39& 2.51&$ -6.30$& 11.0\tablenotemark{b}&  4.8\tablenotemark{b} \\ 
25&    WN5--6b& 0.37& 0.27& 1.95& 1.77& 1.86&$ -4.77$&  4.7&  4.1 \\ 
26&     WC9d?& 0.23& 0.26& 1.95& 1.63& 1.79&$ -6.30$&  9.8\tablenotemark{b}&  4.0\tablenotemark{b} \\ 
27&     WN6& 0.37& 0.11& 2.12& 2.27& 2.19&$ -4.70$&  6.3&  3.3 \\ 
28&     WC8& 0.43& 0.38& 1.86& 1.64& 1.75&$ -4.65$&  4.5&  4.2 \\ 
29&    WN9& 0.13& 0.11& 2.87& 2.66& 2.77&$ -5.92$&  5.2&  3.7 \\ 
30&     WN6& 0.37& 0.11& 1.10& 1.14& 1.12&$ -4.70$&  3.4&  4.9 \\ 
31&     WN5& 0.18& 0.16& 2.26& 2.01& 2.13&$ -4.41$&  3.7&  4.7 \\ 
32& WN9& 0.18& 0.12& 2.01& 2.03& 2.02&$ -6.87$&  6.4&  2.2 \\ 
33&     WN8--9& 0.13& 0.11& 1.85& 1.71& 1.78&$ -5.92$& 10.3&  3.0 \\ 
34&    WN6--7& 0.37& 0.11& 0.85& 0.91& 0.88&$ -4.70$&  3.1&  5.0 \\ 
35&     WN8& 0.13& 0.11& 1.42& 1.34& 1.38&$ -5.92$&  5.8&  3.1 \\ 
36&    WN7& 0.37& 0.09& 0.62& 0.79& 0.70&$ -5.18$&  3.6&  4.7 \\ 
37&     WN6& 0.37& 0.11& 0.59& 0.67& 0.63&$ -4.70$&  5.0&  3.8 \\ 
38&    WN7& 0.37& 0.09& 0.73& 1.03& 0.88&$ -5.18$&  2.4&  5.8 \\ 
39&     WC8& 0.43& 0.38& 1.60& 1.43& 1.51&$ -4.65$&  3.2&  5.2 \\ 
40&     WC9d& 0.23& 0.26& 2.14& 2.13& 2.14&$ -6.30$&  4.9\tablenotemark{b}&  4.2\tablenotemark{b} \\ 
41&     WN7& 0.37& 0.09& 1.80& 1.91& 1.85&$ -5.18$&  6.1&  3.6 \\ 
42&     WC8& 0.43& 0.38& 1.23& 1.00& 1.12&$ -4.65$&  4.1&  4.7 \\ 
43&     WN7& 0.37& 0.09& 0.90& 1.10& 1.00&$ -5.18$&  3.5&  5.2 \\ 
44&     WN8--9& 0.13& 0.11& 2.06& 1.92& 1.99&$ -5.92$&  5.2&  4.2 \\ 
45&    WN7& 0.37& 0.09& 2.44& 2.41& 2.42&$ -5.18$&  6.0&  3.9 \\ 
46&     WN7& 0.37& 0.09& 0.74& 0.97& 0.86&$ -5.18$&  3.7&  5.1 \\ 
47&     WN7& 0.37& 0.09& 1.82& 1.83& 1.83&$ -5.18$&  7.8&  4.0 \\ 
48&     WN7& 0.37& 0.09& 1.82& 1.91& 1.87&$ -5.18$&  6.5&  3.9 \\ 
49&     WC8d?& 0.43& 0.38& 2.04& 1.98& 2.01&$ -4.65$&  4.6\tablenotemark{b}&  4.7\tablenotemark{b} \\ 
50\tablenotemark{a}& if OVe & $0.2$ & $0.1$ & 0.38 & 0.38 & 0.38  & $-4$ & 3.3 &   5.5  \\
51&     WN8--9& 0.13& 0.11& 1.84& 1.75& 1.79&$ -5.92$&  5.6&  4.5 \\ 
52&   WN5--6& 0.18& 0.16& 1.77& 1.65& 1.71&$ -4.41$&  3.5&  5.5 \\ 
53&     WC8& 0.43& 0.38& 2.02& 1.89& 1.96&$ -4.65$&  6.2&  4.9 \\ 
54&     WC8& 0.43& 0.38& 1.90& 1.63& 1.76&$ -4.65$&  6.0&  4.9 \\ 
55&     WN8& 0.13& 0.11& 1.44& 1.39& 1.42&$ -5.92$&  7.7&  5.8 \\ 
56&     WC8& 0.43& 0.38& 1.18& 1.22& 1.20&$ -4.65$&  7.0&  6.1 \\ 
57&     WC7& 0.62& 0.58& 2.10& 1.61& 1.86&$ -4.59$&  7.1&  6.0 \\ 
58&     WC8--9d?& 0.23& 0.26& 2.33& 1.99& 2.16&$ -6.30$&  8.9&  7.1 \\ 
59&     WC7& 0.62& 0.58& 1.59 &  1.18 & 1.39 &$ -4.59$& 3.8  & 5.5  \\ 
60&     WC8d& 0.43& 0.38& 1.89& 1.93& 1.91&$ -4.65$&  2.9\tablenotemark{b}&  6.5\tablenotemark{b} \\ 
61&     WN7& 0.37& 0.11& 1.49& 1.53& 1.51&$ -4.70$&  4.6&  6.4  
\enddata
\tablecomments{The values of $(J-{K_s})_0$, $(H-{K_s})_0$, $M_{K_s}$ were adopted from Crowther et al. (2006b) according to WR subtype (see text). Two independent values of $A_{K_s}$ were calculated and averaged for a final extinction estimate. The distance uncertainties are estimated at $\approx$25\%, but may range as high as $\approx$40\% (see text).}
\tablenotetext{a}{Although the nature of MDM50 is uncertain, the available data are most consistent with an OVe classification and the intrinsic color values in the table (see text).}
\tablenotetext{b}{The distances to these stars should be taken with additional caution, since they are candidates for thermal dust emission. As a result, the adopted colors and $K_s$-band absolute photometry used to make the distance calculation might not be reliable.}
\label{tab:phot2}
\end{deluxetable}

\def\arraystretch{1.7}
\begin{deluxetable}{ccccrr}
\tablecolumns{6}
\tabletypesize{\scriptsize}
\tablecaption{X-ray Detections of WRs}
\tablehead{
\colhead{Star} & \colhead{X-ray Source} & \colhead{$F_X$ (0.5--7.0 keV)}   & \colhead{${N_H}^{IR}$} & \colhead{$F_X^{cor}$ (0.5--7.0 keV)} & \colhead{$L_X$ (0.5--7.0 keV)} \\ [2pt]
\colhead{MDM}  & \colhead{[CXO J]}  & \colhead{[ergs s$^{-1}$ cm$^{-2}$]} & \colhead{[$10^{22}$ cm$^{-2}$]} & \colhead{[ergs s$^{-1}$ cm$^{-2}$]} & \colhead{[ergs s$^{-1}$]}}
\startdata
1  &  102328.8$-$574629  &$6.08^{5.76}_{6.40}\times10^{-14}$ &  2.2 &  (1.8--9.3)$\times10^{-13}$      &   (3--16)$\times10^{32}$                     \\  
3 & 131209.0$-$624326   &$6.68^{6.37}_{6.96}\times10^{-14}$ & 2.6 &  (2.2--12.6)$\times10^{-13}$    &       (1--18)$\times10^{32}$                                  \\ 
4 & 131221.2$-$624012  &  $9.19_{8.21}^{10.18} \times10^{-15}$  & 1.4 &  (2.2--8.6)$\times10^{-14}$    &       (3--11)$\times10^{31}$                         \\ 
5 &131225.4$-$624441 & $2.39_{1.70}^{3.09}\times10^{-15}$ &   2.1  &  (6.9--34.8)$\times10^{-15}$    &     (6--33)$\times10^{30}$                           \\ 
8 &  131228.5$-$624143 & $4.10_{3.89}^{4.31}\times10^{-14}$ & 1.6  & (1.0--4.4)$\times10^{-13}$    &      (5--22)$\times10^{31}$                            \\ 
WR48a & 131239.5$-$624255 & $5.00_{4.98}^{5.03}\times10^{-12}$ &   3.3\tablenotemark{a}  & (1.8--12.8)$\times10^{-11}$ &   (3--19)$\times10^{34}$                              \\ 
30  & 164746.0$-$455904 & $5.48_{4.90}^{6.07}\times10^{-14}$ & 1.8 & (1.5--6.7)$\times10^{-13}$ & (2--9)$\times10^{32}$ \\ 
34 & 181322.4$-$175350 &                         $2.04_{1.69}^{2.39}\times10^{-14}$ &    1.4&       (2.0--19.0)$\times10^{-14}$  &  (2--22)$\times10^{31}$                                                      \\ 
36 & 182606.1$-$130410 &                         $1.09_{0.83}^{1.35}\times10^{-14}$ &   1.1 & (2.4--8.0)$\times10^{-14}$ & (4--12)$\times10^{31}$ \\ 
38 & 183116.4$-$100924 &                           $6.36_{6.04}^{6.68}\times10^{-13}$ &    1.4&   (1.5--5.9)$\times10^{-12}$  & (1--4)$\times10^{32}$                                                               \\ 
45& 184332.5$-$040419 &                           $3.70_{3.40}^{4.00}\times10^{-14}$ &   3.9 &       (1.5--11.8)$\times10^{-13}$  &  (7--51)$\times10^{32}$                                                     \\  
HDM9 & 181922.1$-$160309 &                  $1.26_{1.10}^{1.43}\times10^{-13}$ &    1.2 &        (2.8--10.0)$\times10^{-13}$  &   (4--17)$\times10^{33}$                                                  \\ 
WR19a & 	101853.3$-$580753				 &                  $8.58_{7.59}^{9.59}\times10^{-14}$ &  1.5    &        (2.1--8.6)$\times10^{-13}$  &   (1--10)$\times10^{32}$  \\
\enddata
\tablecomments{Basic X-ray photometry data were obtained from the \textit{Chandra} point source catalog (Evans et a. 2010). $F_X$ is the total flux within the 0.5--7.0 keV energy band, and includes 90\% confidence values. ${N_H}^{IR}$ is the interstellar absorption column derived by inserting the infrared extinction values in Table 5 into the relations of Cardelli et al. (1989) and Predehl \& Schmitt (1995).  $F_X^{cor}$ and $L_X$ are the absorption-corrected fluxes and luminosities, respectively, derived using the PIMMS software package. The spread in the individual $F_X^{cor}$ and $L_X$ values were derived assuming that the X-ray sources can be approximated by a single-temperature thermal plasma having $kT$ in the range of 0.6--2 keV, which is generally a safe approximation for both single WRs and those in colliding-wind binaries. We estimate that the derived $L_X$values are accurate to 1 dex. The table includes X-ray data for new WRs, as well as for WR48a, HDM9 from \ca{had07}(\cy{had07}) and WR19a, the latter two of which were not previously recognized as X-ray sources. Stars MDM1, MDM3, MDM4, MDM5, MDM8, and WR48a lie in the field of the Danks 1 and 2 double cluster shown in Figure \ref{fig:danks}. WR48a, a known colliding-wind X-ray binary (WC6+O), is included in the table for comparison; it is 2--3 orders of magnitude brighter in X-rays than any of the neighboring sources.}
\tablenotetext{a}{Using the method in Section 6, we derived the extinction of $A_{K_s}$=2.05 mag for WR48a (WC6+O) using its observed 2MASS photometry ($J$=8.74 , $H$=6.80, and $K_s$=5.09 mag ) and assuming the intrinsic colors for WC5--7 stars from Crowther et al. (2006); the same calculations were performed for WR19a ($J$=9.08 , $H$=8.13, and $K_s$=7.50 mag ), for which we obtained  $A_{K_s}$=0.96 mag and $D=3.1$ kpc. The corresponding value of ${N_H}^{IR} $ was calculated in the same manner as the other sources. We use the accepted average distance to Danks 1 and 2 of  3.5 kpc for WR48a.}
\label{tab:x}
\end{deluxetable}
\clearpage
\newpage

\begin{deluxetable}{lc}
\tablecolumns{2}
\tablewidth{0pc}
\tabletypesize{\scriptsize}
\tablecaption{Total Galactic WR Census}
\tablehead{
 \colhead{Reference} & \colhead{N (WR)}} 
\startdata
Van der Hucht (2001, 2006)  & 171 WN, 10 WN/WC, 113 WC, 4 WO\\
Kurtev et al. (2007) &                         4 WN \\
Hadfield et al. (2007) &                   11 WN,                        4 WC\\
Hyodo et al. (2008) &                                                            1 WC\\
Messineo et al. (2009) &                   2 WN,                        2 WC\\
Liermann et al. (2009) &                                                      2 WC\\
Mikles et al. (2006) &                          1 WN \\
Hanson et al. (2010a)  &                         1 WN \\
Gvaramadze et al. (2009) &              1 WN\\
Gvaramadze et al. (2010) &              1 WN\\
Mauerhan et al. (2007) &                   2 WN\\
Mauerhan et al. (2009) &                   9 WN,                       3 WC\\
Mauerhan et al. (2010a) &                 5 WN,                       6 WC\\
Mauerhan et al. (2010b) &                 7 WN,                       2 WC\\
Mauerhan et al. (2010c) &                 2 WN,                       1 WC\\
Motch et al. (2010)      &                     1 WN      \\
Roman-Lopes et al. (2011) &             1 WN,                       1 WC\\
Shara et al. (2009) &         15(13)\tablenotemark{a} WN, 26 WC\\ 
Wachter et al. (2010) &                        5 WN,                        1 WC\\
Anderson et al. (2011) &                     3 WN \\
This work &                                          38 WN                        22 WC \\                           
\textbf{Total} &                                                (278 WN)$+$(10 WN/WC)$+$(184 WC)$+$(4 WO)=\textbf{476}  
\enddata
\tablecomments{The table includes the WRs cataloged in van der Hucht (2001, 2006) and other references that contain all subsequent discoveries. Note that some of the listed works contain references therein that credit the original discoveries made by authors unlisted in this table.}
\tablenotetext{a}{2 WN stars were co-discovered by Mauerhan et al. (2009) and Shara et al. (2009); so we only count 39 WRs from the latter author.}
 \label{tab:wr_census}
\end{deluxetable}

\def\arraystretch{1.7}
\begin{landscape}
\begin{deluxetable}{lccrrrrrrrrccc}
\setlength{\tabcolsep}{0.06in}
\tablecolumns{14}
\tablewidth{0pc}
\tabletypesize{\scriptsize}
\tablecaption{Photometry of Cluster Neighbors to WC8 Stars MDM52 and MDM53}
\tablehead{
 \colhead{Star} & \colhead{Source} & \colhead{R.A.} & \colhead{Decl.}  & \colhead{$J$} & \colhead{$H$} & \colhead{$K_s$} &\colhead{$(J-K)_0$} &\colhead{$(H-K)_0$} &\colhead{$A_{K_{s}}^{J-K_s}$} &\colhead{$A_{K_{s}}^{H-K_s}$} &\colhead{$\overline{A_{K_S}}$} &\colhead{$M_{K_s}$} & \colhead{Spectral}  \\ [2pt]
 \colhead{} & \colhead{(2MASS J)} & \multicolumn{2}{c}{(deg, J2000)}  & \colhead{(mag)} & \colhead{(mag)} & \colhead{(mag)} & \colhead{(mag)} & \colhead{(mag)} & \colhead{(mag)} & \colhead{mag} & \colhead{mag} & \colhead{mag} &\colhead{Type} }
\startdata
A &19012504$+$0351392 &285.354335&  	3.860890 &13.491&  11.367  &10.294 &  \nodata  &  $-0.08$ &  \nodata  & 2.10 & 2.10 &$-5.73$ & B0--3 (I--II) \\  
B &19012634$+$0351471 &285.359753 & 	3.863086 &12.829&  10.733   &9.630 &   \nodata  &  $-0.08$ & \nodata  & 2.15 & 2.15 & $-6.45$& B0--3 (I--II)  \\ 
C &19012726$+$0351461 &285.363620 & 	3.862817 &13.143&  11.074  &10.102 &   0.79       &  0.18       & 1.51 & 1.44 &1.48&  \nodata &M  \\ 

D &19012736$+$0351587 &285.364040 & 	3.866331 &13.590&  11.409 & 10.252 &  	\nodata  & $ -0.08 $& \nodata & 2.25  & 2.25 & $-5.93 $& B0--3 (I--II) \\ 

E &19012738$+$0352203 &285.364104 & 	3.872317 &13.412&  10.871  & 9.673 &   0.79         &   0.18      & 1.98 & 1.85 &1.92&  \nodata &M \\ 
F &19012804$+$0351581 &285.366835 & 	3.866142 &12.504&  10.281  & 9.115 & \nodata     &  $-0.08$ & \nodata &  2.27 & 2.27 & $-7.08$ & B0--3 (I--II) \\ 
G &19012889+0351541 & 285.370405  &	3.865032 &14.136 & 11.977  & 10.927 &  \nodata &  $-0.08$  & \nodata  & 2.06   & 2.06  & $-5.06$ & B0--3 (I--II)  \\ 
H &19013009$+$0351595& 285.375405 & 	3.866535& 13.101&  10.806  & 9.831 &   0.79&  0.18 &1.66 &1.45 &1.56	 &  \nodata & M 	
\enddata
\tablecomments{Photometric data are from the 2MASS  catalog. Extinction values were derived after adopting the intrinsic colors of B0--3 I stars from Bibby et al. (2008). $M_{K_s}$ values were then calculated assuming a distance 6.1 kpc, which is the average distance of the two WC8 stars MDM53 and MDM54 from Table \ref{tab:phot2}. The extinction values of the late-type stars were derived after assuming the intrinsic colors of M0Ia stars from Elias et al. (1985); these colors were adopted to demonstrate that the photometry of the late-type stars is inconsistent with them being red supergiants within the cluster. The data suggest, rather, that they are foreground stars.}
 \label{tab:newclusterphot}
\end{deluxetable}
\clearpage
\end{landscape}

\clearpage

\begin{landscape}
\begin{figure*}[t]
\centering
\subfigure[]{
\includegraphics[scale=0.58]{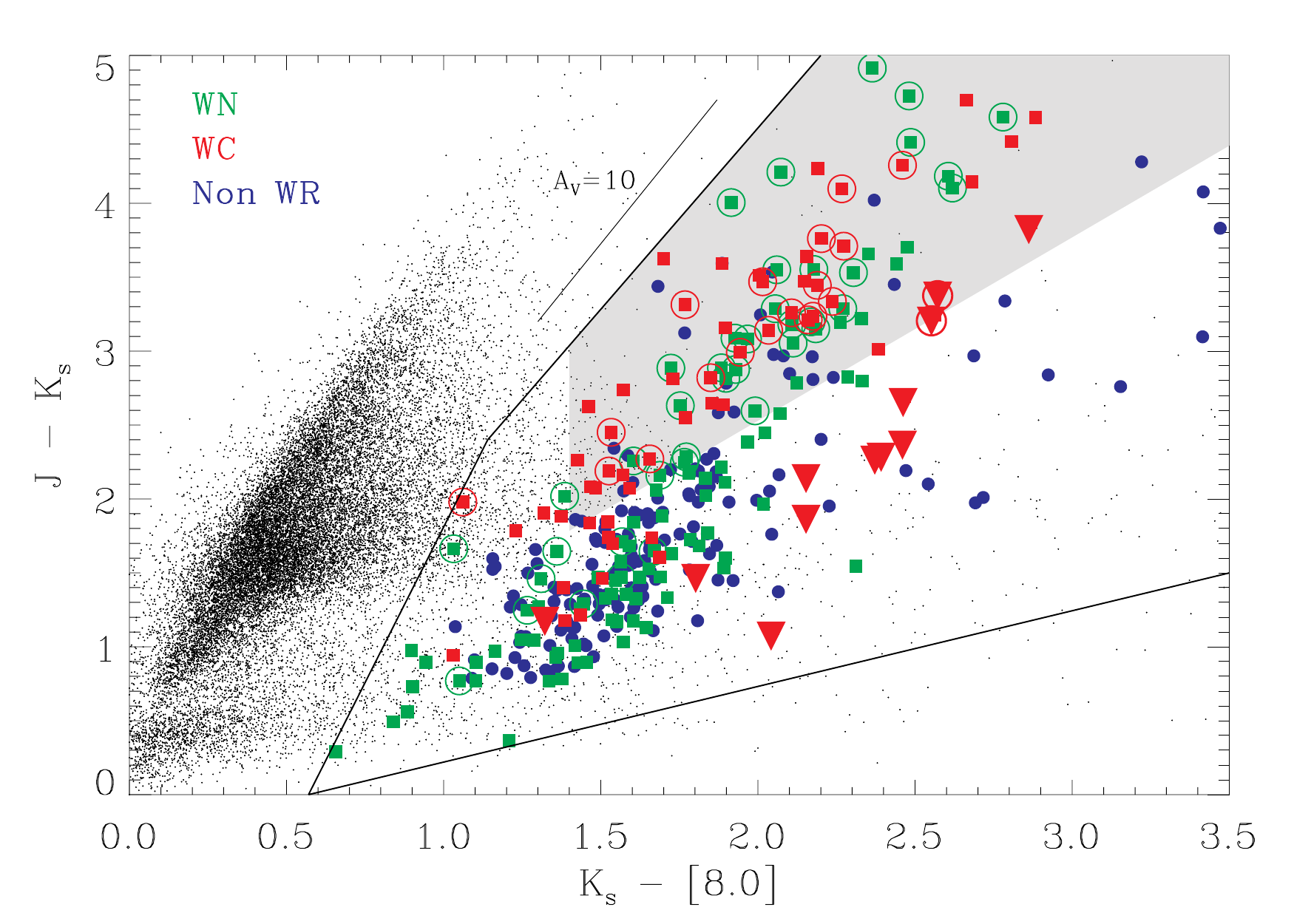}}
\subfigure[]{
\includegraphics[scale=0.58]{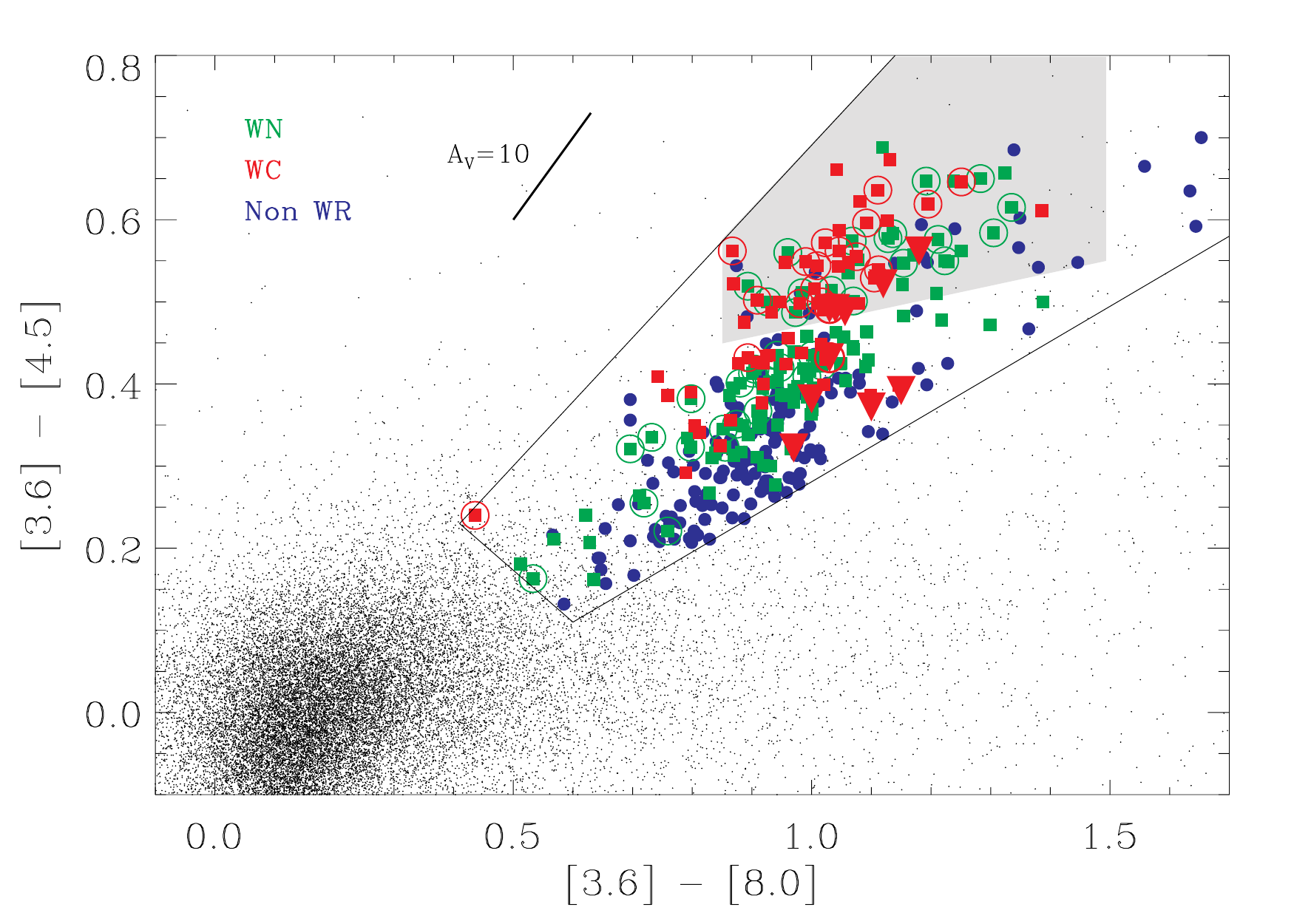}}
\subfigure[]{
\includegraphics[scale=0.58]{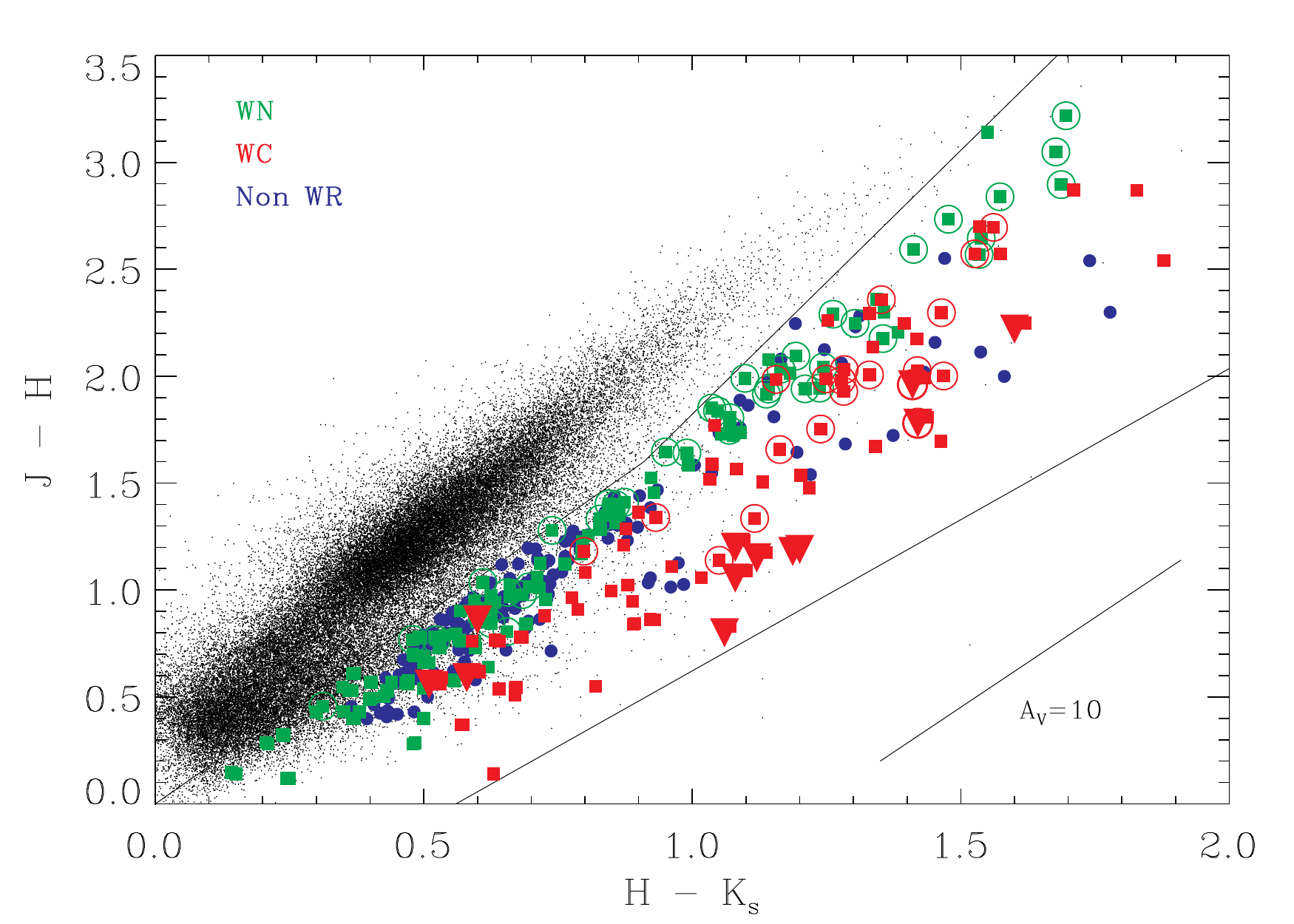}}
\subfigure[]{
\includegraphics[scale=0.58]{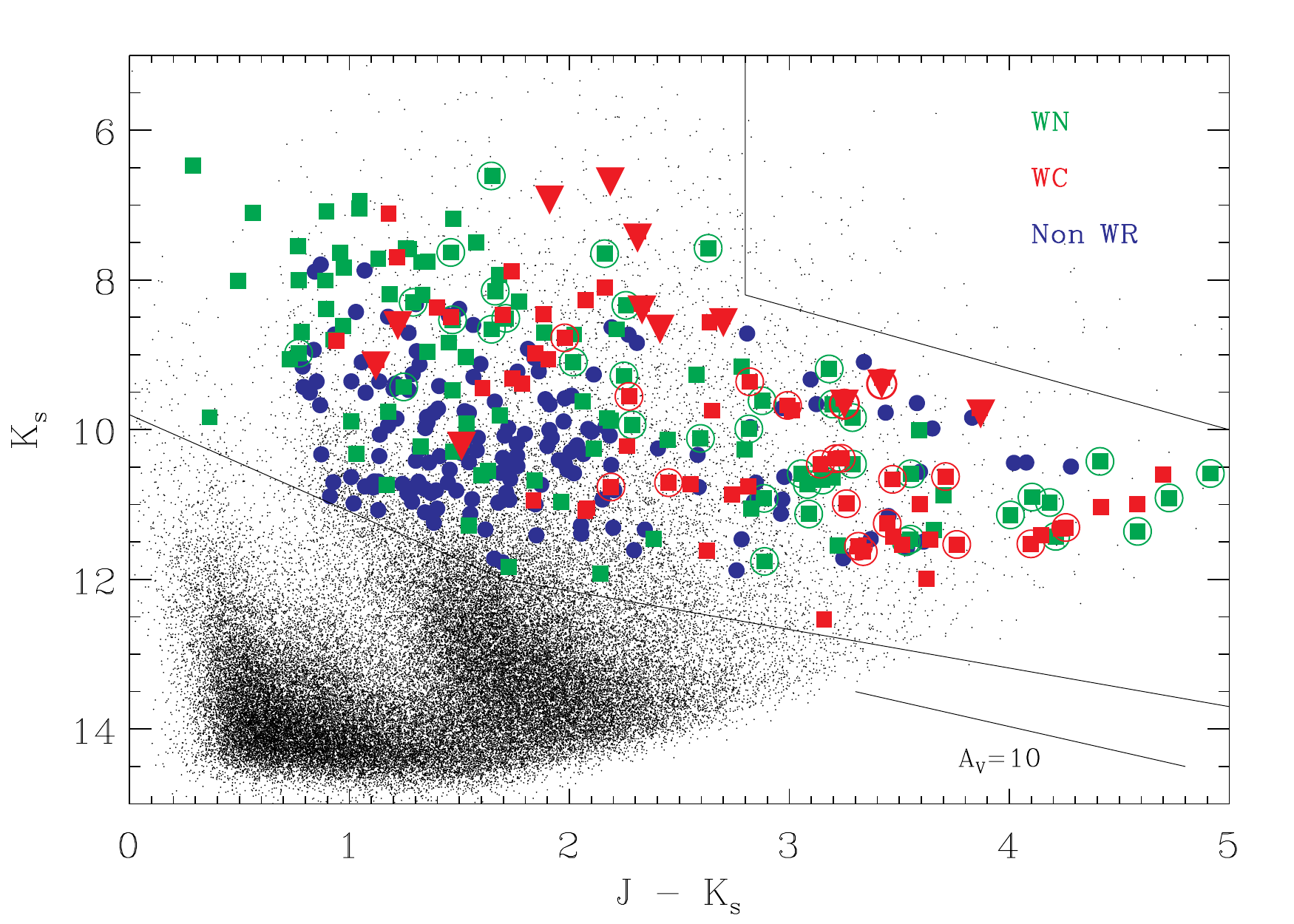}}
\caption[]{\linespread{1}\normalsize{Infrared color-color and color-magnitude diagrams illustrating the color selection criteria of our survey. The solid black lines define the \textit{broad color space}, discussed in the text.  WRs that lie within the GLIMPSE survey area are represented by green and red square symbols. Dusty WC stars are represented by red inverted triangles.  Our latest discoveries (this paper) are marked with an additional, open circle. Non-WRs detected in our survey, which are mainly Be stars, are represented by blue filled circles. The small black points are stars randomly drawn from the GLIMPSE point-source catalog. The grey shaded region in the upper two plots is the WR ``sweet spot," discussed in the text; the confirmation rate of bona fide WRs is relatively high within this region, at $>$50\%.}}
\label{fig:cc}
\end{figure*}
\clearpage
\end{landscape}

\clearpage

\begin{figure*}[t]
\centering
\includegraphics{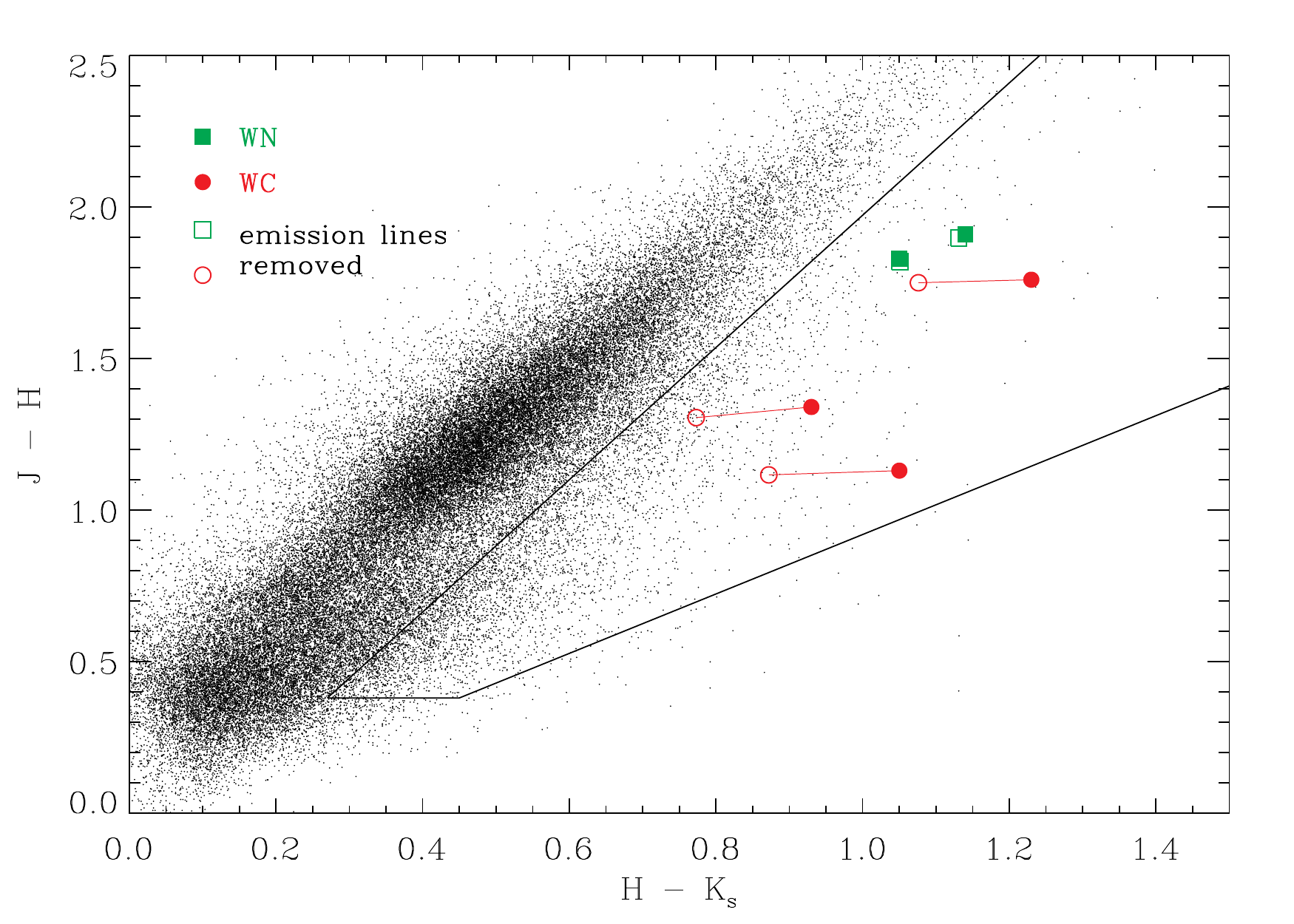}
\caption[]{\linespread{1}\normalsize{Color-color diagram similar to Figure 1c, illustrating the effect that line emission has on the infrared excess of WRs.  The colored lines connect the observed colors (filled symbols) of selected WN and WC stars to the adjusted colors that result from the removal of emission lines (open symbols). WN stars MDM33 (WN9, left) and MDM41 (WN6, right) show little color shift as a result of emission lines, while WC stars MDM42 (WC8, left), MDM56 (WC8, center) and MDM59 (WC6, right), show significant color shift that mainly results from strong emission lines of C {\sc iii} and C {\sc iv} in the $K$ band. The remaining excess relative to the photospheric emission from normal dwarfs and giants (small black points), after the removal of emission lines, is free-free emission from the ionized winds and, possibly, hot dust for some WC stars.}}
\label{fig:cc_lines}
\end{figure*}

\begin{figure*}[t]
\centering
\subfigure[]{
\includegraphics[scale=0.6]{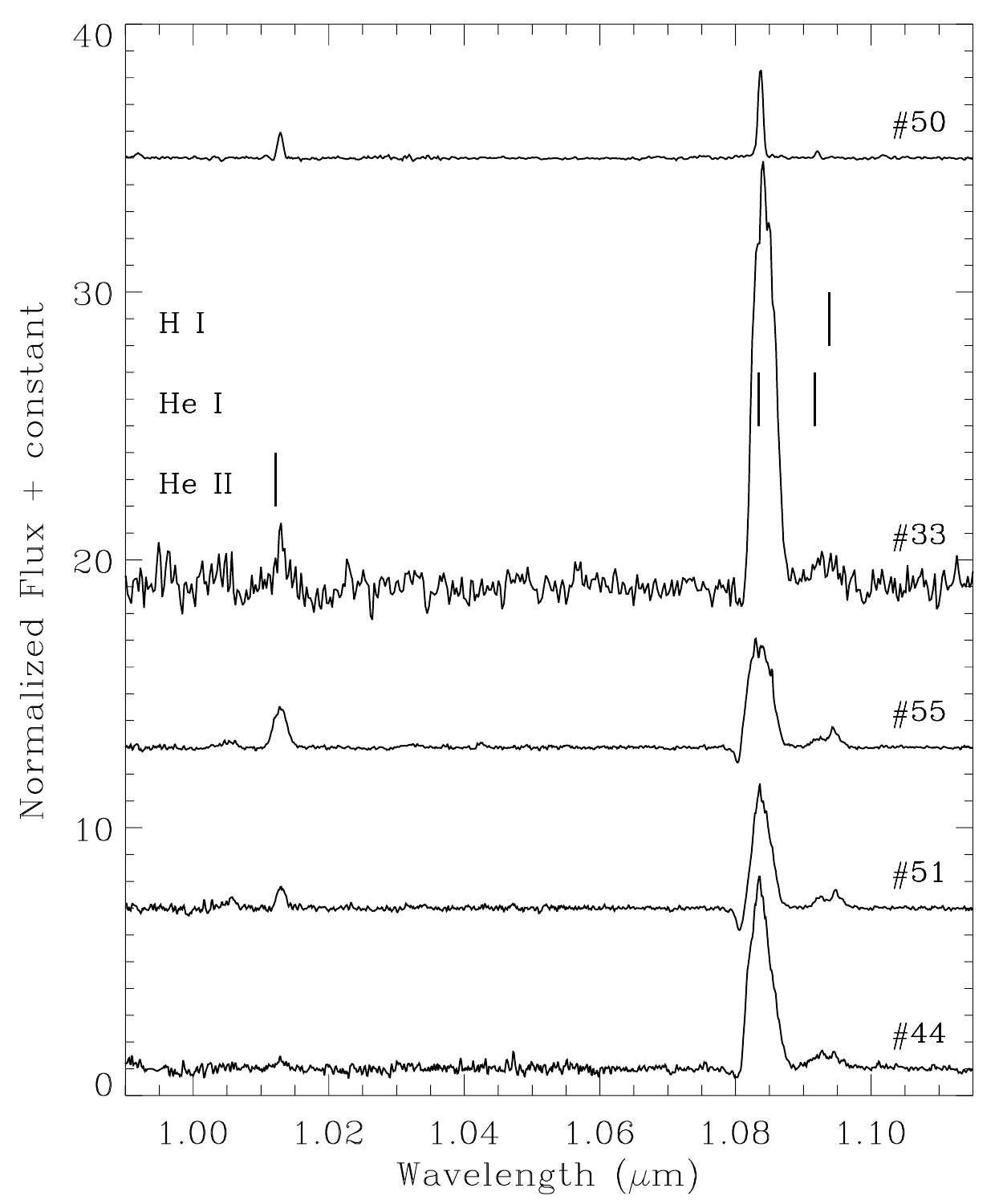}}
\subfigure[]{
\includegraphics[scale=0.6]{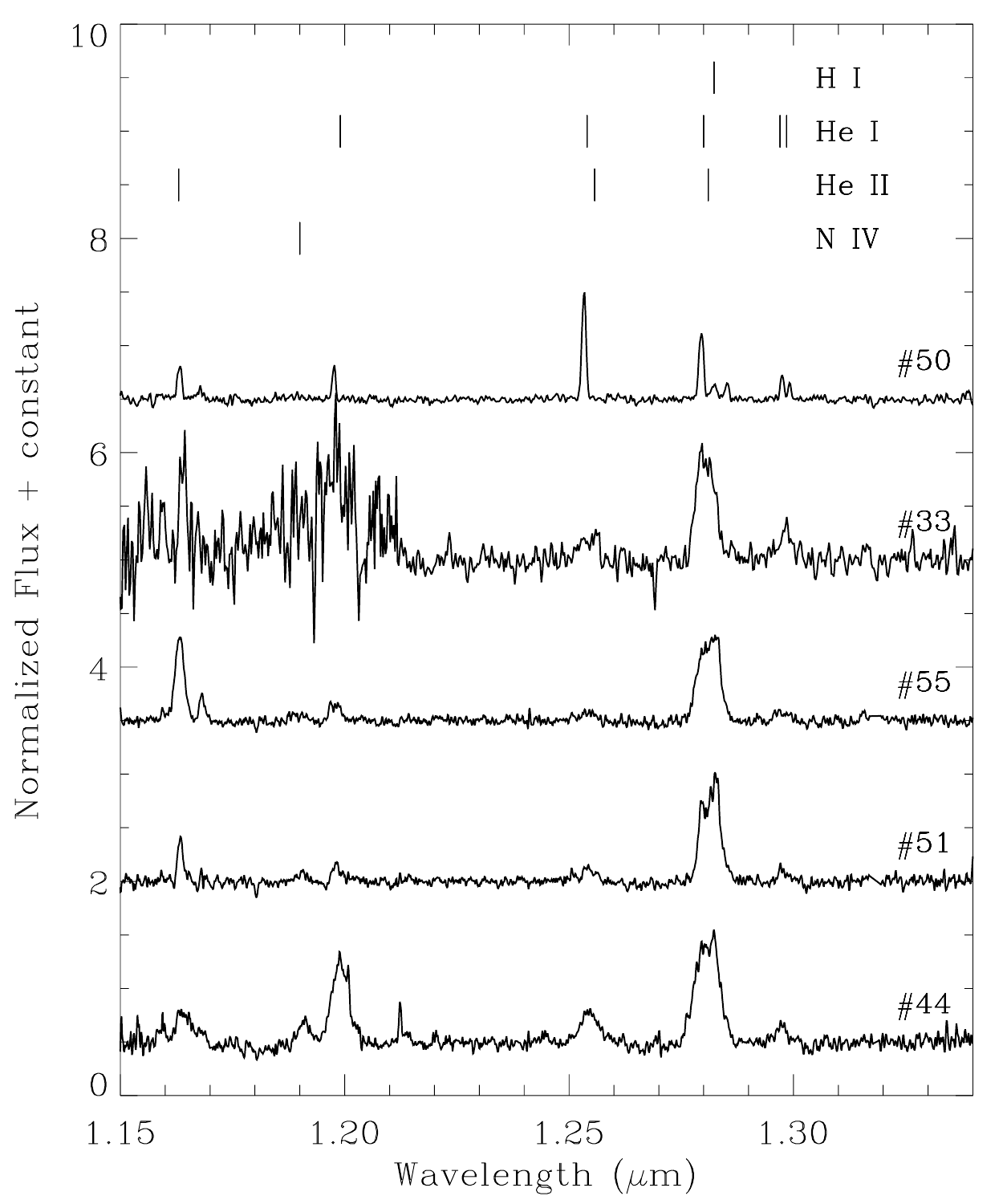}}
\subfigure[]{
\includegraphics[scale=0.6]{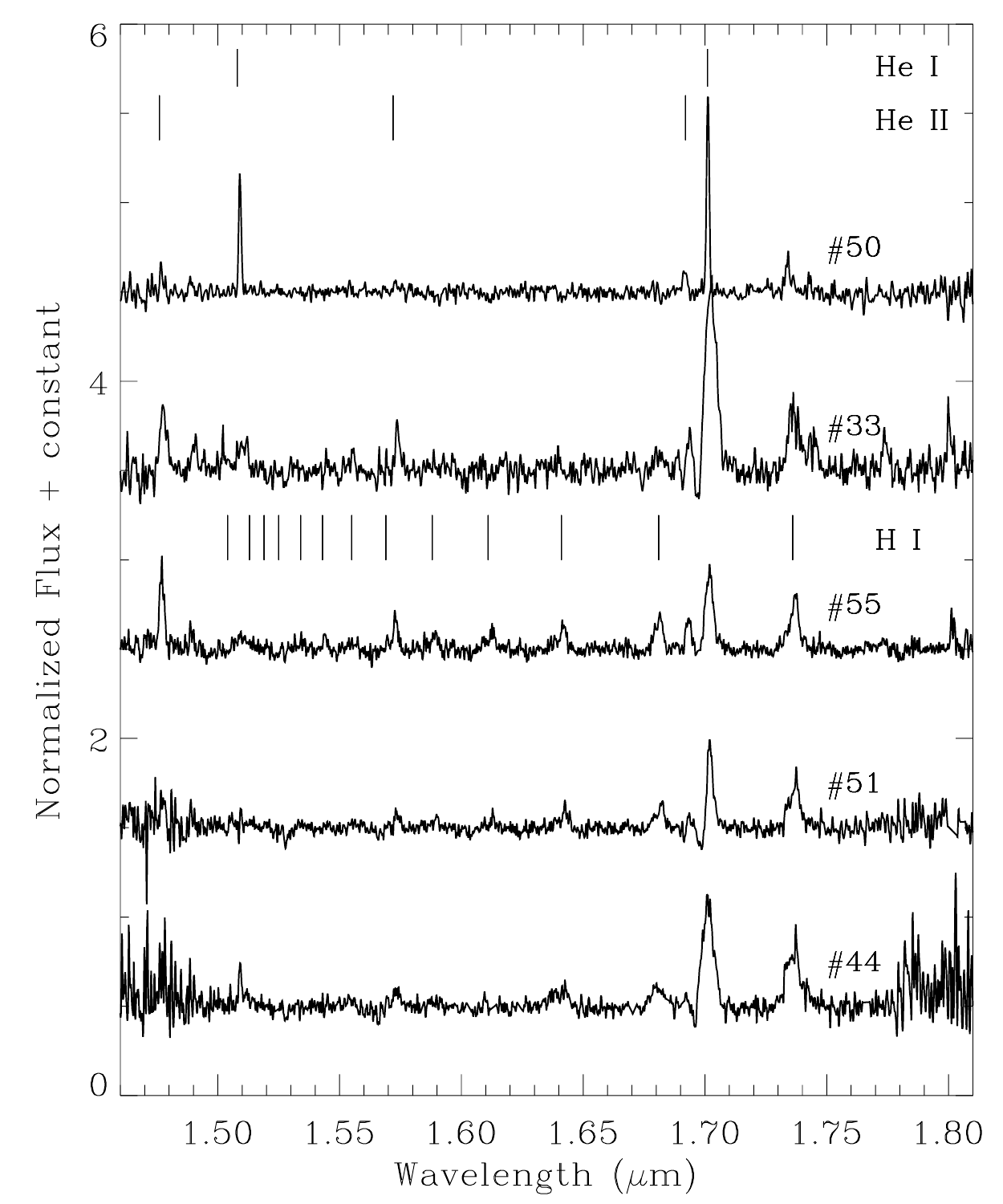}}
\subfigure[]{
\includegraphics[scale=0.6]{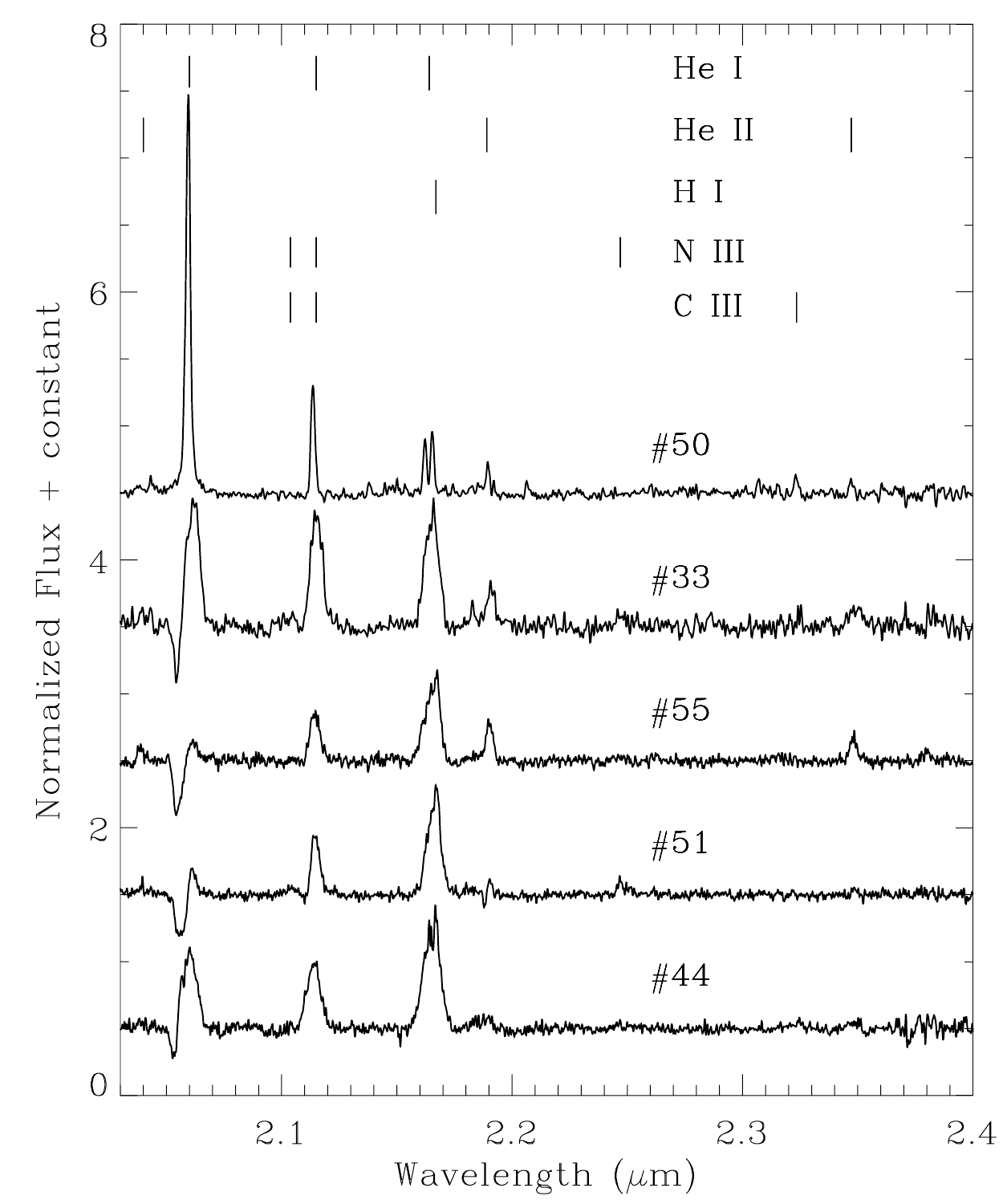}}
\caption[]{\linespread{1}\normalsize{Near-infrared spectra of WN stars at $R\approx1200$--2500 (see text regarding MDM50).}}
\label{fig:wnl_med}
\end{figure*}

\clearpage

\begin{figure*}[t]
\centering
\subfigure[]{
\includegraphics[scale=0.55]{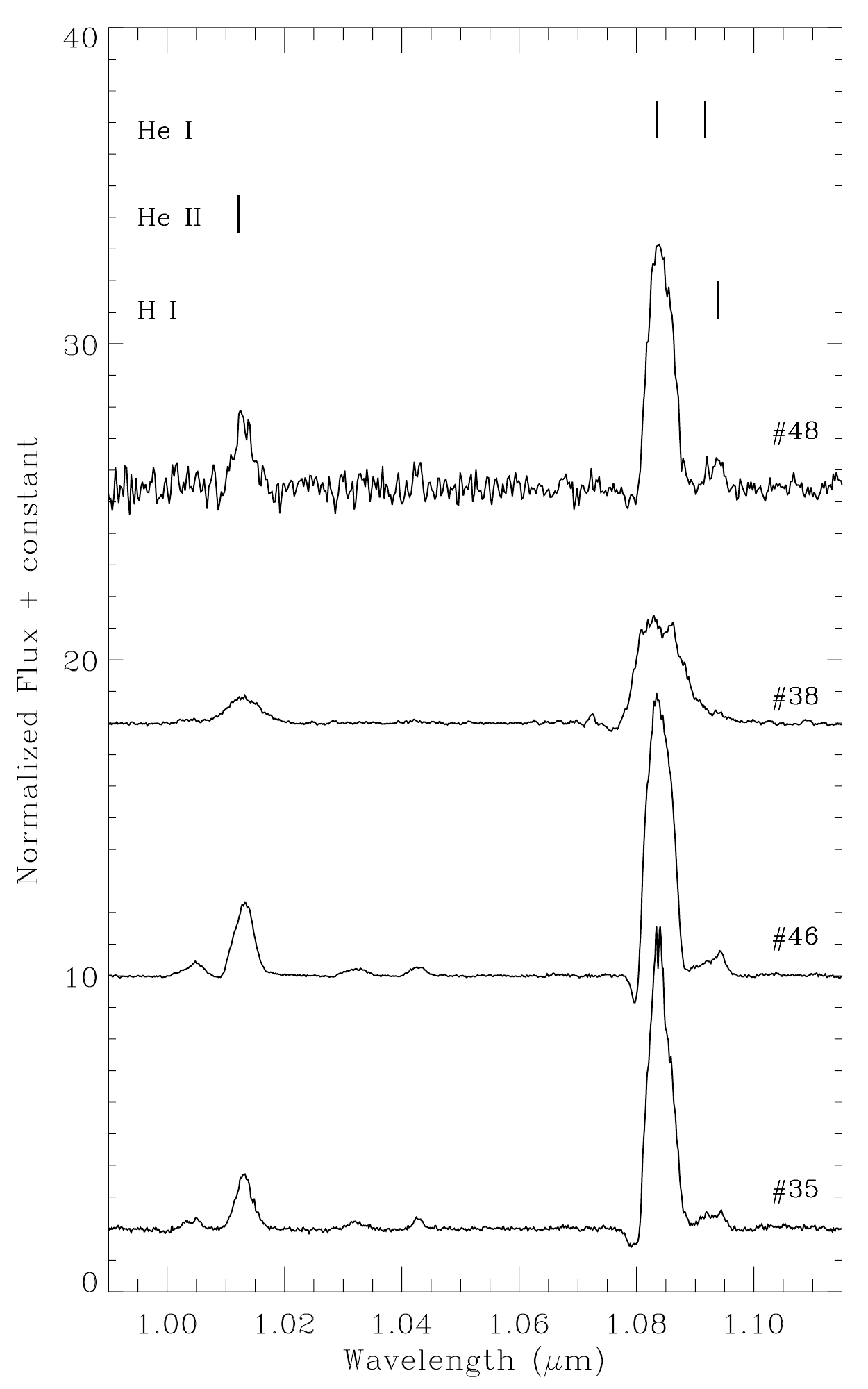}}
\subfigure[]{
\includegraphics[scale=0.55]{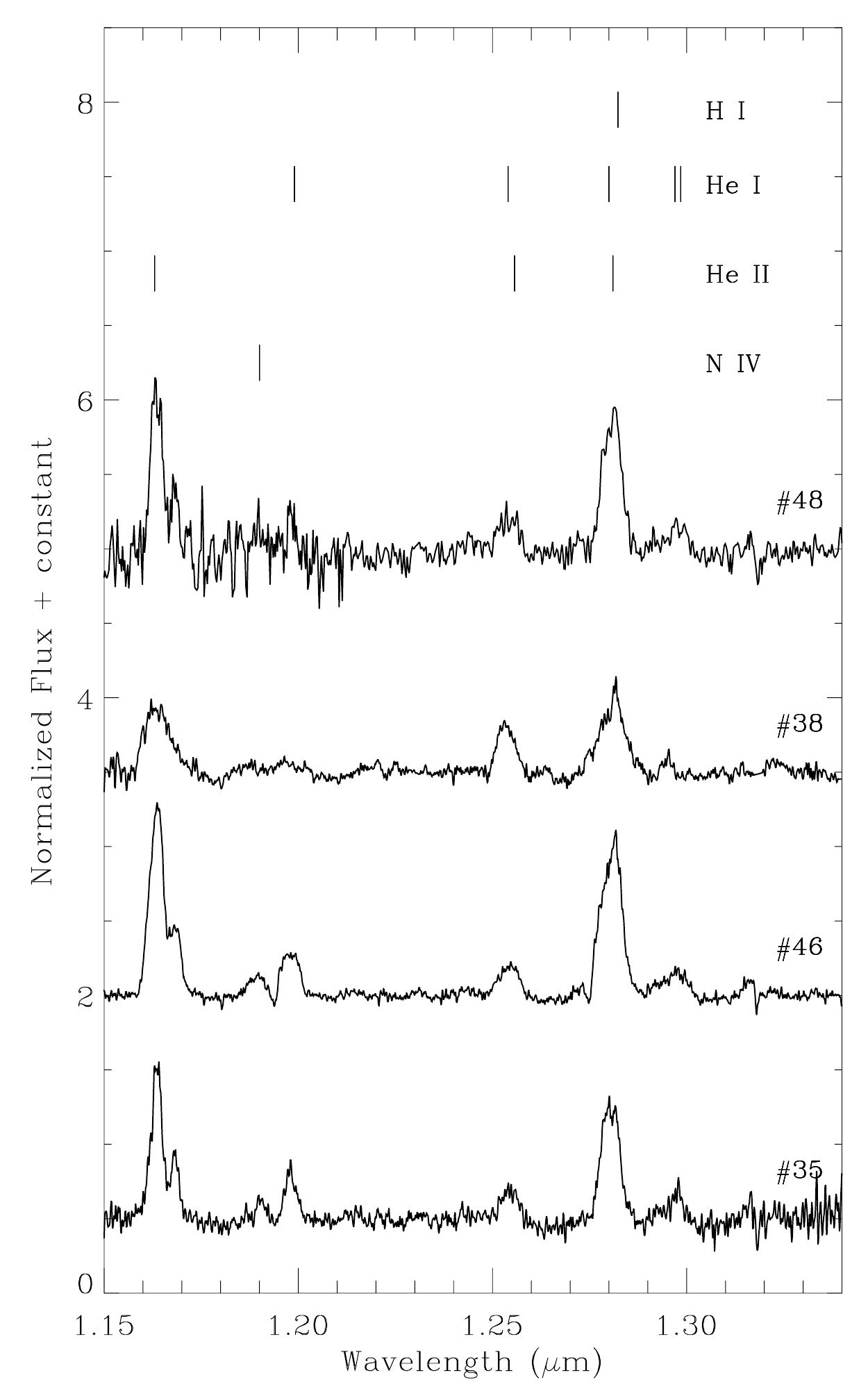}}
\subfigure[]{
\includegraphics[scale=0.55]{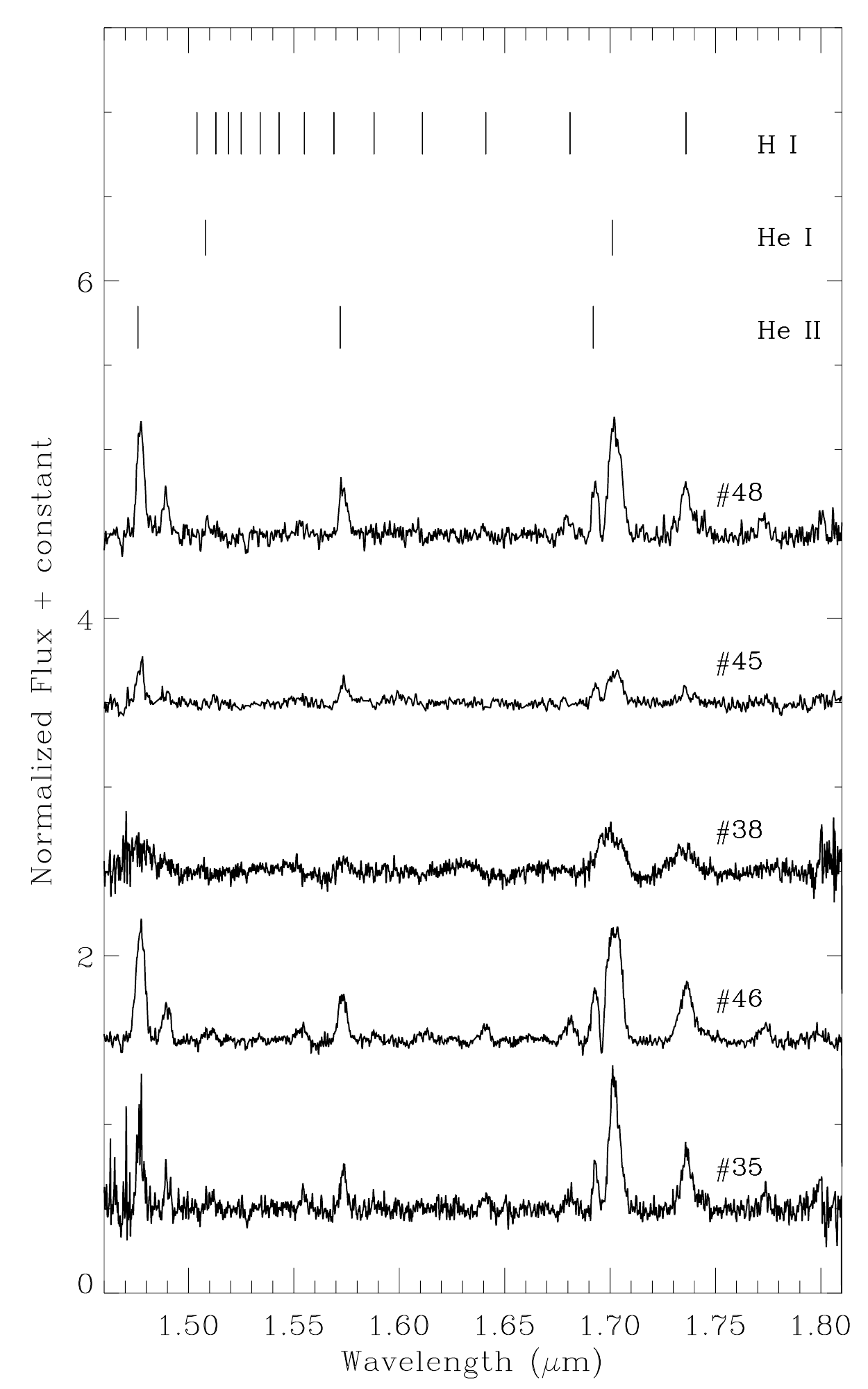}}
\subfigure[]{
\includegraphics[scale=0.55]{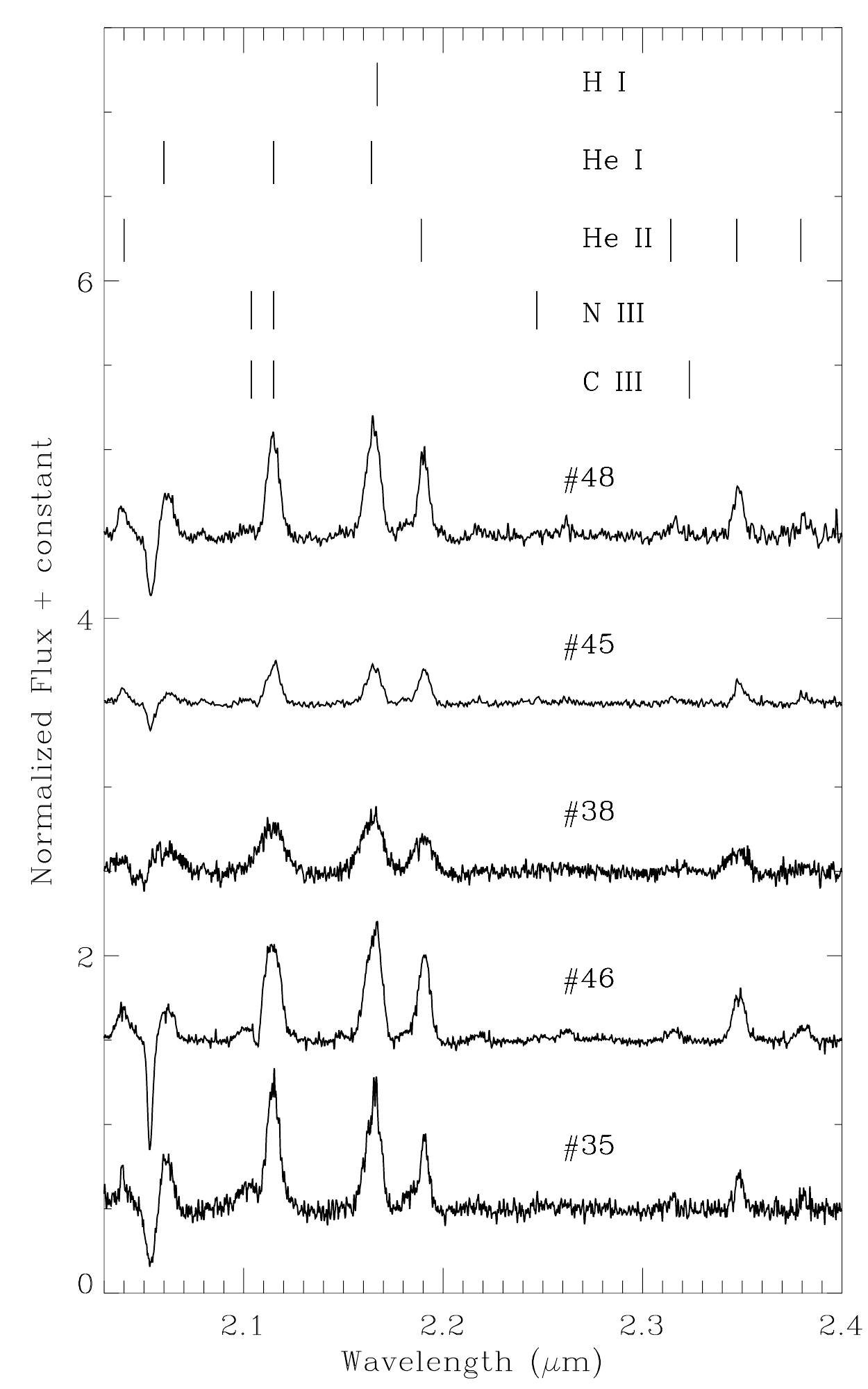}}
\caption[]{\linespread{1}\normalsize{Near-infrared spectra of WN stars at $R\approx1200$--2500.}}
\label{fig:wn7_med}
\end{figure*}

\clearpage

\begin{figure*}[t]
\centering
\subfigure[]{
\includegraphics[scale=0.55]{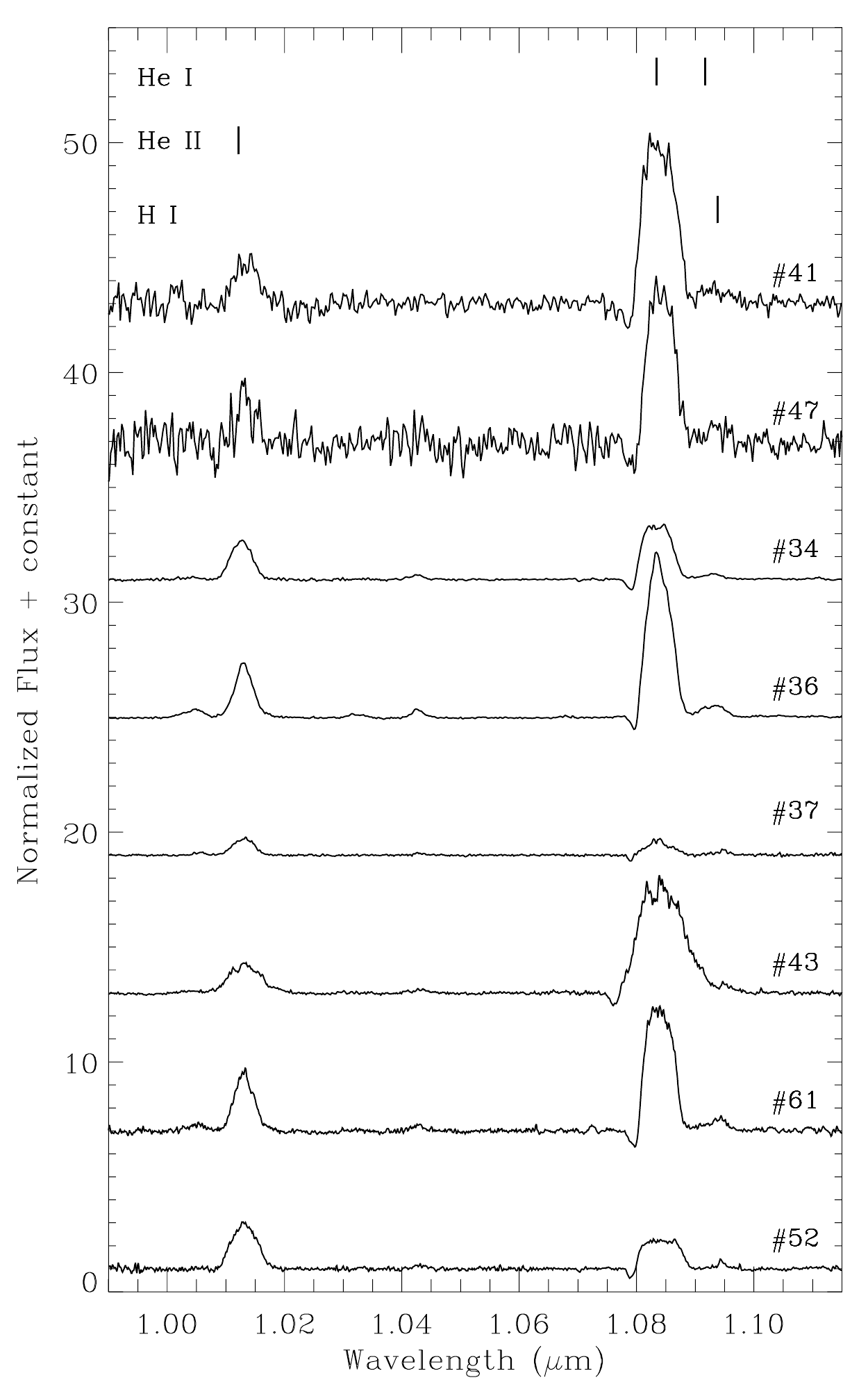}}
\subfigure[]{
\includegraphics[scale=0.55]{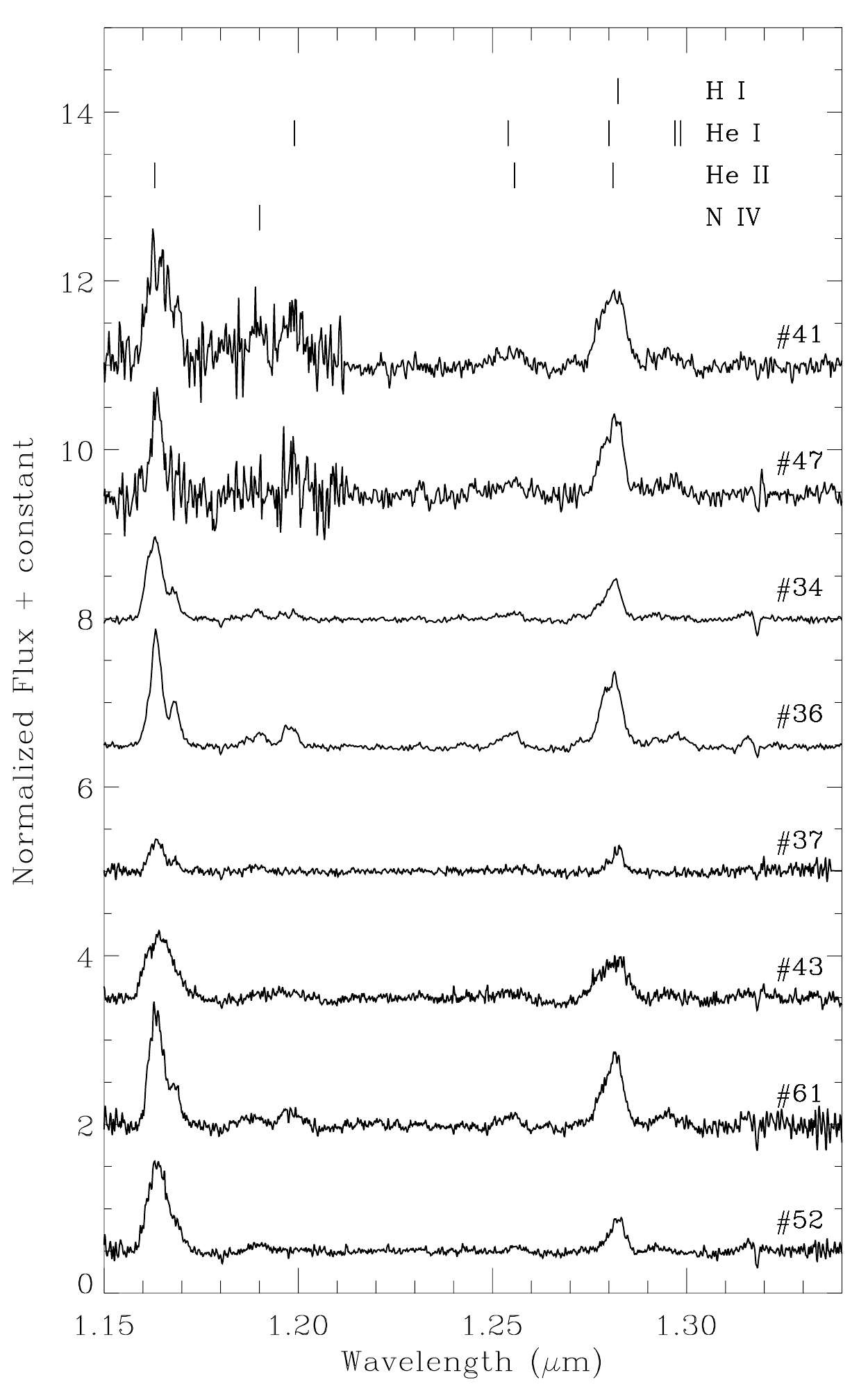}}
\subfigure[]{
\includegraphics[scale=0.55]{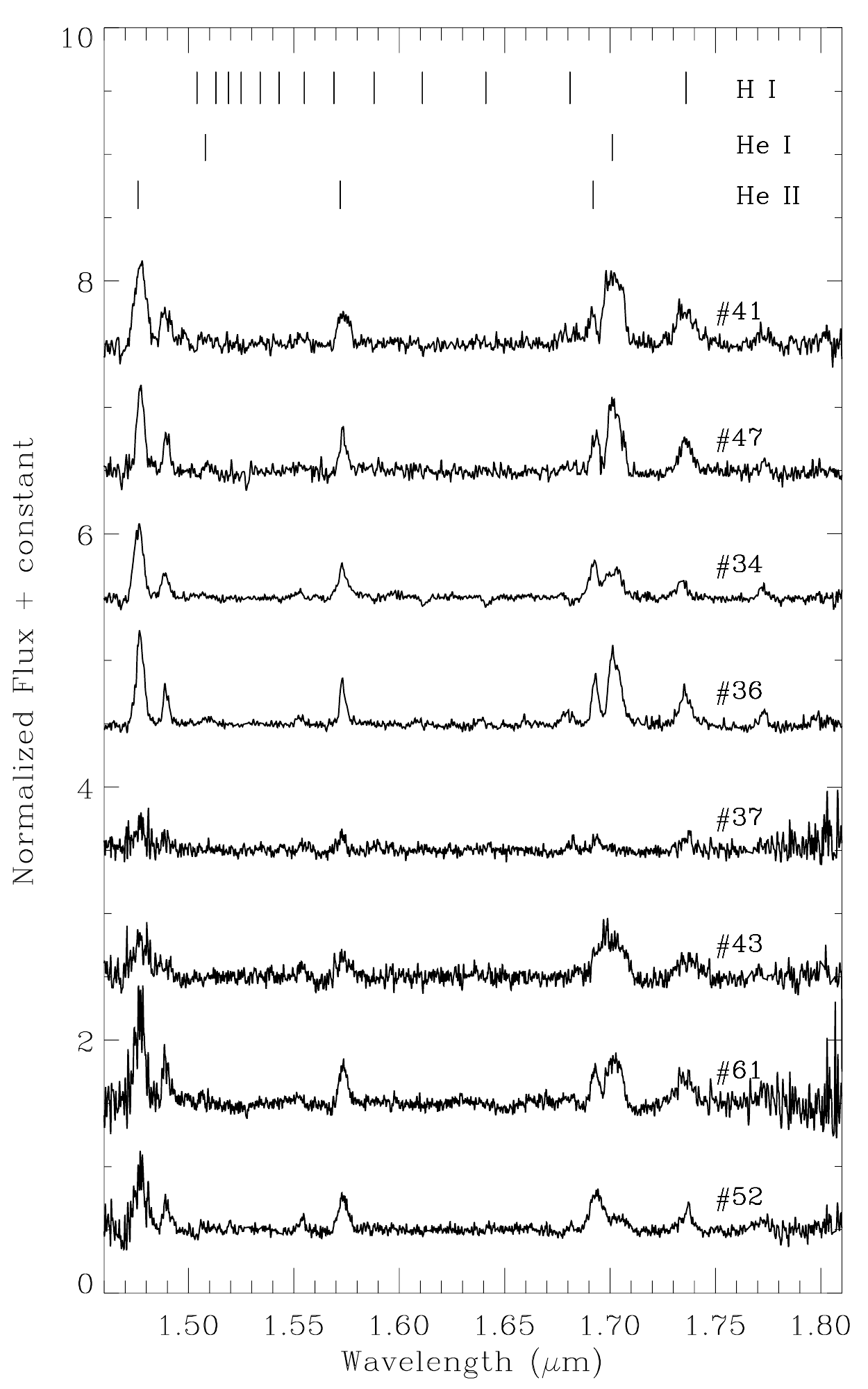}}
\subfigure[]{
\includegraphics[scale=0.55]{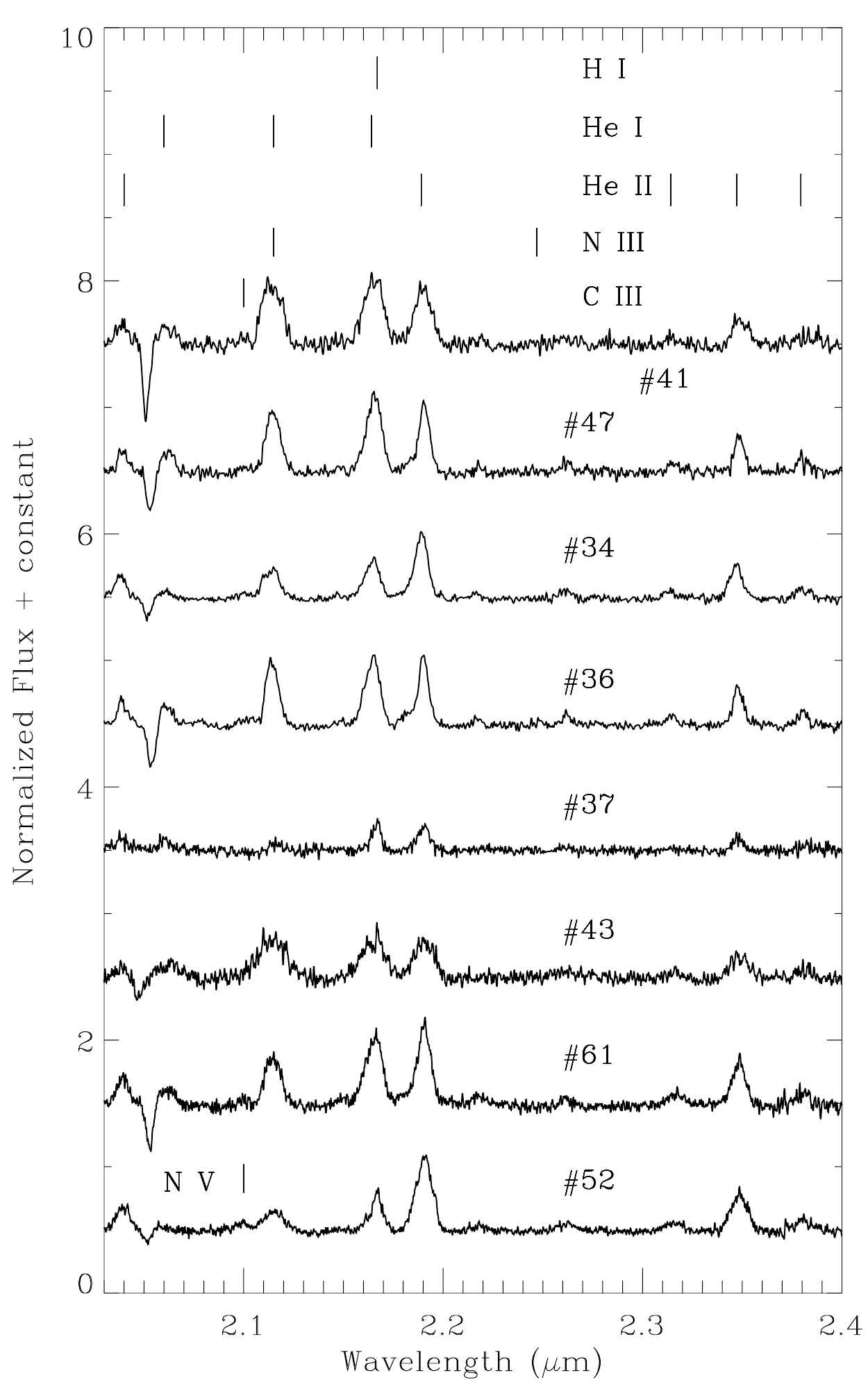}}
\caption[]{\linespread{1}\normalsize{Near-infrared spectra of WN stars at $R\approx1200$--2500.}}
\label{fig:wne_med}
\end{figure*}

\clearpage

\begin{figure*}[t]
\centering
\subfigure[]{
\includegraphics[scale=0.6]{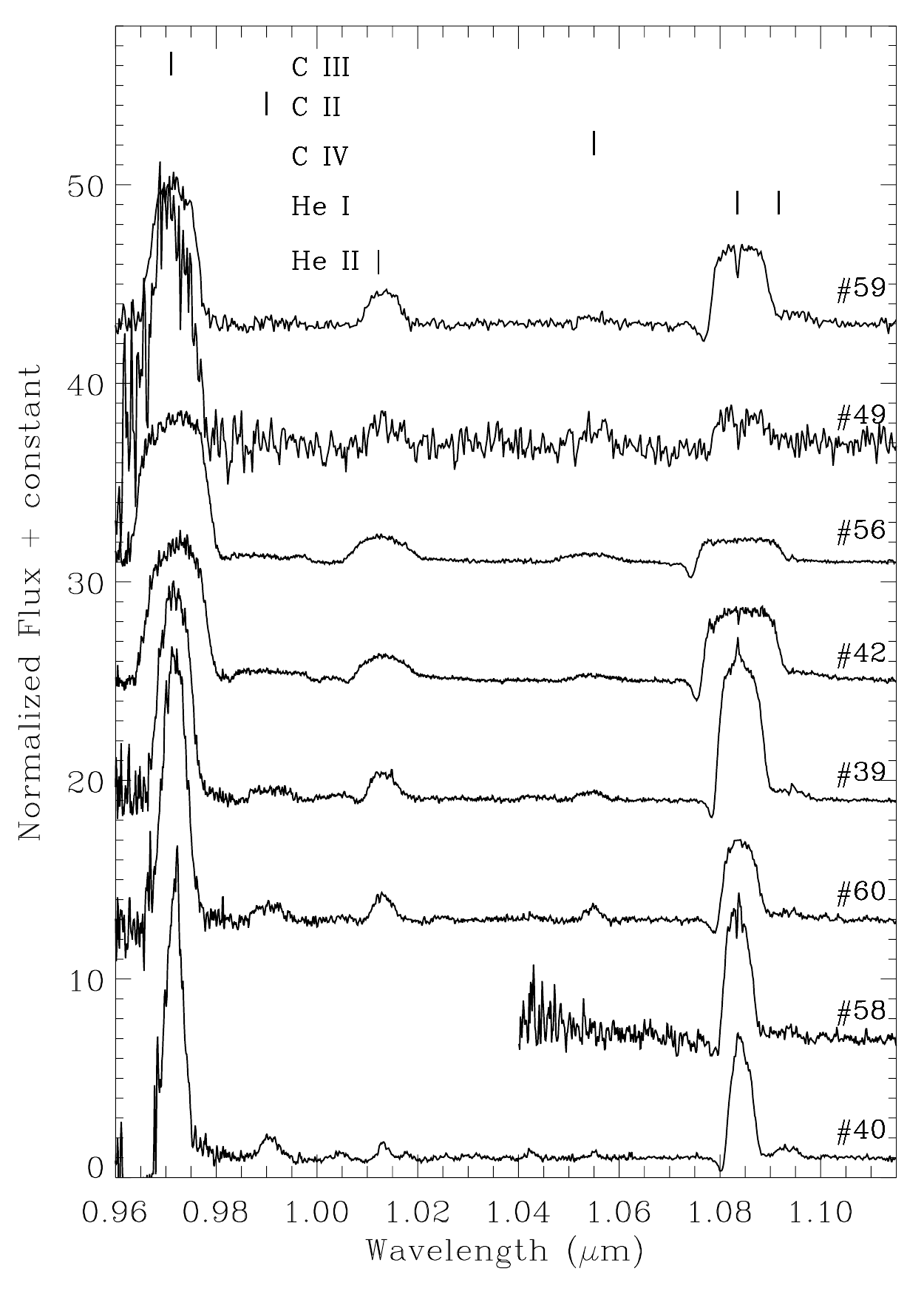}}
\subfigure[]{
\includegraphics[scale=0.6]{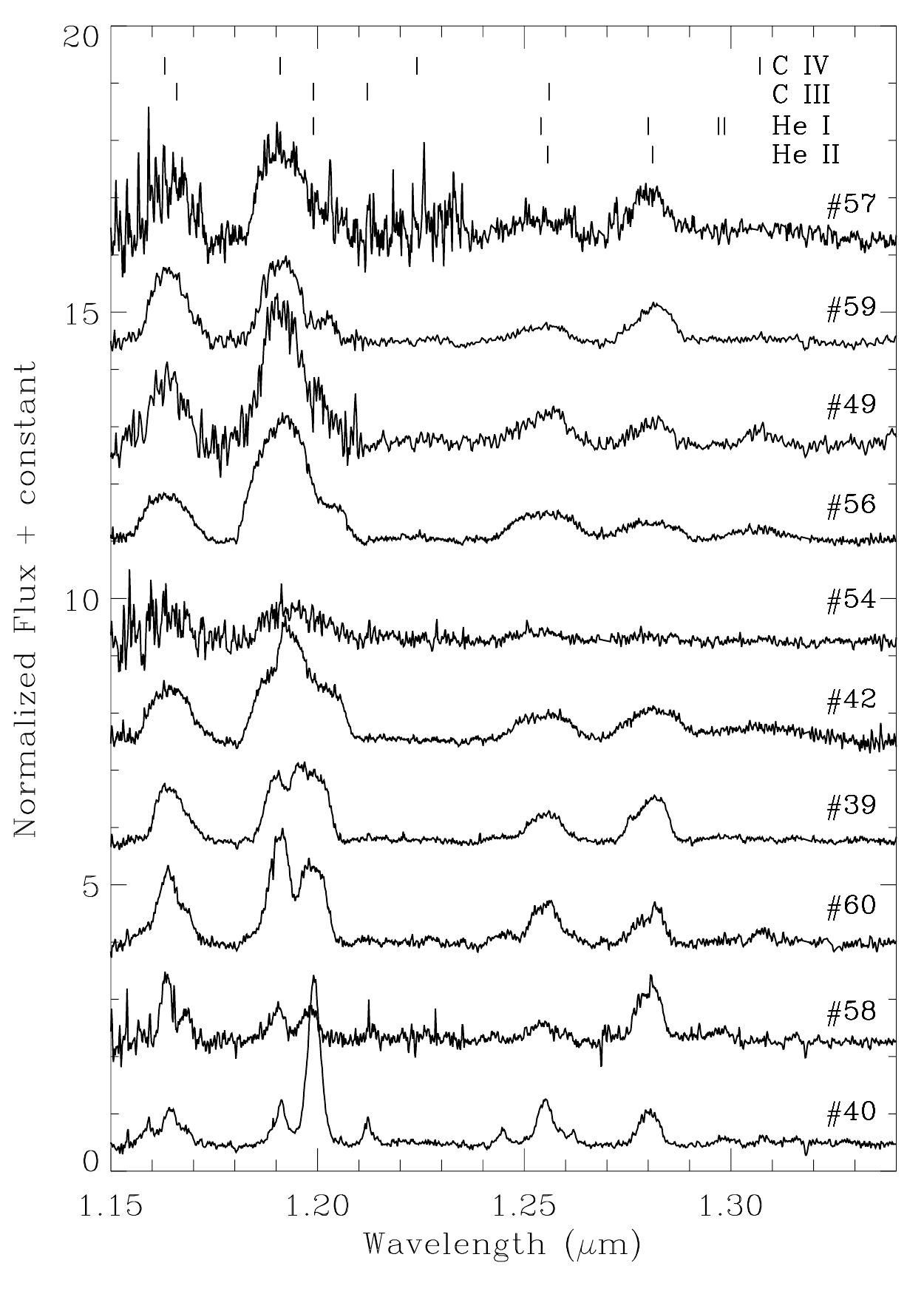}}
\subfigure[]{
\includegraphics[scale=0.6]{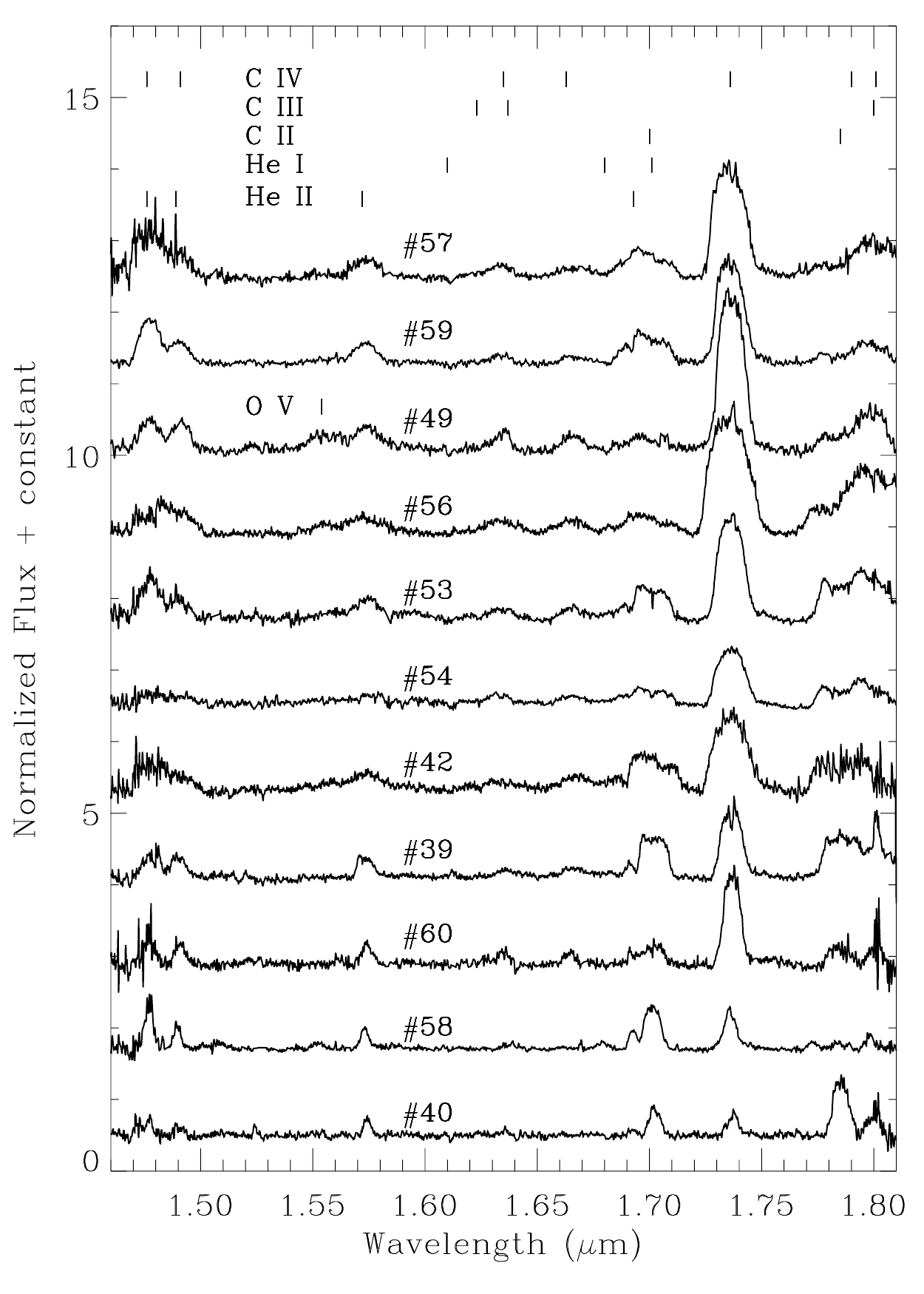}}
\subfigure[]{
\includegraphics[scale=0.6]{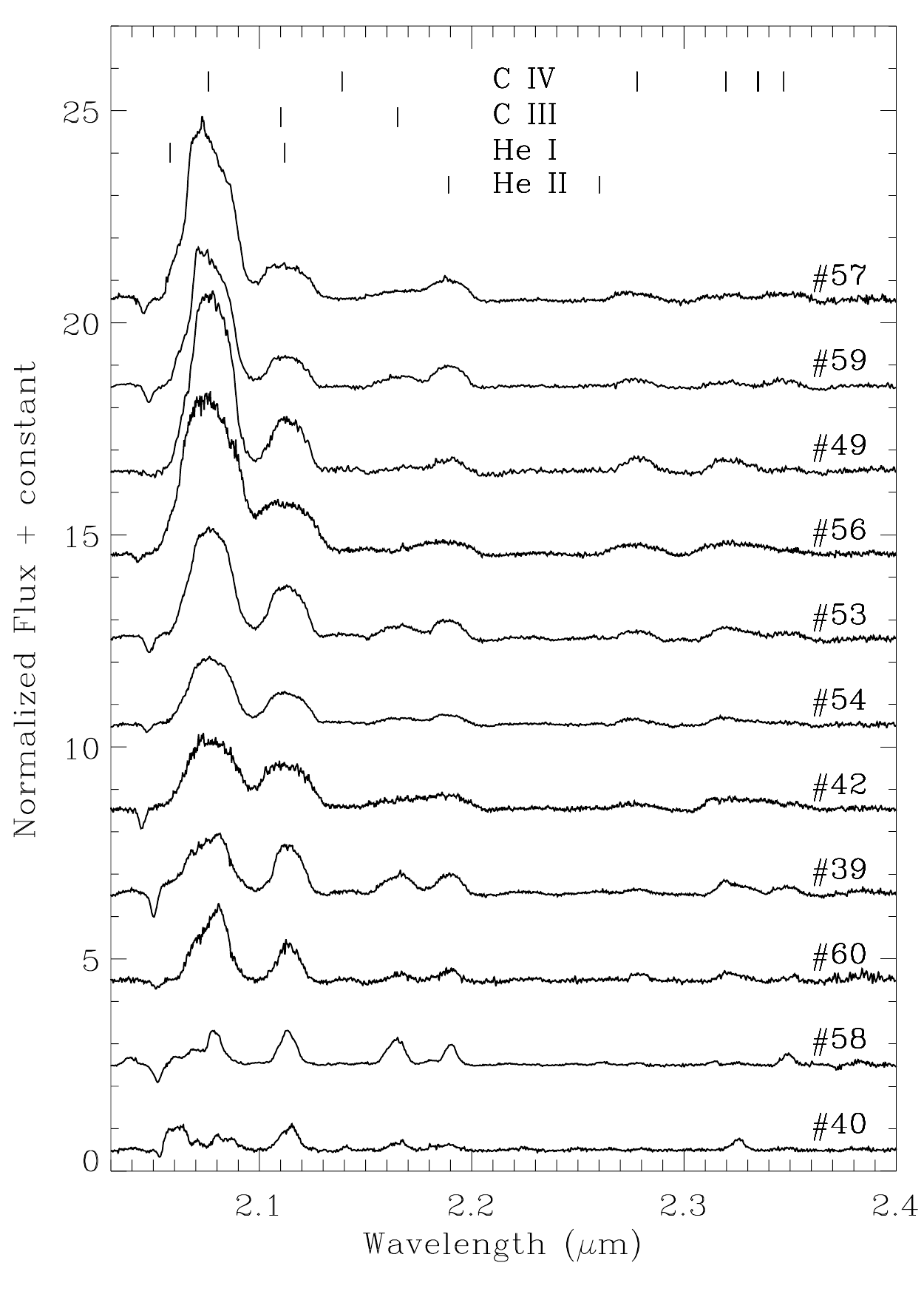}}
\caption[]{\linespread{1}\normalsize{Near-infrared spectra of WC stars at $R\approx1200$--2500.}}
\label{fig:wc_med}
\end{figure*}

\clearpage

\begin{figure*}[t]
\centering
\subfigure[]{
\includegraphics[scale=0.6]{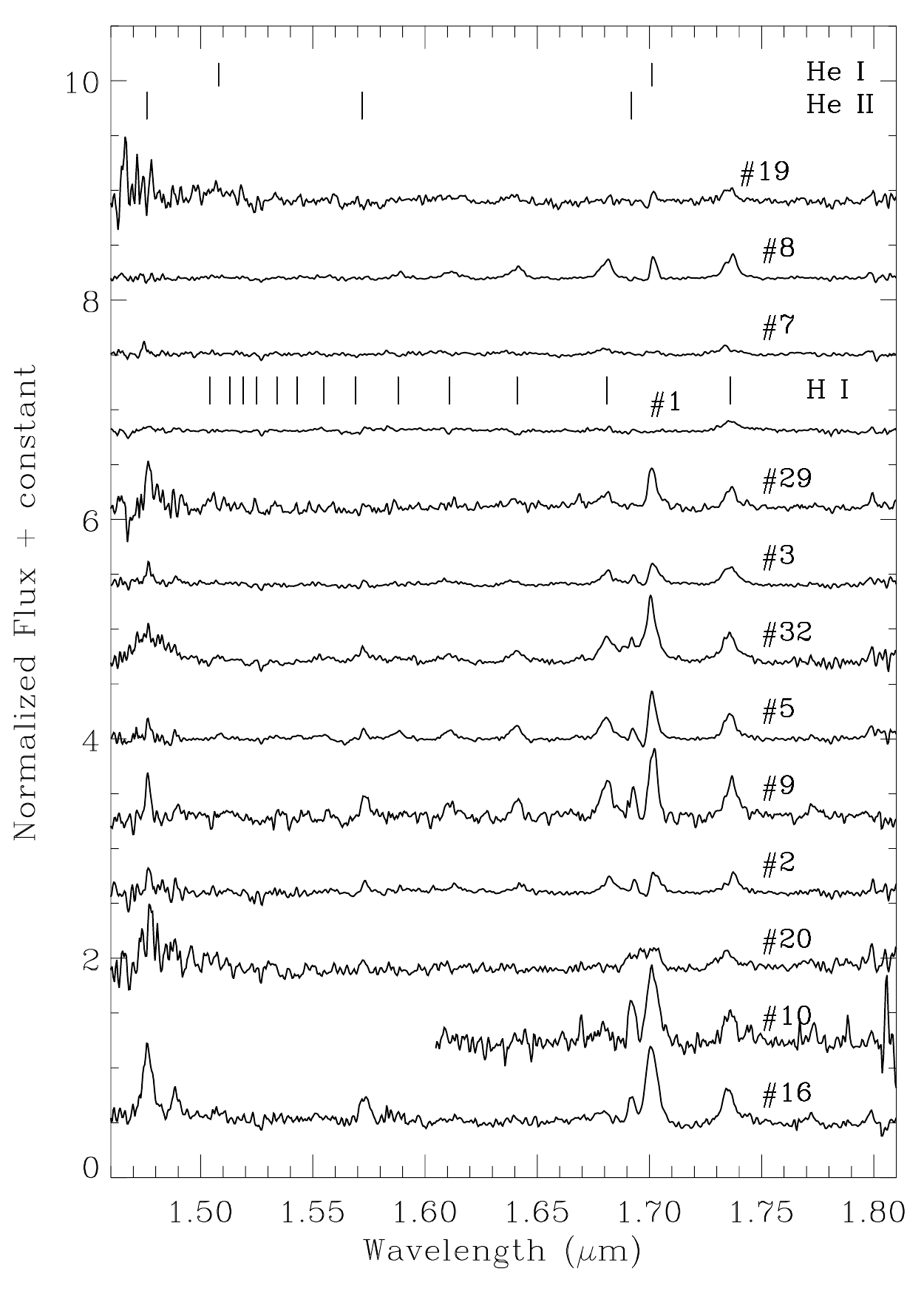}}
\subfigure[]{
\includegraphics[scale=0.6]{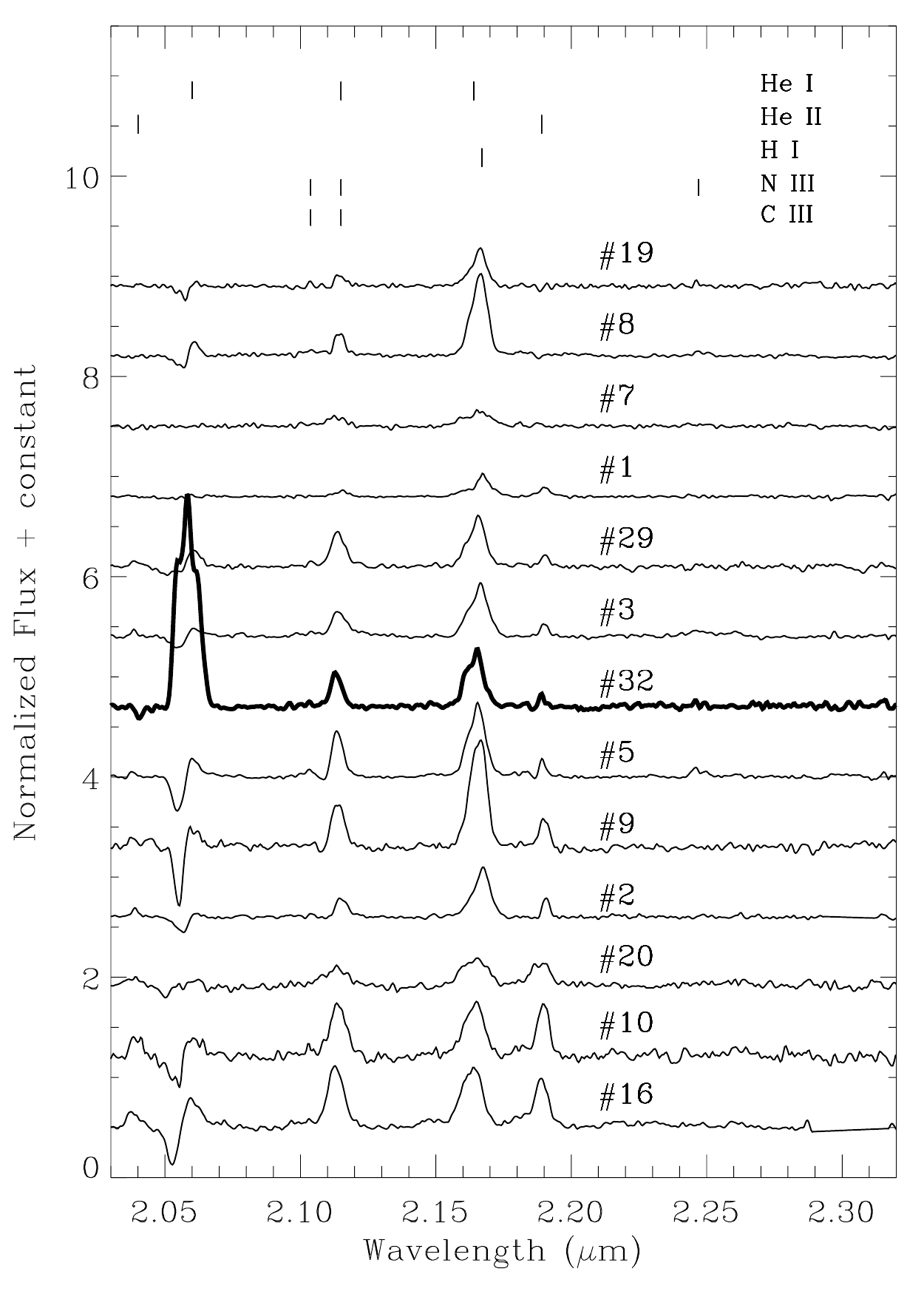}}
\caption[]{\linespread{1}\normalsize{$H$ and $K$ spectra of WN stars at $R\approx800$. Bold face is used for the $K$-band spectrum of MDM32, so as not to confuse the relatively strong He emission line near $\lambda$2.06 {\micron} from this star with the other spectra in the figure.}}
\label{fig:wnl_low}
\end{figure*}

\clearpage

\begin{figure*}[t]
\centering
\subfigure[]{
\includegraphics[scale=0.6]{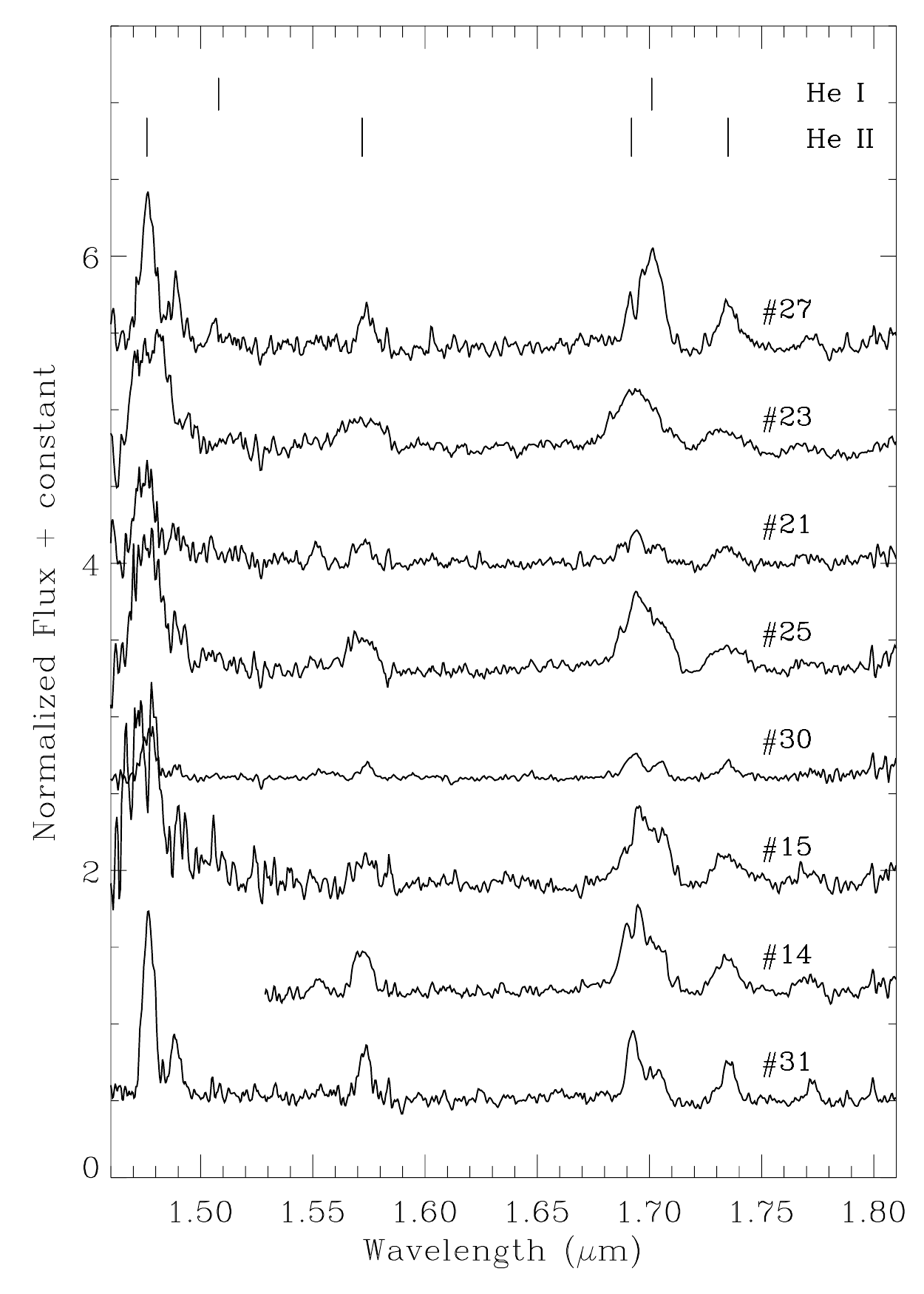}}
\subfigure[]{
\includegraphics[scale=0.6]{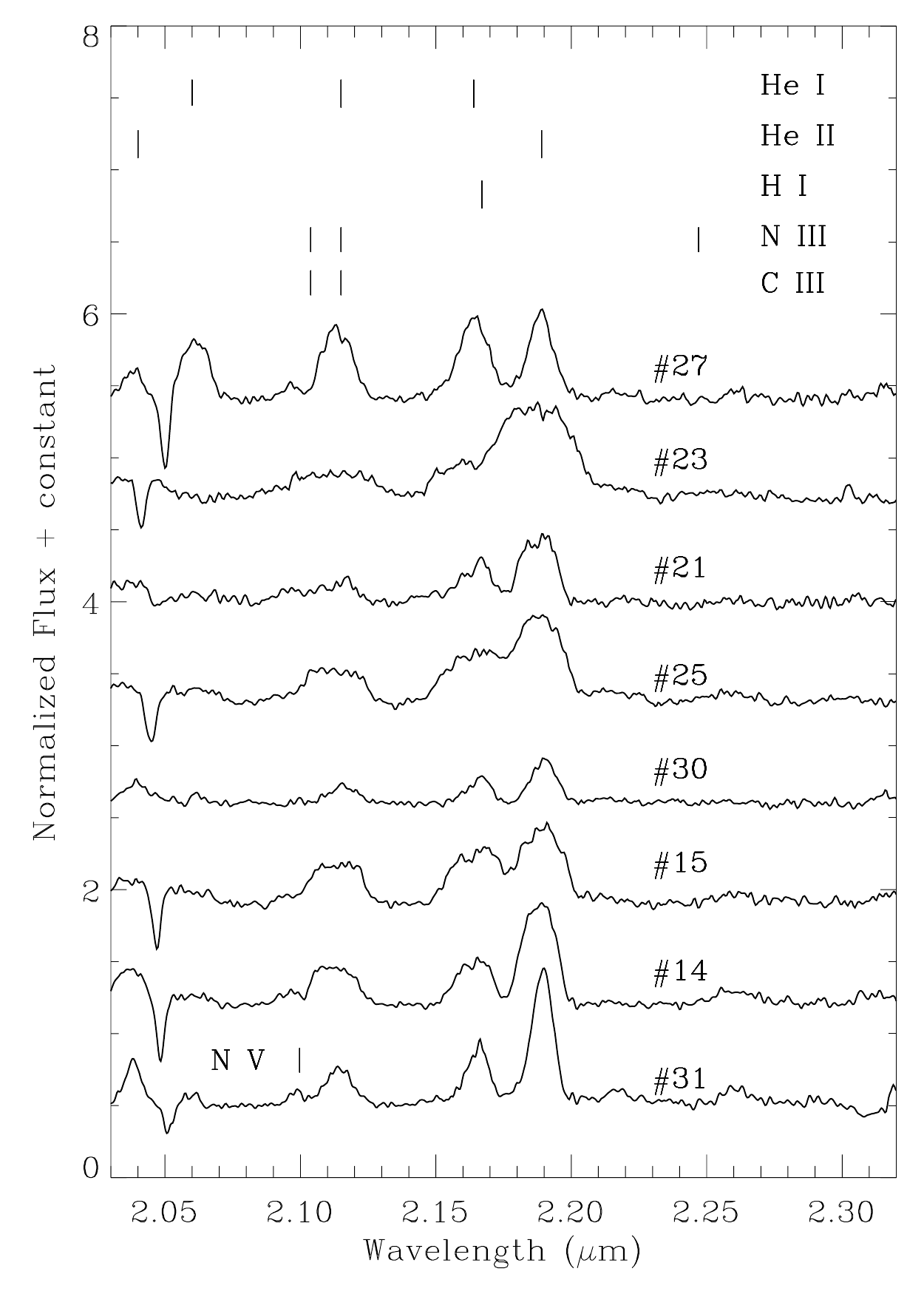}}
\caption[]{\linespread{1}\normalsize{$H$ and $K$ spectra of WN stars at $R\approx800$. }}
\label{fig:wne_low}
\end{figure*}

\clearpage

\begin{figure*}[t]
\centering
\subfigure[]{
\includegraphics[scale=0.6]{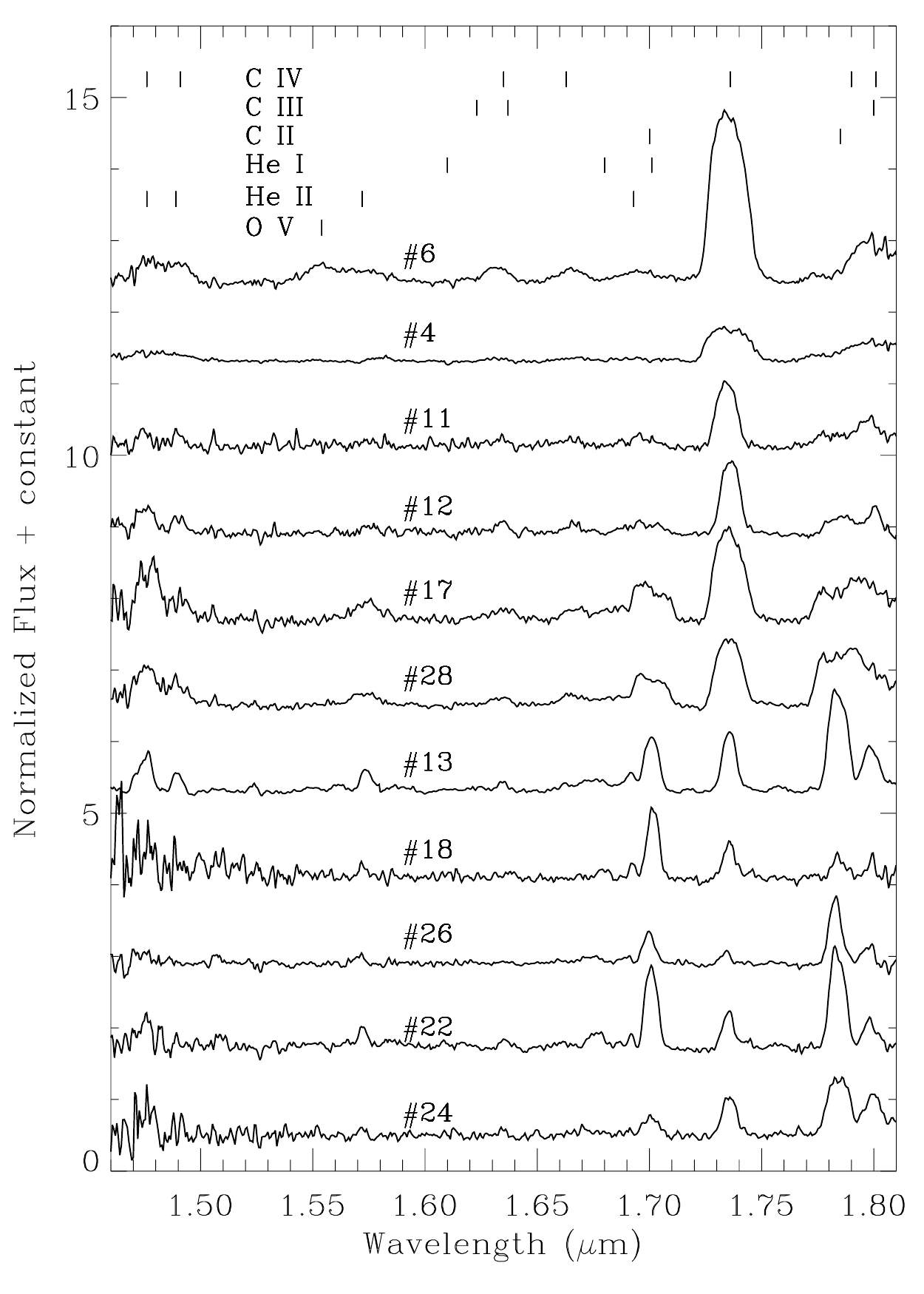}}
\subfigure[]{
\includegraphics[scale=0.6]{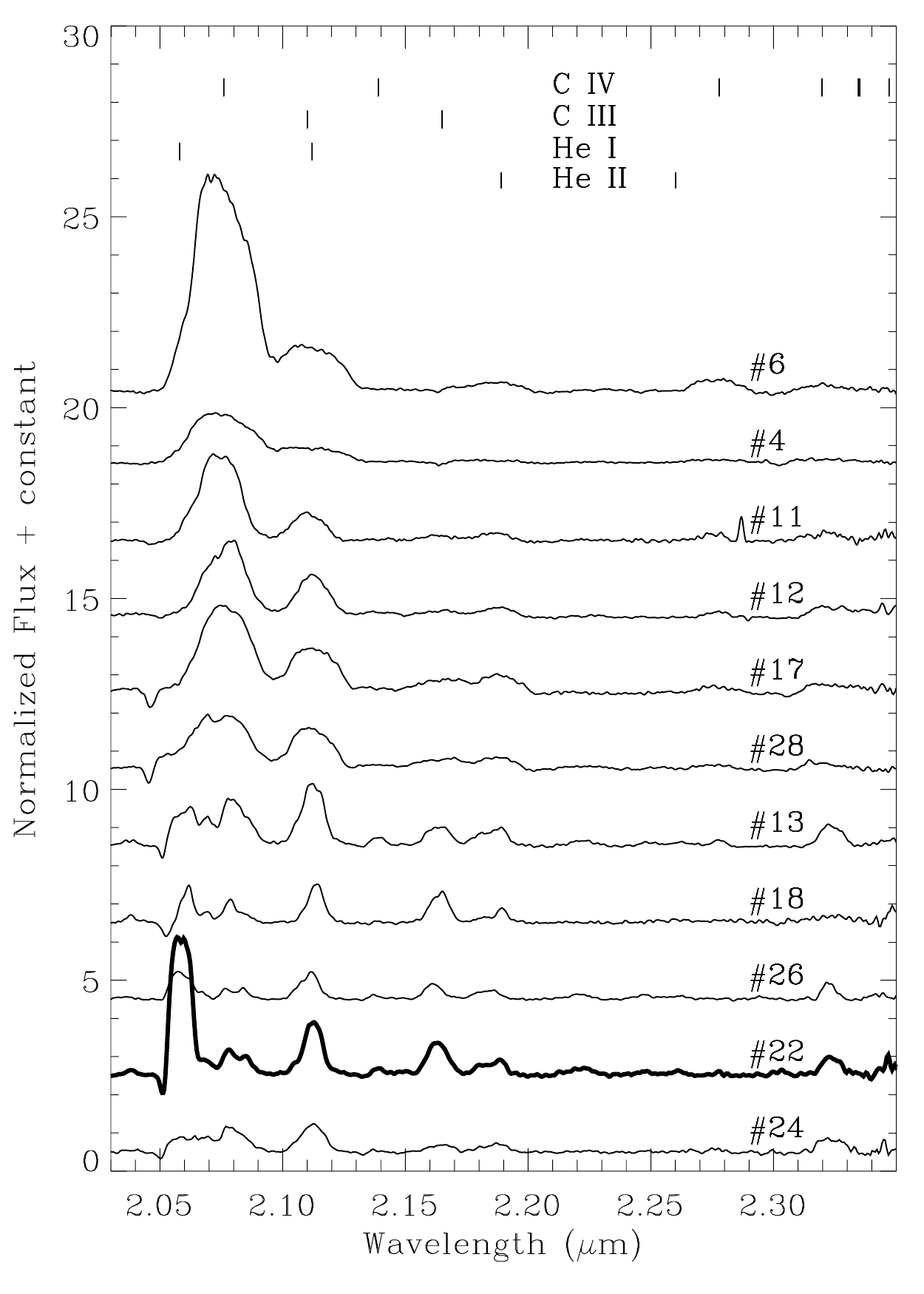}}
\caption[]{\linespread{1}\normalsize{$H$ and $K$ spectra of WC stars at $R\approx800$. Bold face is used for the $K$-band spectrum of MDM22, so as not to confuse the relatively strong emission line near $\lambda$2.06 {\micron} from this star with the other spectra in the figure.}}
\label{fig:wc_low}
\end{figure*}

\clearpage

\begin{figure*}[t]
\centering
\includegraphics[scale=0.95]{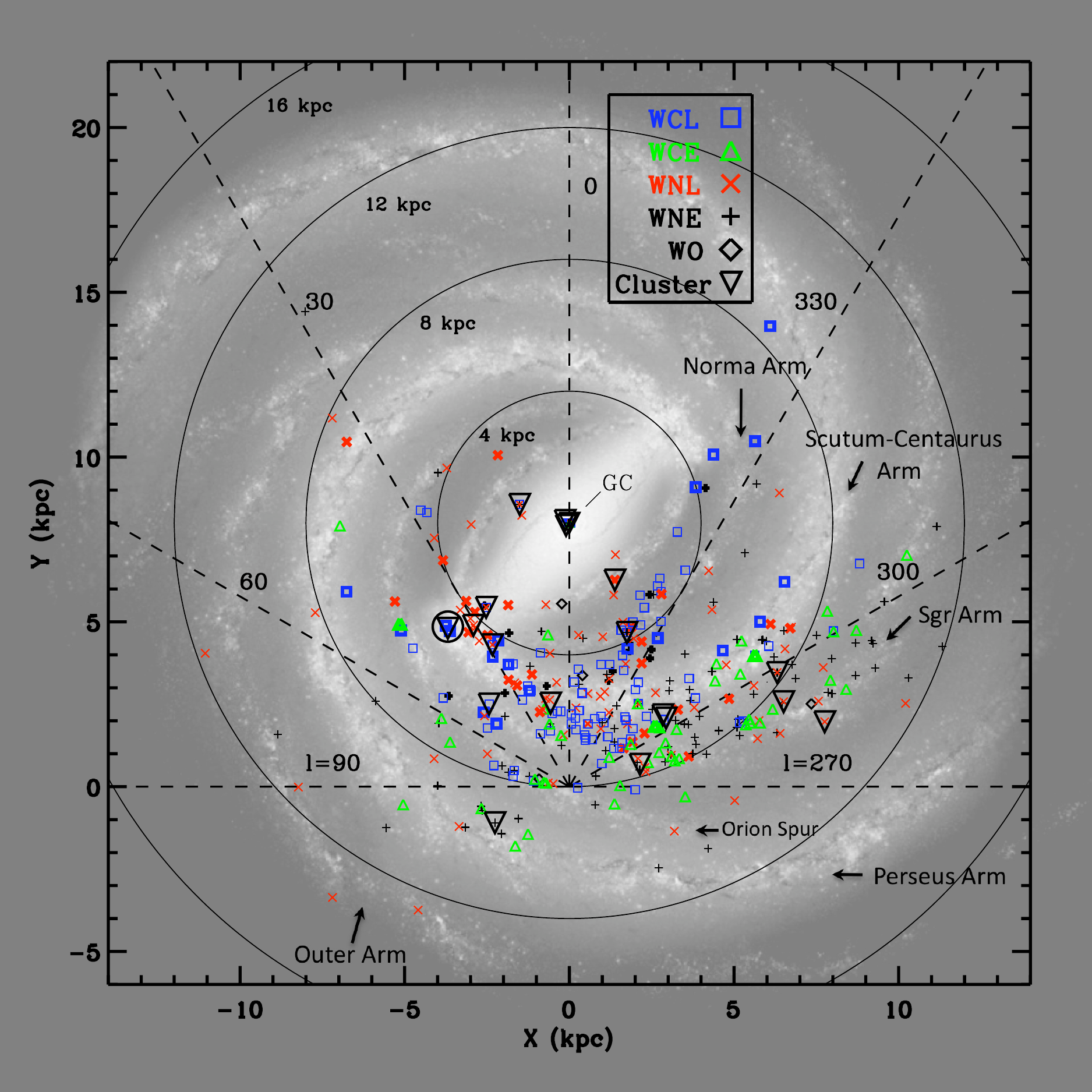}
\caption[]{\linespread{1}\normalsize{Galactic distribution of WRs and selected WR-bearing clusters plotted over the Galactic model of Churchwell et al. (2009). New identifications are marked as bold symbols. The figure axes are in units of kpc, with the location of the Sun at the origin. Dotted lines mark several values of constant Galactic longitude ($l$). The GLIMPSE survey area encompasses $\mid{l}\mid \le65^{\circ}$ on both sides of the Galactic center.  The following clusters are listed clockwise from left with respect to the Sun: Cas OB5 (Negueruela 2003) ;The Quartet cluster (Messineo et al. 2009);  G37.51$-$0.46 (this work; circled);  G28.46$+$0.32 (Mauerhan et al. 2010c); W43 (Blum et al. 1999); GLIMPSE20 (Messineo et al. 2009); [MFD2008] (Messineo et al. 2008); Sgr 1806$-$20 (Bibby et al. 2008); Quintuplet, Arches and Central parsec of the Galactic center; [DBS2003] 179 (Borissova et al. 2008);  Westerlund 1 (Crowther et al. 2006); Danks 1 and 2 (Bica et al. 2004); GLIMPSE30 (Kurtev et al. 2007); Trumpler 16 (Walborn 1973); NGC 3603 (Moffat 1983); and Westerlund 2 (Rauw et al. 2007). The Galactic center  contains $\approx$20\% of the known Galactic WRs. }}
\label{fig:wrdist}
\end{figure*}

\clearpage

\begin{figure*}[t]
\centering
\includegraphics[scale=0.7]{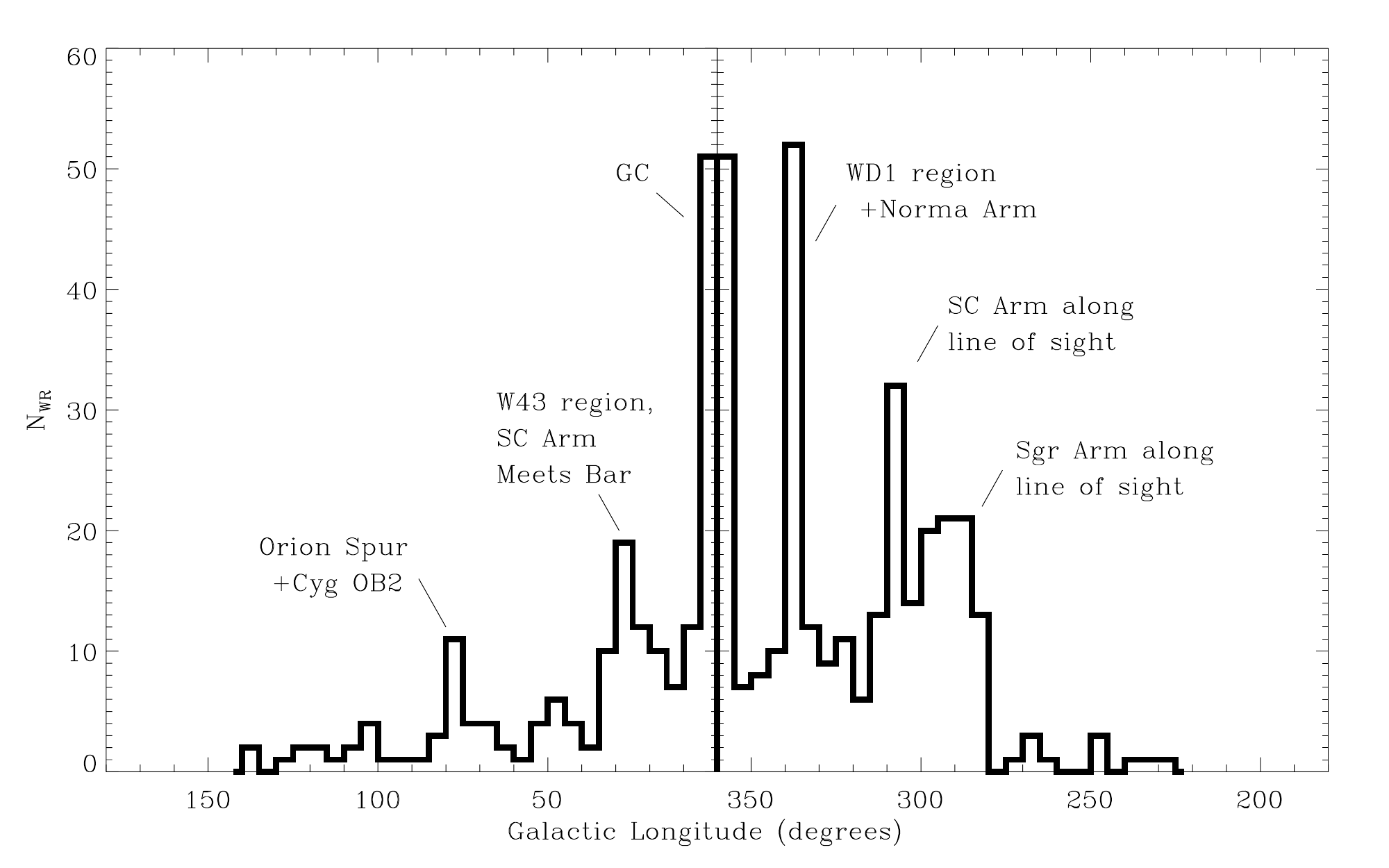}
\includegraphics[scale=0.7]{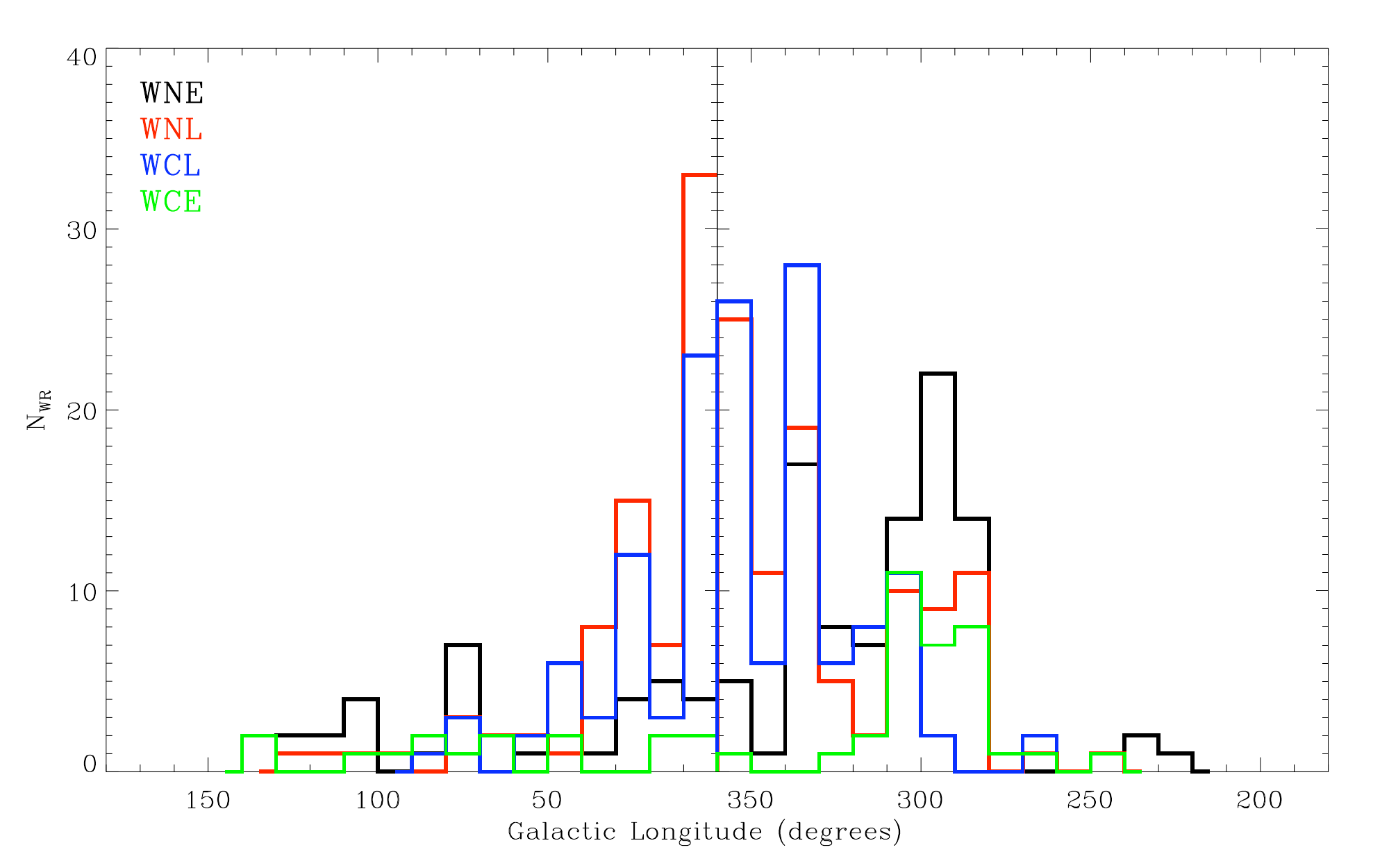}
\caption[]{\linespread{1}\normalsize{Galactic longitudinal distribution for known WRs included in Table \ref{tab:wr_census}. The histogram in the upper panel is for all WRs, including those outside of the GLIMPSE survey area, and has bins of 5$^{\circ}$ longitude. The histogram in the lower panel shows the subtype distributions for the same WRs, and has bins of 10$^{\circ}$ longitude.}}
\label{fig:lplot}
\end{figure*}

\clearpage

\begin{figure*}[t]
\centering
\includegraphics[scale=0.8]{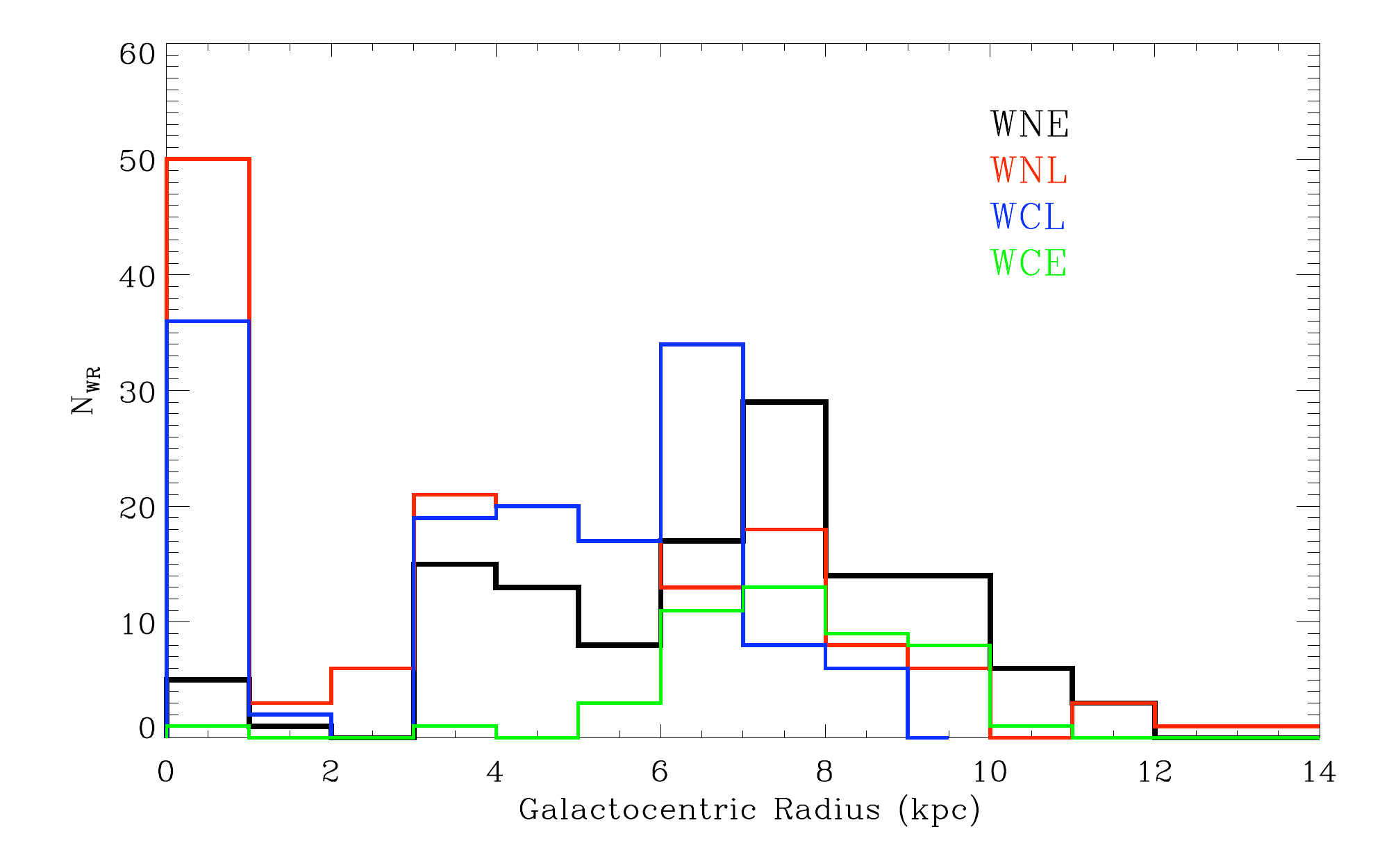}
\caption[]{\linespread{1}\normalsize{Galactocentric radial distribution of all known WRs included in  \ref{tab:wr_census}.}}
\label{fig:rg}
\end{figure*}

\clearpage

\begin{figure*}[t]
\centering
\includegraphics[scale=1]{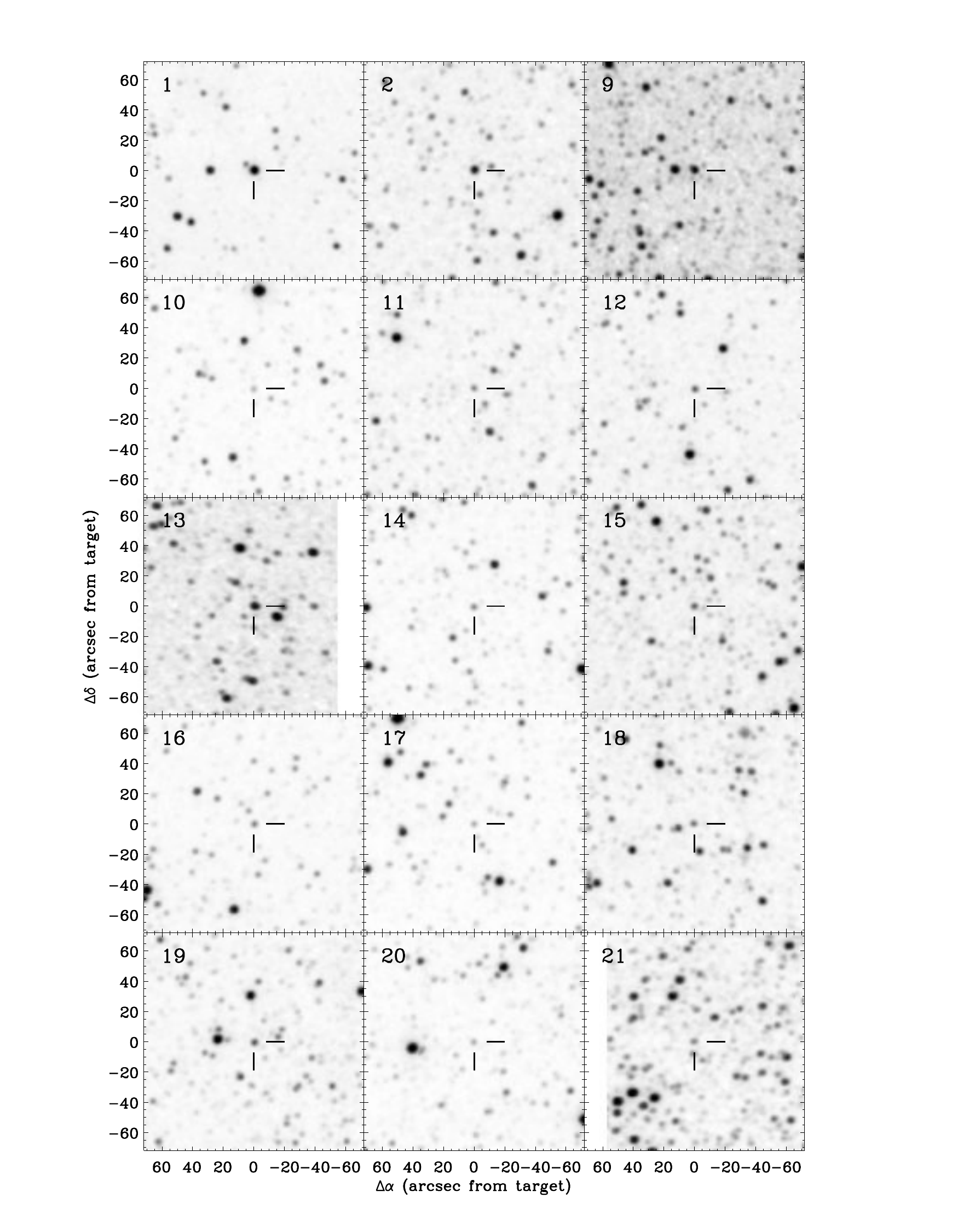}
\caption[]{\linespread{1}\normalsize{2MASS $K_s$-band field images of new WRs. The numbers at the upper left of the individual images represent the names introduced in Table 1 (e.g., frame 1 is for star MDM1). }}
\label{fig:k1}
\end{figure*}

\clearpage

\begin{figure*}[t]
\centering
\includegraphics[scale=1]{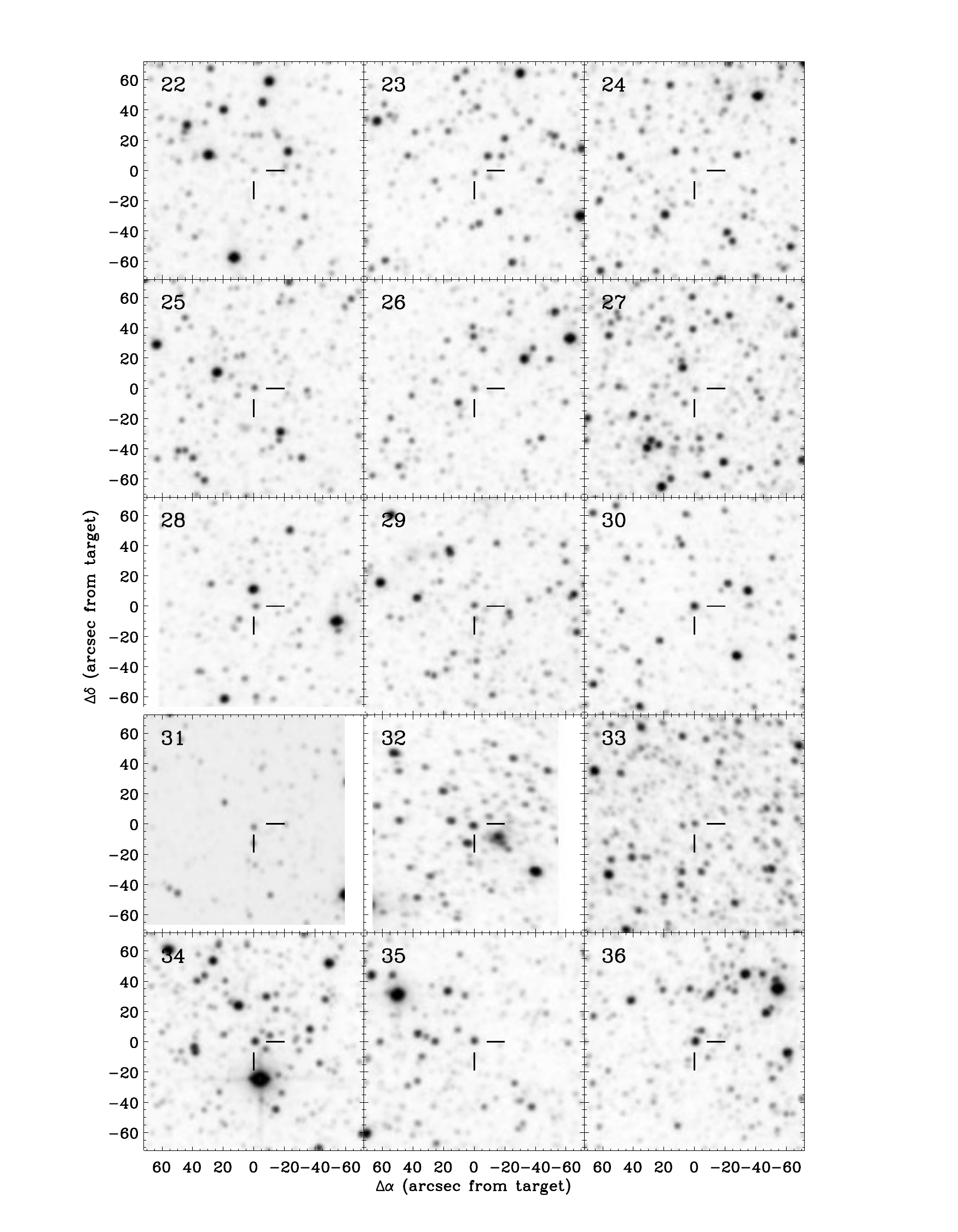}
\caption[]{\linespread{1}\normalsize{2MASS $K_s$-band field images of new WRs (con't).}}
\label{fig:k2}
\end{figure*}

\clearpage

\begin{figure*}[t]
\centering
\includegraphics[scale=1]{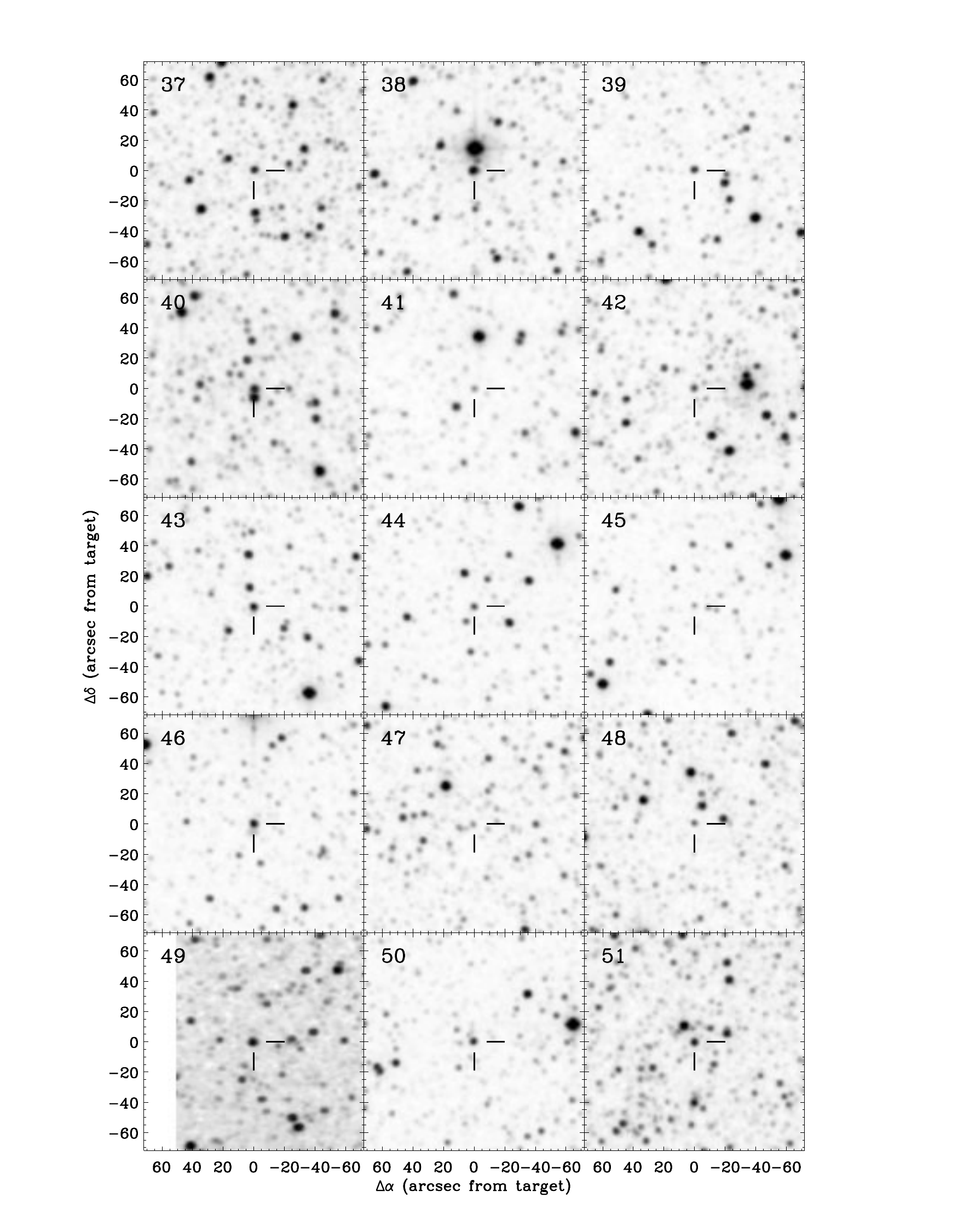}
\caption[]{\linespread{1}\normalsize{2MASS $K_s$-band field images of new WRs (con't). }}
\label{fig:k3}
\end{figure*}

\clearpage

\begin{figure*}[t]
\centering
\includegraphics[scale=1]{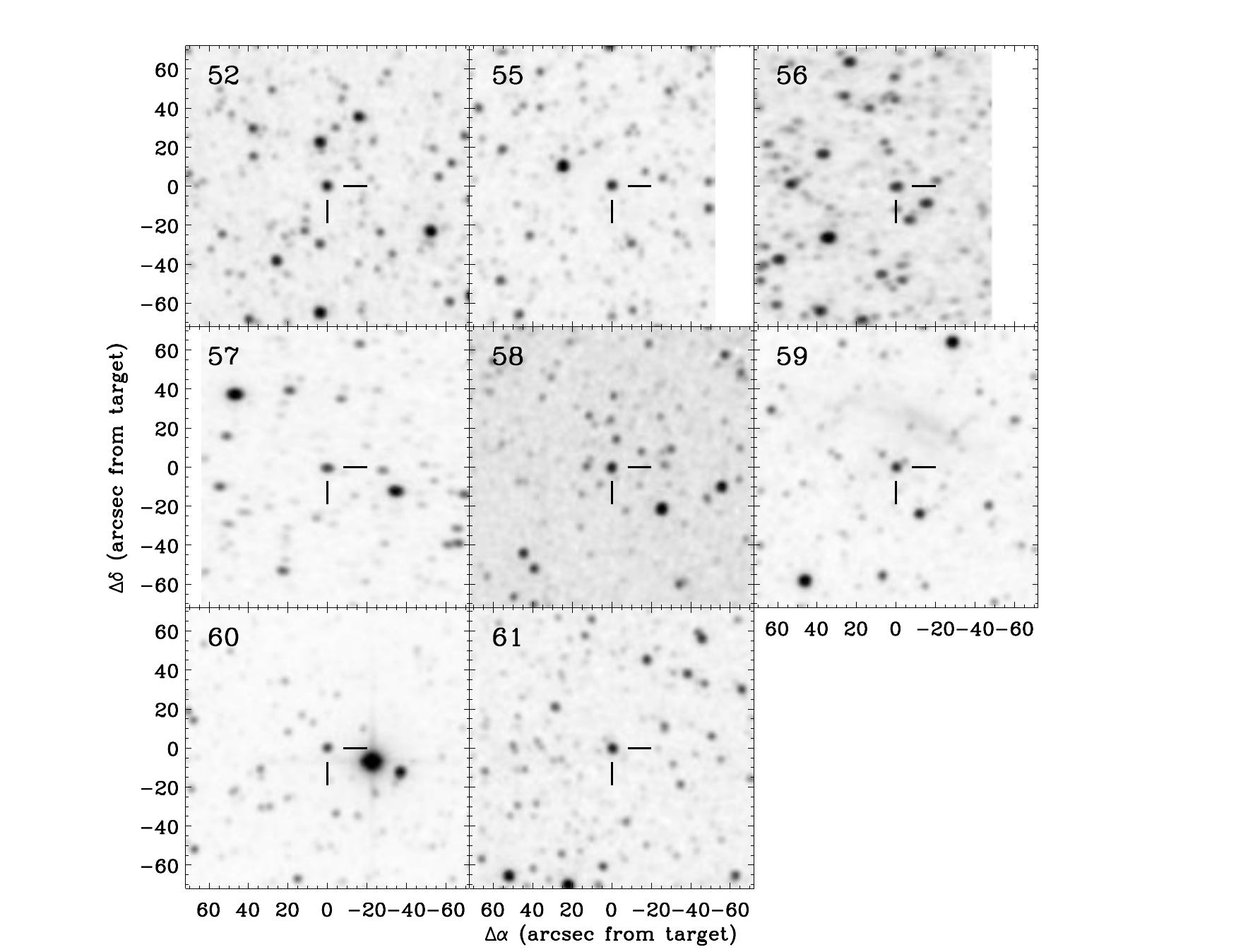}
\caption[]{\linespread{1}\normalsize{2MASS $K_s$-band field images of new WRs con't. Note the faint nebulosity surrounding the WC8 star MDM59, which is presumably wind-swept interstellar medium, as indicated by \textit{Spitzer}/IRAC mid-infrared images of this field (not presented here).}}
\label{fig:k4}
\end{figure*}

\clearpage

\begin{figure*}[t]
\centering
\includegraphics[scale=0.95]{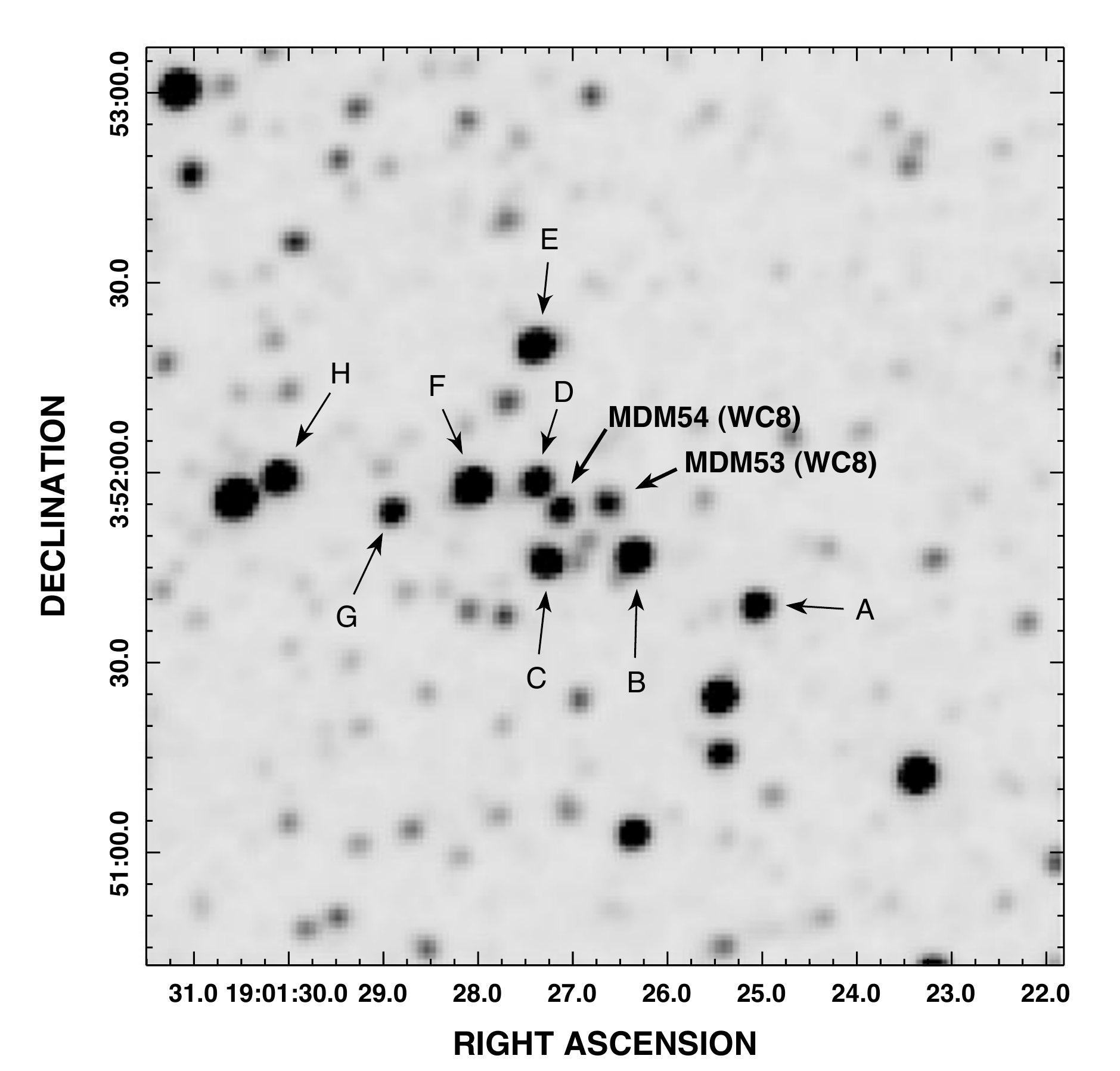}
\caption[]{\linespread{1}\normalsize{2MASS $K_s$-band field image of the cluster containing the WC8 stars MDM53 and MDM54, and the spectroscopically-confirmed, neighboring OB supergiants listed in Table \ref{tab:newclusterphot} (Stars A, B, D, F, and G; see spectra in Figure 18). Stars C, E, and H are late-type stars; their relatively low extinction values suggest that they lie in the foreground. }}
\label{fig:newcluster}
\end{figure*}

\clearpage

\begin{figure*}[t]
\centering
\includegraphics[scale=1]{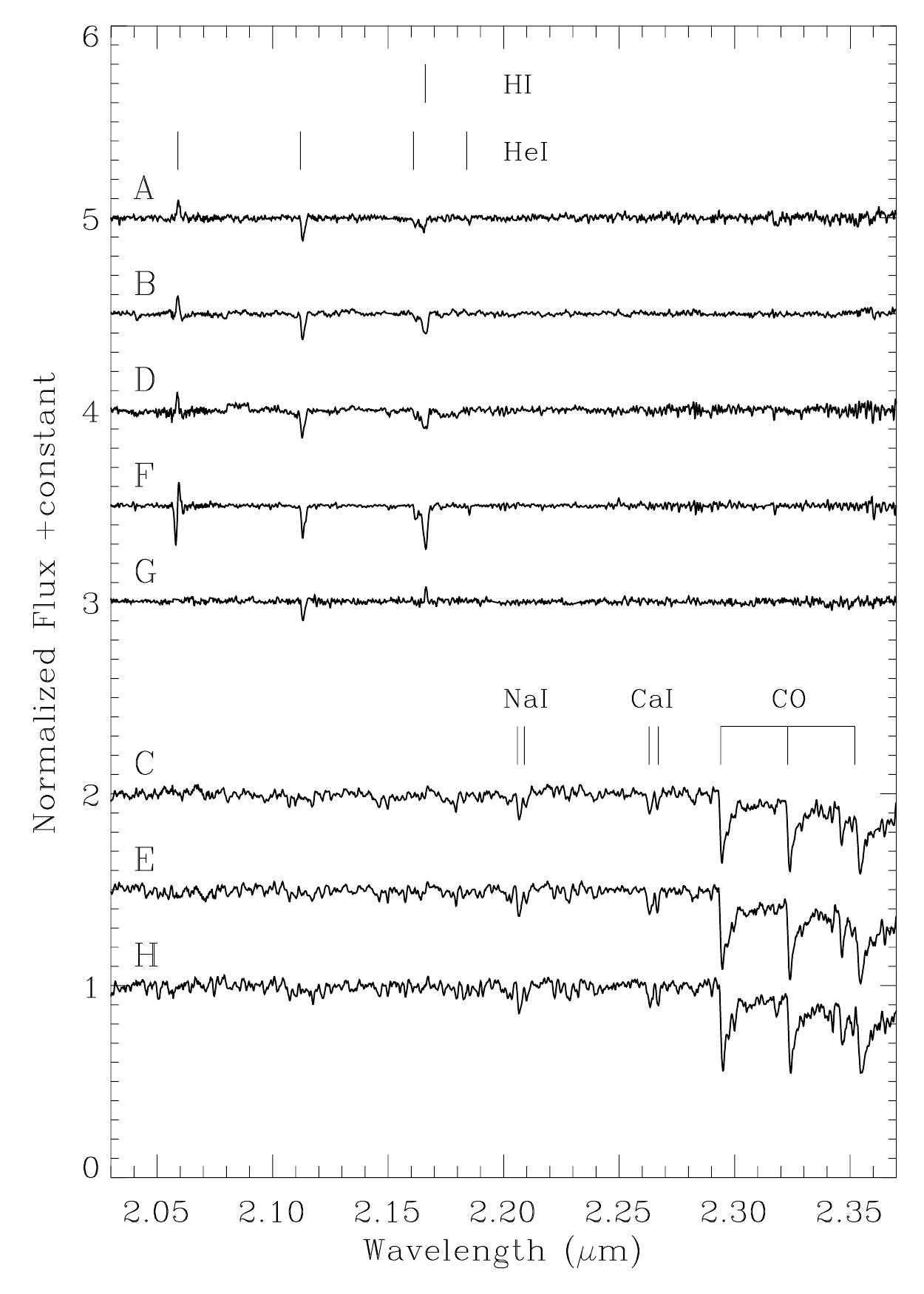}
\caption[]{\linespread{1}\normalsize{$K$-band spectra of selected stars in the vicinity of the clustered WC8 stars MDM53 and MDM54. Stars A, B, D, F, and G have spectral morphologies similar to known B0--3 I--II stars from Bibby et al. (2008). The spectra of stars C, E, and H exhibit features of late-type M stars; their relatively low extinction values imply that they lie in front of the cluster. }}
\label{fig:newclusterspec}
\end{figure*}

\clearpage

\begin{figure*}[t]
\centering
\includegraphics[scale=0.95]{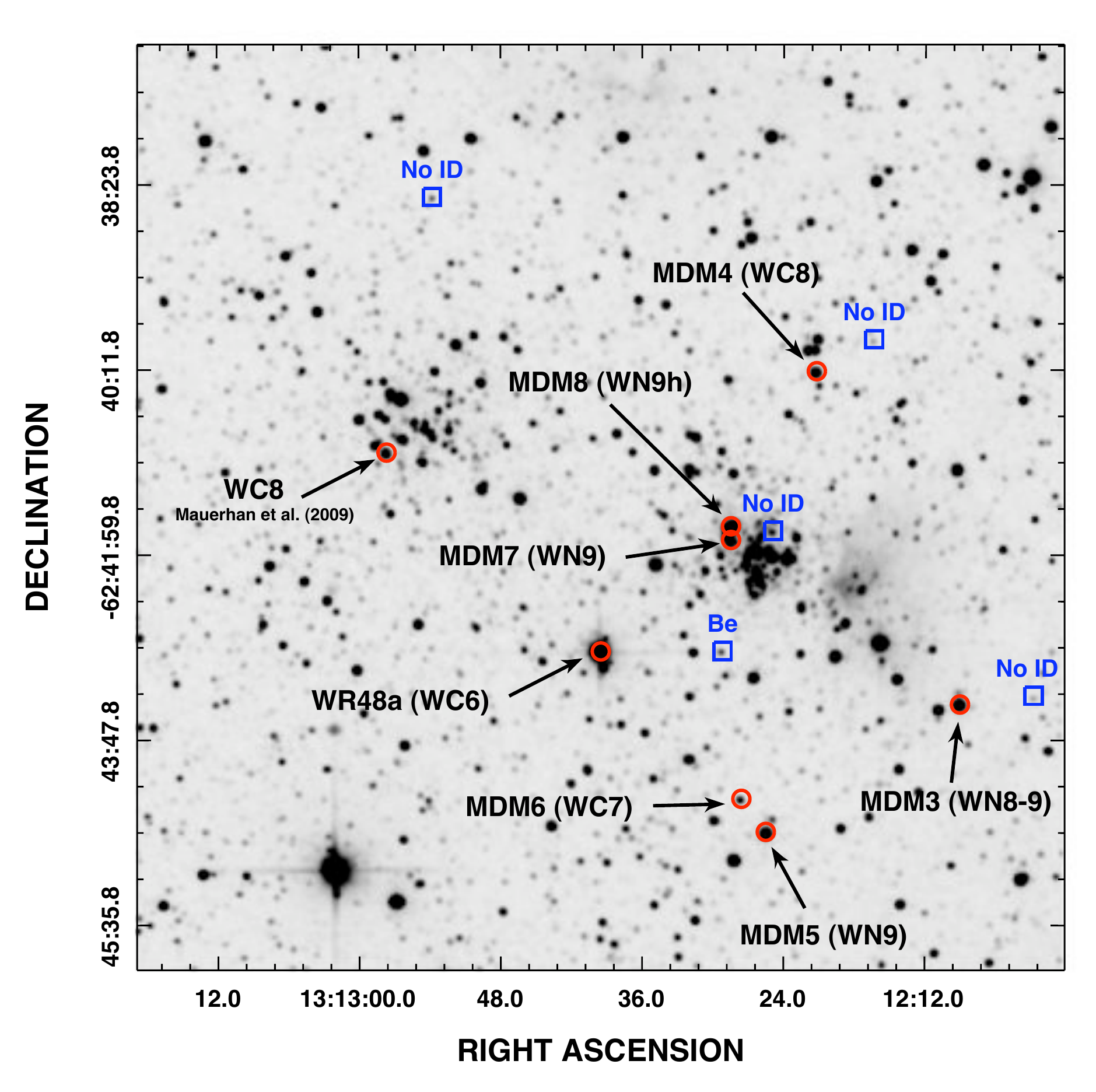}
\caption[]{\linespread{1}\normalsize{2MASS $K_s$-band image of the field containing the clusters Danks 1 (right) and 2 (left). The newly identified WRs are marked, along with the two previously identified WRs in the region. Unidentified WR candidates in the field are also marked, although they are significantly fainter than confirmed WRs associated with the clusters.}}
\label{fig:danks}
\end{figure*}

\clearpage

\begin{appendix}
\begin{figure*}[h]
\setcounter{figure}{0}
\renewcommand{\thefigure}{\Alph{figure}.\arabic{figure}}
\centering
\includegraphics[scale=0.8]{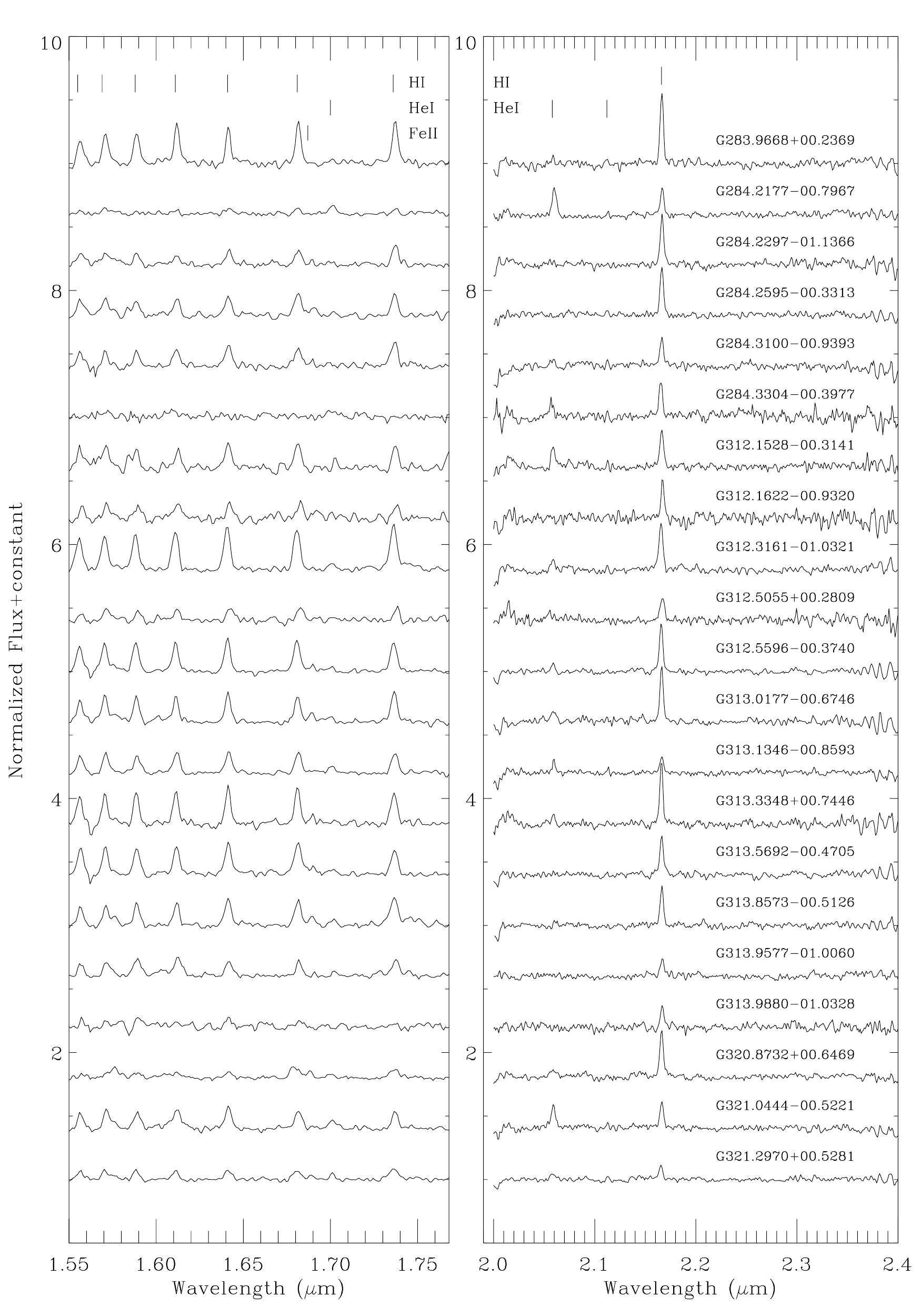}
\caption[]{\linespread{1}\normalsize{Example $H$ and $K$ spectra of non-WR ``contaminants" that have selected by our method, which are dominated by a series of H {\sc i} emission lines. We conclude these are Be stars, 
based on the similarities with the Be spectral atlases of Clark \& Steele (2000) and Steele \& Clark (2001). Be stars mimic WRs in infrared color space, owing to similarities between the physical parameters of the 
free-free emitting Be disks and WR winds (see text).}}
\label{fig:be}
\end{figure*}

\end{appendix}

\end{document}